\documentclass{article}
\usepackage{epsfig}
\usepackage{amssymb}
\usepackage{verbatim}
\def\ds{\displaystyle}

\def\un{\underline}
\def\res{\mathop{\mathrm {res}}\limits_}

\renewcommand{\theequation}{\arabic{section}.\arabic{subsection}.\arabic{equation}}
\makeatletter
\@addtoreset{equation}{subsection}
\newtheorem{theorem}{Theorem}[section]
\newtheorem{examp}{Example}[section]
\newtheorem{coroll}{Corollary}[section]
\newtheorem{examps}{Examples}[section]

\newtheorem{lemma}{Lemma}[section]
\newtheorem{remark}{Remark}[section]
\def\mod{\,\hbox{mod}\,}
\newtheorem{remarks}[remark]{Remarks}
\newtheorem{proposition}{Proposition}[section] 
\newtheorem{definition}{Definition}[section]

\def\le{\left}
\def\m{\mathop}

\def\Amat{\mathop{\mathbb A}}
\def\ri{\right}
\def\br{\begin{remark}}
\def\1{{\bf 1}}
\def\er{\end{remark}}
\def\bt{\begin{theorem}}
\def\et{\end{theorem}}
\def\bc{\begin{coroll}}
\def\ec{\end{coroll}}
\def\brs{\begin{remarks}.\\ \rm\small\begin{enumerate}}
\def\ers{\end{enumerate}\end{remarks}}
\def\bx{\begin{examp}\small}
\def\ex{\end{examp}}
\def\bl{\begin{lemma}}
\def\el{\end{lemma}}
\def\bxs{\begin{examps}. \rm\begin{enumerate}}
\def\exs{\end{enumerate}\end{examps}}
\def\bd{\begin{definition}}
\def\ed{\end{definition}}
\def\bp{\begin{proposition}}
\def\ep{\end{proposition}}
\def\be{\begin{equation}}
\def\ee{\end{equation}}

\def\bea{\begin{eqnarray}}
\def\eea{\end{eqnarray}}
\def\beas{\begin{eqnarray*}}
\def\eeas{\end{eqnarray*}}
\def \pa{\partial}
\def\C{{\mathbb C}}
\def\A{\mathop{\mathbf a}}
\def\B{\mathop{\mathbf b}}

\def\R{{\mathbb R}}
\def\N{{\mathbb N}}

\def\Z{{\mathbb Z}}
\def\a{{\alpha}}
\def\b{{\beta}}
\date{}
\begin{document}
\baselineskip 16pt plus 1pt minus 1pt
\begin{titlepage}
\begin{flushright}
CRM-2852 (2002)\\
\hfill Saclay-T02/097\\
nlin.SI/0208002
\end{flushright}
\vspace{0.2cm}
\begin{center}
\begin{Large}\fontfamily{cmss}
\fontsize{17pt}{27pt}
\selectfont
\textbf{Differential systems for biorthogonal polynomials\\[-1pt]
appearing in 2-matrix models and the associated \\[8pt] 
Riemann--Hilbert problem
}\footnote{
Work supported in part by the Natural Sciences and Engineering Research Council
of Canada (NSERC) and the Fonds FCAR du Qu\'ebec.}
\end{Large}\\
\vspace{1.0cm}
\begin{large} {M.
Bertola}$^{\dagger\ddagger}$\footnote{bertola@crm.umontreal.ca}, 
 { B. Eynard}$^{\dagger
\star}$\footnote{eynard@spht.saclay.cea.fr}
 and {J. Harnad}$^{\dagger \ddagger}$\footnote{harnad@crm.umontreal.ca}
\end{large}
\\
\bigskip
$^{\dagger}$ {\em Centre de recherches math\'ematiques,
Universit\'e de Montr\'eal\\ C.~P.~6128, succ. centre ville, Montr\'eal,
Qu\'ebec, Canada H3C 3J7} \\
\smallskip
$^{\ddagger}$ {\em Department of Mathematics and
Statistics, Concordia University\\ 7141 Sherbrooke W., Montr\'eal, Qu\'ebec,
Canada H4B 1R6} \\ 
\smallskip
$^{\star}$ {\em Service de Physique Th\'eorique, CEA/Saclay \\ Orme des
Merisiers F-91191 Gif-sur-Yvette Cedex, FRANCE } \\
\bigskip
\bigskip
{\bf Abstract}
\end{center}
\hrule\vskip 10pt
We consider biorthogonal polynomials that arise in the study of
a generalization of two--matrix Hermitian models with two 
polynomial potentials $V_1(x)$, $V_2(y)$ of any degree, with arbitrary 
complex coefficients.
Finite consecutive subsequences of biorthogonal polynomials (``windows''), of lengths
equal to the degrees of the potentials $V_1$ and $V_2$,
 satisfy systems of ODE's with polynomial coefficients as well as
PDE's (deformation equations) 
with respect to the coefficients of the
potentials and recursion relations connecting consecutive windows. 
A compatible sequence of fundamental systems of solutions is constructed for
these equations. The (Stokes) sectorial asymptotics of these fundamental
systems are derived through saddle-point integration and  the Riemann-Hilbert
problem characterizing the differential equations is deduced. 

\vskip 8pt
\hrule
\end{titlepage}


\section{Introduction}

In \cite{BEH, Needs} the differential systems satisfied by sequences
of biorthogonal polynomials associated to $2$-matrix models were
studied, together with the deformations induced by changes in the
coefficients of the potentials determining the orthogonality measure. For
ensembles consisting  of pairs of $N\times N$ hermitian matrices $M_1$ and
$M_2$, the  $U(N)$  invariant probability measure is taken to be of the form:
\be
{1 \over \tau_N}d\mu(M_1,M_2):= {1\over \tau_N} \exp{\frac 1 \hbar {\rm tr}\,
\left( -V_1(M_1) - V_2(M_2) + M_1 M_2 \right)} dM_1 dM_2  \ .
\label{twomatrixmeas}
\ee
where $dM_1 dM_2$ is the standard Lebesgue measure for pairs of Hermitian
matrices and the {\em potentials} $V_1$ and $V_2$ are chosen to be 
polynomials of degrees  $d_1+1$, $d_2+1$ respectively, with real coefficients.
The overall positive  small parameter $\hbar$ in the exponential is  
taken of order $N^{-1}$ when considering the large $N$ limit, but in 
the present context it will just play the role of Planck's constant in
the string equation.  

Using the Harish-Chandra-Itzykson-Zuber formula, one can reduce the 
computation of the corresponding partition function to an integral 
over only the eigenvalues of the two matrices 
\be 
\tau_N:= \int\int d\mu(M_1,M_2) \propto \int\prod_{i=1}^N{\rm 
d}x_i{\rm d}y_i \Delta(x)\Delta(y){\rm e}^{-\frac 1\hbar\sum_{j=1}^{N} 
V_1(x_j)+V_2(y_j) -x_jy_j}\ , 
\ee 
and then express all spectral statistics in terms of the associated
biorthogonal polynomials, in the same spirit as orthogonal polynomials are used 
in the spectral statistics of one-matrix models \cite{Mehta}. 
In this context, what is meant by biorthogonal polynomials is a pair of 
sequences of monic polynomials 
\be 
\pi_n(x) = x^n + \cdots , \qquad \sigma_n(y)=y^n + \cdots, \qquad n\in \N 
\ee 
which are mutually dual with respect to the associated coupled measure
 on the product space
\be 
\int_\R\!\!\int_\R\!\!\! {\rm d}x\, {\rm d}y \,\, \pi_n(x)\sigma_m(y) {\rm 
e}^{-\frac 1\hbar( V_1(x)+ V_2(y) -xy)} = h_n\delta_{mn} . 
\ee

 In this work, we use essentially the same definition of 
 orthogonality, but extend it to the case of polynomials $V_1$ and $V_2$ with 
 arbitrary (possibly complex) coefficients, with the contours of integration 
  no longer restricted to  the real axis, but taken as curves in the
  complex plane starting and ending at $\infty$, chosen so that the integrals 
  are convergent.  The orthogonality relations determine the two families 
  uniquely, if they exist \cite{McLaughlin, Needs}.
 
 It was shown in  \cite{BEH, Needs} that the finite consecutive subsequences
 of lengths $d_2 +1$ and $d_1+1$ respectively, within the sequences of dual 
 quasi-polynomials:
 \bea
&& \psi_n(x) = \frac 1{\sqrt{h_n}} \pi_n(x){\rm e}^{-\frac 1 \hbar V_1(x)} \ ,
\ \  \phi_n(y) = \frac 1{\sqrt{h_n}} \sigma_n(y) {\rm e}^{-\frac 1
\hbar V_2(y)}\ ,
\eea
beginning (or ending) at the points $n=N$, satisfy compatible overdetermined 
systems of first order differential equations with polynomial coefficients
of degrees $d_1$ and $d_2$, respectively, as well as recursion relations
for consecutive values of $N$.  In fact, certain quadruples of 
Differential--Deformation--Difference equations
 (DDD for short) were derived for these ``windows'', as well as for their 
 Fourier Laplace transforms, in which the deformation parameters were taken to
 be the coefficients of the potentials $V_1$ and $V_2$. 
It was shown in  \cite{BEH, Needs} that these systems are Frobenius 
compatible and hence admit joint fundamental systems of solutions.

In the present work we  explicitly construct such fundamental systems in terms 
of certain integral transforms applied to the biorthogonal 
polynomials. The main purpose is to derive the  Riemann--Hilbert problem 
characterizing the sectorial asymptotic behavior at $x=\infty$ or $y=\infty$.

The ultimate purpose of this analysis is to deduce in a rigorous way the 
double--scaling limits $N\to \infty,\ \hbar N=\mathcal O(1)$ of the partition 
function and spectral statistics, for which the corresponding large $N$ 
asymptotics of the biorthogonal polynomials are required.
(See \cite{ZJDFG,Kazakov} and references therein for further background on 
$2$-matrix models, and \cite{KazakoVDK, eynard, eynardchain, eynardmehta, 
McLaughlin} for other more recent developments.) 
The study of the large $N$ limit of matrix integrals is of considerable 
interest in physics, since many physical systems having a large number of 
strongly correlated degrees of freedom (quantum chaos, mesoscopic 
conductors, $\dots$) share the statistical properties of the 
spectra of random matrices.   Also, the large $N$ expansion of a random matrix
integral (if it exists) is expected to be the generating functional of
discretized surfaces, and therefore random matrices provide a powerful tool for
studying statistical physics on a random surface. (The $2$-matrix model was
first introduced in  this context, as the Ising model on a random surface
\cite{Kazakov}.)

  It has been understood for some time that the $1$-matrix model is not
general enough, since it cannot represent all models of statistical physics
(e.g., it contains only the $(p,2)$ conformal minimal models).
In order to recover the missing conformal models ($(p,q)$ with $p$ and $q$ 
integers), it is necessary to introduce at least a two-matrix model 
\cite{KazakoVDK}. The $1$-matrix models are actually included in the
 $2$-matrix models, since if one takes $d_2=1$, and integrates over the
Gaussian   matrix $M_2$, one sees that the $1$-matrix model follows, and hence
 may be seen as a particular case.
 
It should also be mentioned that most of the results about the $2$-matrix 
model (in particular those derived in the present work) can easily be extended 
to multi-matrix models  without major modifications (see the appendix of
\cite{BEH}). Indeed, the multi-matrix model is not expected to be very 
different from the  $2$-matrix case \cite{KazakoVDK}. (In particular, it
contains the same  conformal models.)

The present paper is organized as follows:  in Section \ref{due}, we present 
the required formalism for biorthogonal polynomials, beginning with the
systems of differential and recursion relations they satisfy, and recalling the
main definitions and results of  \cite{BEH}. We then derive the fundamental
systems of solutions to the overdetermined systems for the ``windows'' of
biorthogonal polynomials in two ways: one, by exploiting the recursion matrices
$Q,P$ for the biorthogonal polynomials, which satisfy the string equation, and
another by giving explicit integral formulas for solutions and showing their 
independence when taken over a suitably defined homology basis of inequivalent
integration paths.
   
In Section \ref{asymptotic} we use saddle-point integration methods to 
deduce the asymptotic form of these fundamental systems of solutions within 
the various Stokes sectors, and from these, to deduce the Stokes matrices and
jump discontinuities at $\infty$. The full formulation of the matrix
Riemann-Hilbert problem characterizing these solutions is given in
Theorem \ref{RHpr}.
\vskip 12pt

\noindent
{\it Acknowledgements:}  The authors would like to thank A. Kapaev for
helpful comments concerning nonsingular integral representations of the
fundamental systems of solutions to eq. (\ref{ODEs1Psi}), which resulted in the
addition of appendix A and the discussion at the end of Sec. \ref{explicit}. 

\section{Differential systems for biorthogonal polynomials and fundamental
solutions }
\label{due}
\setcounter{equation}{0}
\subsection{Biorthogonal polynomials, recursion relations and differential
systems}

The definitions and notation to be used  generally follow 
\cite{BEH}, with some minor modifications, and will be recalled here.
Let us fix two polynomials which will be  referred to  as the
``potentials'',
\bea
V_1(x) = u_0 + \sum_{K=1}^{d_1+1}\frac {u_K}{K} x^K\ , \qquad
V_2(y) = v_0 + \sum_{J=1}^{d_2+1}\frac {v_J}{J} y^J\ .
\eea 
In terms of these potentials we define a {\em bimoment functional}, i.e., a
bilinear pairing between polynomials $\pi(x)$ and $\sigma(y)$ by means
of the following formula
\be
\big(\pi,\sigma\big):= \int_{\Gamma^{(x)}}\int_{\Gamma^{(y)}}{\rm d}x{\rm d}y\,{\rm
e}^{-\frac 1 \hbar(V_1(x)+V_2(y)-xy)} \pi(x)\sigma (y)\ .
\ee
The contours of integration $\Gamma^{(x)}$,  $\Gamma^{(y)}$ in the complex $x$ and $y$
planes remain to be specified. They will generally be chosen to begin and end
at $\infty$, approaching it asymptotically in any direction that assures
convergence. More generally, linear combinations of such integrals along
various inequivalent contours may also be chosen.  In fact there are precisely
$d_1$ (homologically) independent choices for the individual contours in the $x$
plane and $d_2$ in the $y$-plane, due to the choice of the integrands
\cite{Marco}. The necessary and sufficient condition for the convergence
of these integrals is that the contours 
approach $\infty$ in such a way that 
\be
\Re(V_1(x))\mathop{\longrightarrow}_{x\to\infty}  +\infty ,  \quad
\Re(V_2(y))\mathop{\longrightarrow}_{y\to\infty}  +\infty . 
\ee

For $k=0, \dots 2d_1+1$ (mod $2d_1+2$) and $l=0,\dots 2d_2+1$ (mod $2d_2+2$),
let us define the sectors
\bea
&&\mathcal S_k^{(x)}:=\le\{x\in \C\ ,\ {\rm arg}(x)\in
\le(\vartheta_x+\frac{(2k-1)\pi}{2(d_1+1)},
\vartheta_y+\frac{(2k+1)\pi}{2(d_1+1)} \ri)\ri\},
\label{sectx}  \\
&& \qquad \qquad \vartheta_x:=- \arg(u_{d_1+1})/(d_1+1)\ ,   \cr
&&\mathcal S_l^{(y)}:=\le\{y\in \C\ ,\ {\rm arg}(y)\in
\le(\vartheta_y+\frac{(2l-1)\pi}{2(d_2+1)},\vartheta_y+
\frac{(2l+1)\pi}{2(d_2+1)} \ri)\ri\},
\label{secty}\\
&& \qquad \qquad \vartheta_y:= -\arg(v_{d_2+1})/(d_2+1)\ .\nonumber
\eea
\bd\label{defwedgecontours}
  (See Fig. 1)
The wedge contours in the $x$-plane (resp. $y$-plane) $\Gamma_i^{(x)}$
(resp. $\Gamma_j^{(y)}$) for $i=0,\dots d_1$ (resp. $j=0,\dots d_2$),
and the anti-wedge contours $\tilde\Gamma_i^{(x)}$
(resp. $\tilde\Gamma_j^{(y)}$) for $i=0,\dots d_1$ (resp. $j=0,\dots
d_2$) are defined as follows:\\ 
$\bullet$ $\Gamma_i^{(x)}$ (resp. $\Gamma_j^{(y)}$) comes from
$\infty$ within the sector $\mathcal S^{(x)}_{2i-2}$ (resp. $\mathcal
S^{(y)}_{2j-2}$) and returns to infinity in the sector $\mathcal
S^{(x)}_{2i}$ (resp. $\mathcal S^{(y)}_{2j}$).\\ 
$\bullet$ $\tilde\Gamma_i^{(x)}$ (resp. $\tilde\Gamma_j^{(y)}$) comes
from $\infty$ within the sector $\mathcal S^{(x)}_{2i-1}$
(resp. $\mathcal S^{(y)}_{2j-1}$) and returns to infinity in the sector
$\mathcal S^{(x)}_{2i+1}$   (resp. $\mathcal S^{(y)}_{2j+1}$).\\ 
They have the property that $\Re V_1(x)\to\infty$ on the wedge
contours $\Gamma_i^{(x)}$ and $\Re V_1(x)\to-\infty$ on the anti-wedge
contours $\widetilde \Gamma_i^{(x)}$ as $|x|\to \infty$ (and similarly for
$V_2(y)$). 
\ed
\begin{figure}[!ht]
\parbox{9cm}{
\begin{center}
\epsfysize 8cm
\epsffile{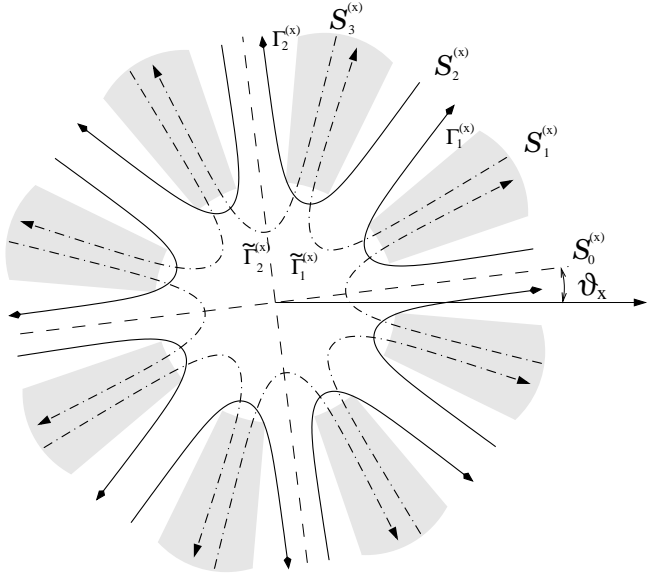}
\end{center}}
\parbox[t]{6.8cm}{\caption
{An example of wedge (solid) and anti-wedge (line-dot-line)
contours in the $x$-plane for a case with $d_1=7$ and real positive
leading coefficient 
in $V_1$.The white sectors are the even numbered ones and the gray sectors  the
odd-numbered ones. (Only the first few are labelled).}}
\end{figure}
Note that since there are no singularities in the finite region of the
$x$- or $y$-planes for the integrands we are considering, we can
deform the contours $\Gamma_j^{(x)}$ and $\Gamma_k^{(y)}$ arbitrarily
in the finite part of the $x$ and $y$ planes.
We could therefore also take such contours as  straight rays coming
from the $(2k-2)$-th sector combined with another ray going to infinity within
the $2k$-th sector. Moreover, by the Cauchy theorem, we have the
homological relation
\be
\sum_{i=0}^{d_1}\Gamma_i^{(x)} \equiv 0, \quad 
 \sum_{j=0}^{d_2}\Gamma_j^{(y)} \equiv 0
\ ,\ \hbox{homologically}\ .\label{homology}
\ee
There are therefore are only $d_2$ homologically independent
contours  $\Gamma_k^{(y)}$ and $d_1$ of type $\Gamma_k^{(x)}$. (Similar
remarks apply to the anti-wedge contours which will be needed later in
this article.)

In general, for any complex $d_1 \times d_2$ matrix with elements
$\{\varkappa^{i,j}\}_{i=1, \dots d_1 \atop j=1,\dots d_2}$ define the homology
class of contours $\varkappa \Gamma$ to be
\bea
\varkappa \Gamma := \sum_{ij} \varkappa^{i,j} \Gamma_i^{(x)} \times
\Gamma_j^{(y)}
\eea
For brevity, we will denote the corresponding integral operator  as follows.
\bea
\sum_{i=1}^{d_1} \sum_{j=1}^{d_2}
\varkappa^{i,j}\int_{\Gamma_i^{(x)}\times
\Gamma_j^{(y)}} := \int_{\varkappa\Gamma}\ .
\eea

 For a given $\varkappa$, we denote the corresponding bilinear pairing as:
\bea
&& \big(\pi,\sigma\big)_{\varkappa}:=\int_{\varkappa\Gamma} {\rm d}x\, {\rm d}y\ 
\pi(x){\rm e}^{-\frac 1 \hbar (V_1(x)+V_2(y)-xy)}
\sigma (y) \ ,
\eea
and define two sequences of  monic  polynomials
$\pi_n(x),\sigma_n(y)$ of degree $n$ such that they are biorthogonal 
with respect to this  pairing 
\bea
&& \big(\pi_n,\sigma_m\big)_{\varkappa} =h_n\delta_{nm} \ ,\\
&& \pi_n(x) = x^n +...\ , \  \sigma_n(y) = y^n+...\ ,
\eea
The  matrix $\varkappa^{i,j}$ must be chosen such that the $N\times N$ finite
submatrices  of the matrix of bimoments
\be
\{B_{ij}:=\big(x^i,y^j\big)\}_{i,j=0,\dots,N-1}
\ee
are nonsingular for all $N$; i.e., the nondegeneracy condition that ensures the
existence of the  biorthogonal polynomials is given by
\be
\Delta_N(\varkappa):=
 \det\le[\big(x^i,y^j\big)_\varkappa\ri]_{i,j=0,\dots,N}
 \neq 0\qquad \forall N\in \N\ .
\ee
Since these $\Delta_N(\varkappa)$'s form a denumerable sequence of homogeneous
polynomials in the coefficients $\varkappa^{i,j}$, this will hold on the
complement of a denumerable set of hypersurfaces, and in {\it this} sense the
permissible choices of $\varkappa$'s  are ``generic'', whatever the choice
of potentials $V_1$ and $V_2$.  

We introduce the corresponding normalized quasipolynomials and
combine them into semi-infinite column vectors (``wave-vectors'')
$\ds{\m{\Psi}_\infty(x) , \ \m{\Phi}_\infty(y)}$ as follows.
\bea
&& \psi_n(x) := \frac 1{\sqrt{h_n}} \pi_n(x){\rm e}^{-\frac 1 \hbar V_1(x)} \ ,
\ \  \phi_n(y) := \frac 1{\sqrt{h_n}} \sigma_n(y) {\rm e}^{-\frac 1
\hbar V_2(y)}\ ,\cr
&&\m{\Psi}_\infty(x) := [\psi_0(x),...,\psi_n(x),...]^t\ ,\qquad
\m{\Phi}_\infty(y):=[\phi_0(y),...,\phi_n(y),...]^t\ .
\eea
In matrix notation the biorthogonality reads 
\be
\int_{\varkappa\Gamma}\!\!\!\!\!   {\rm d}x {\rm d}y
\,{\rm e}^{\frac{xy}\hbar } \m{\Psi}_\infty(x) {\m{\Phi}_\infty}^t (y) = \1 \ ,
\ee
where $\1$ denotes the semi-infinite unit matrix.

We denote the Fourier-Laplace transforms of the wave-vectors along
the  wedge contours by 
\bea
{\m{\un\Psi}_\infty}^{(i)}(y) := \int_{\Gamma_i^{(x)}} \!\!\! {\rm d}x\,
{\rm e}^{\frac{xy}\hbar } {\m{\Psi}_\infty}^t(x)\ ,\ \ i=0,\dots,d_1\label{FLT0}\\
{\m{\un\Phi}_\infty}^{(j)}(x) := \int_{\Gamma_j^{(y)}} \!\!\! {\rm d}y\,
{\rm e}^{\frac{xy}\hbar } {\m{\Phi}_\infty}^t(y)\ ,\ \
j=0,\dots,d_2,\label{FLT}
\eea
viewed as row vectors, with components denoted $\un\psi_n^{(i)}(y)$ and
$\un\phi_n^{(j)}(x)$ respectively.
Because of eq. (\ref{homology}) only $d_1$ (or $d_2$) are linearly independent 
\be
\sum_{i=0}^{d_1}{\m{\un\Psi}_\infty}^{(i)}(y)= \sum_{j=0}^{d_2}
{\m{\un\Phi}_\infty}^{(j)}(x) \equiv 0.
\ee

The recursion relations for the biorthogonal
polynomials are expressed by the following matrix equations for the wave-vectors
\cite{BEH}:
\bea
&&x\m{\Psi}_\infty (x) = Q\m{\Psi}_\infty (x)\ , \qquad
\hbar\pa_x \m{\Psi}_\infty (x) = -P \m{\Psi}_\infty (x)\ ,\label{matrixreca}\\
&&y\m{\Phi}_\infty (y) = P^t\m{\Phi}_\infty (y)\ , \qquad
\hbar \pa_y \m{\Phi}_\infty (y) = -Q^t \m{\Phi}_\infty (y)\
,\label{matrixrecb} 
\eea
where the matrices $Q$ and $P$ have finite band sizes $d_2+1$ and $d_1+1$ 
respectively, with $Q$ having only one nonzero diagonal above the principal
one and $P$ only one below
\bea
&& Q :=\le[
\begin{array}{ccccc}
 \a_0(0) &  \gamma(0) & 0 & 0 &\cdots \\
 \a_1(1) & \a_0(1) &\gamma(1) & 0 &\cdots \\
  \vdots &
\ddots
 &\ddots
 &  
\ddots
 &  \ddots
\cr
\a_{d_2}(d_2) & \a_{d_2\!-\!1}(d_2) & \cdots & \a_0(d_2)& \gamma(d_2)
 \cr
0 &
\ddots&
\ddots&
\ddots&
\ddots
\end{array}  
\ri]\label{Qdef}\\
 && P := \le[
\begin{array}{ccccc}
 \b_0(0) &   \b_1(1) & \cdots & \b_{d_1}(d_1)& \cdots\\
\gamma(0) & \b_0(1) & \b_1(2)&\ddots& \b_{d_1}(d_1\!+\!1)\\
0 & \gamma(1) & \b_0(2) & \ddots &\!\!\!\!\!\!\!\!\!\!\!\!\!\!\!\!\ddots \\
0 &0&  \gamma(2) & \b_0(3) &\!\!\!\!\!\!\!\!\!\!\!\!\!\!\!\!\ddots\\
\vdots& \ddots & \ddots &\ddots &\!\!\!\!\!\!\!\!\!\!\!\!\!\!\!\! \ddots
\end{array}  
\ri]\ .
\label{Pdef}
\eea
These semi-infinite matrices satisfy the string equation
\be
[P,Q]= \hbar \1\ .\label{comm}
\ee
(See \cite{BEH, ZJDFG, eynard} for simple proofs of these assertions.)
For the dual sequences of Fourier--Laplace transforms, a simple
integration by parts in eqs. (\ref{matrixreca}), (\ref{matrixrecb}) gives
\bea
&&x\m{\un\Phi}_\infty\!^{(j)} (x) = \m{\un\Phi}_\infty\!^{(j)} (x)Q\ , \qquad
\hbar\pa_x \m{\un\Phi}_\infty\!^{(j)} (x) = \m{\un\Phi}_\infty\!^{(j)}
(x)P\ ,\ \ j=0,\dots,d_2\label{matrixrecduala}\\
&&y\m{\un\Psi}_\infty\!^{(k)} (y) = \m{\un\Psi}_\infty\!^{(k)} (y)P^t\ , \qquad
\hbar\pa_y \m{\un\Psi}_\infty\!^{(k)} (y) =  \m{\un\Psi}_\infty\!^{(k)}
(y)Q^t\ ,\ \ k=0,\dots,d_1\ .
\label{matrixrecdualb}
\eea
Notice that integration by parts is allowed due to the exponential
decay of the integrand along the chosen contours.

We recall from \cite{BEH} the definition of dual sequences of windows.
\bd
\label{dualwindows}
Call a {\bf window of size $\mathbf {d_1+1}$ or $\mathbf {d_2+1}$} any  subset
of 
$d_1+1$ or $d_2+1$  consecutive elements of type $\psi_n$, $\un \phi_n$, 
 $\phi_n$ or $\un\psi_n$, with the notations 
\bea
&&\m{\Psi}_N :=[\psi_{N\!-\!d_2},\dots,\psi_{N}]^t \ , \quad  N\geq d_2, \qquad
\m{\Phi}_N :=[\phi_{N\!-\!d_1},\dots,\phi_{N}]^t  \ ,\quad  N\geq d_1 \\
&&\m{\Psi}^N :=[\psi_{N\!-\!1},\dots,\psi_{N\!+\!d_1\!-\!1}]^t \ , \quad  N\geq 0,   \qquad
\m{\Phi}^N :=[\phi_{N-1},\dots,\phi_{N\!+\!d_2\!-\!1}]^t \ , \qquad N\geq 1 \\ 
&&{\m{\un\Psi}_N}\!^{(j)}
:=[\un\psi^{(j)}_{N\!-\!d_2},\dots,\un\psi^{(j)}_{N}]
\ \quad  N\geq d_2,  
\qquad
{\m{\un\Phi}_N}^{(k)}
:=[\un\phi^{(k)}_{N\!-\!d_1},\dots,\un\phi^{(k)}_{N}]  \ 
,\quad  N\geq d_1 \\ 
&&{\m{\un\Psi}^N}\!\,^{(j)}
:=[\un\psi^{(j)}_{N\!-\!1},\dots,\un\psi^{(j)}_{N\!+\!d_1\!-\! 1}]\ , \quad  N\geq 1,     
\qquad   
{\m{\un\Phi}^N}\!\,^{(k)}
:=[\un\phi^{(k)}_{N\!-\! 1},\dots,\un\phi^{(k)}_{N\!+\!d_2\!-\! 1}]\ , \qquad N\geq 1
\ . 
\eea
\ed
Note the difference in positioning and size of the windows in the various
cases, and the fact that  the barred quantities are defined as row
vectors while the unbarred ones are column vectors. The notation we are using 
here differs slightly from that used in \cite{BEH}, in which, e.g.,
$\ds{\m{\un\Phi}^N}$ here would be denoted
$\ds{\m{\un\Phi}^{N\!+\!1}}$ in \cite{BEH}.
\bd
Each of the two pairs of windows
 $(\ds{\m{\Psi}_N}, \  \ds{\m{\un{\Phi}}^{N} })$
 and $(\ds{\m{\Phi}_N}, \ \ds{\m{\un{\Psi}}^{N}})$ of dimensions
$d_2+1$ and $d_1+1$ respectively, will be called {\bf dual windows}
(and these are defined for $N\geq d_2$ and $N\geq d_1$ respectively).
\ed

We now recall the notion of {\em folding}, as defined in
\cite{BEH}. We give here only the main statements without proofs and refer the
reader to ref. \cite{BEH} for further details.
Let us introduce the sequence of companion--like  matrices $\ds{\A_N(x)}$
and $\ds{ \m{\un\A}^N(x)}$ of size $d_2+1$
\bea
\A_N(x) & := & 
\le[\begin{array}{cccc} 0  & 1 & 0 &\!\!\! \!\!\!\!\!\! \!\!\! 0 \cr
0 & 0 & \ddots &\!\!\! \!\!\!\!\!\! \!\!\!0\cr
0 & 0 & 0 &\!\!\! \!\!\!\!\!\! \!\!\!1\cr
\!\!\frac {-\alpha_{d_2}(N)} {\gamma(N)}\!\! &
\cdots
&\!\! \frac {-\alpha_1(N)}
{\gamma(N)} \!\! &\!\!  \frac{(x-\alpha_0(N))}{\gamma(N)} \!\!  
\end{array}\ri] \ ,  \quad N\geq d_2 \ ,  \label{aNdef}\\
 \m{\un\A}^{N}(x) & := & \le[
\begin{array}{cccc} 
  \frac{x\!-\!\a_0\!(N)}{\gamma(N\!-\!1)}&1&0&0\\[5pt]
 \frac{-\a_1\!(N\!+\!1)}{\gamma(N\!-\!1)}&0&
^{\ds{^{\ds{\cdot}}}}\hbox{}
 ^{\ds{\cdot}} \cdot&0\\
 \vdots &0&0&1\\[5pt]
   \frac{-\a_{d_2}\!(N\!+\!d_2)}{\gamma(N\!-\!1)}&0&0&0
\end{array} \ri]\ ,\ N\geq 1,
\label{abarN}
\eea
and also the analogous sequence of matrices $\ds{\B_N(y)}$
and $\ds{ \m{\un\B}^N(y)}$ of size $d_1+1$
\bea
\B_N(y) & := & 
\le[\begin{array}{cccc} 0  & 1 & 0 &\!\!\! \!\!\!\!\!\! \!\!\! 0 \cr
0 & 0 & \ddots &\!\!\! \!\!\!\!\!\! \!\!\!0\cr
0 & 0 & 0 &\!\!\! \!\!\!\!\!\! \!\!\!1\cr
\!\!\frac {-\beta_{d_1}(N)} {\gamma(N)}\!\! &
\cdots
&\!\! \frac {-\beta_1(N)}
{\gamma(N)} \!\! &\!\!  \frac{(y-\beta_0(N))}{\gamma(N)} \!\!  
\end{array}\ri] \ ,  \quad N\geq d_1 \ ,  \label{bNdef}\\
 \m{\un\B}^{N}(y) & := & \le[
\begin{array}{cccc} 
  \frac{y\!-\!\b_0\!(N)}{\gamma(N\!-\!1)}&1&0&0\\[5pt]
 \frac{-\b_1\!(N\!+\!1)}{\gamma(N\!-\!1)}&0&
^{\ds{^{\ds{\cdot}}}}\hbox{}
 ^{\ds{\cdot}} \cdot&0\\
 \vdots &0&0&1\\[5pt]
   \frac{-\b_{d_1}\!(N\!+\!d_1)}{\gamma(N\!-\!1)}&0&0&0
\end{array} \ri]\ ,\ N\geq 1. 
\label{bbarN}
\eea
The first equations in (\ref{matrixreca}), (\ref{matrixrecb}) and
(\ref{matrixrecduala}), (\ref{matrixrecdualb}) imply  the 
following
\bl \label{abNrecursions} 
The sequences of matrices  $\ds{\m{\A}_N}$, $\ds{\m{\B}_N}$ and
$\ds{\m{\un\A}^N}$, $\ds{\m{\un\B}^N}$ implement the 
shift  $N\mapsto N+1$ and $N\mapsto N-1$ in the windows of
quasi-polynomials and Fourier--Laplace transforms in the sense that  
\bea
 \A_N\m{\Psi}_N(x) =
\m{\Psi}_{N\!+\!1}(x) \ ,\qquad
\m{\un\Phi}^{N}(x)  =
\m{\un\Phi}^{N\!+\!1}(x)\m{\un\A}^N \ ,\label{ladder1}\\
\B_N\m{\Phi}_N(y) =
\m{\Phi}_{N\!+\!1}(y) \ ,\qquad
\m{\un\Psi}^{N}(y)  =
\m{\un\Psi}^{N\!+\!1}(y)\m{\un\B}^N \ . \label{ladder2}
\eea
 and in general 
\bea
\m{\Psi}_{N\!+\!j} = \A_{N+j-1}\cdots\A_{N}
\m{\Psi}_N\ ,\qquad \m{\un\Phi}^{N}  =
\m{\un\Phi}^{N\!+\!j}\m{\un\A}^{N+j-1}\cdots \m{\un\A}^{N}\ ,\label{folding}\\
\m{\Phi}_{N\!+\!j} = \B_{N+j-1}\cdots\B_{N}
\m{\Phi}_N\ ,\qquad \m{\un\Psi}^{N}  =
\m{\un\Psi}^{N\!+\!j}\m{\un\B}^{N+j-1}\cdots \m{\un\B}^{N}\ ,\label{folding1}
\eea
where $\ds{\m{\un\Psi}^{N}\ ,\m{\un\Phi}^{N} }$ here denotes a window in any of the
Fourier--Laplace transforms defined in eqs. (\ref{FLT0}), (\ref{FLT}).
\el
(In this lemma, and in what follows, if no superscript distinguishing the
integration path in the Fourier-Laplace transform in eqs. (\ref{FLT0}),
(\ref{FLT}) is present in $\ds{\m{\un\Psi}^{N},\ \m{\un\Phi}^{N}}$,
 this means that the result being discussed holds
for all cases.)

  Equations (\ref{ladder1}), (\ref{ladder2}) will henceforth be referred
to as the ``ladder relations''.
 We refer to the  process of expressing any $\psi_n(x)$ or
 $\un\phi_n(x)$ by means of linear 
combinations of elements in a specific window with polynomial coefficients, as 
{\em folding} onto the specified window. 
We also recall that  matrices $\ds{\A_N(y)}$ and $\ds{\m{\un\A}^N(y)}$,
$\ds{\B_N(y)}$ and $\ds{ \m{\un\B}^N(y)}$ are invertible (see \cite{BEH}). The
opposite shifts are therefore implemented by the inverse matrices
and the folding may take place in either direction with 
respect to polynomial degrees.

\bl(From \cite{BEH}, with the adapted notation.)
\label{D1s}
The windows of quasi-polynomials $\ds{\m{\Psi}_N(x), \m{\Phi}_N(y)}$
 and Fourier--Laplace
transforms  $\ds{\m{\un \Phi}^N(x),\m{\un \Psi}^N(y)}$
 satisfy the following differential systems
\bea
-\hbar \frac \pa{\pa x} \m{\Psi}_N(x) \!\!\!&=&\!\!\! \m{D_1}^N (x)
 \m{\Psi}_N(x)  \ ,  \quad N\ge d_2+1 \ , \label{ODEs1Psi}\\
\hbar \frac \pa{\pa x}\m{\un\Phi}^{N}(x)\!\!\! & =&\!\!\!
\ds{\m{\un\Phi}^{N}(x) {\m{\un D}^{N}}_1(x)} \ , 
\quad N\ge d_2+1 \ , \label{ODEs1}
\eea
\bea
-\hbar \frac \pa{\pa y} \m{\Phi}_N(y) \!\!\!&=&\!\!\! \m{D_2}^N (y)
 \m{\Phi}_N(y)  \ ,  \quad N\ge d_1+1 \ , \label{ODEs2Phi}\\
\hbar \frac \pa{\pa y}\m{\un\Psi}^{N}(y)\!\!\! & =&\!\!\!
\ds{\m{\un\Psi}^{N}(y) {\m{\un D}^{N}}_2(y)} \ , 
\quad N\ge d_1+1 \ , \label{ODEs2}
\eea
where 
\bea
 &&
 \m{D_1}^N(x) := 
 \stackrel{N}{\b}_{-1}\m{\A}_{N-1}\!^{-1} +
\stackrel{N}{\b_0} + \sum_{j=1}^{d_1}\stackrel{N}{\b_j}\A_{N+j-1}
\A_{N+j-2}\cdots
\A_N\ \in\ gl_{d_2+1}[x]  \ . \label{defD1}\\
&&\m{\underline D_1}^N(x)  := \ds{\m{\un\A}^{N}\!^{-1}
{\m{\underline \b}^{N}}\!_{-1}  + {\m{\underline\b_0}^{N}}
+\sum_{j=1}^{d_1}
\m{\un\A}^{N-1}\m{\un\A}^{N-2}\cdots
\m{\un\A}^{N-j}
{\m{\underline\b_j}^{N}} } \ \in\ gl_{d_2+1}[x]\ ,\\
&& \stackrel{N}\b_j := 
{\rm diag}\le[\b_j(N\!+\!j\!-\!d_2),\b_j(N\!+\!j\!-\!
d_2\!+\!1),\dots,\b_j(N\!+\!j) \ri] \ ,\ 
j=0, \dots d_1\ , \label{alphajN}\\
&&\stackrel{N}{\b}_{-1} :=
{\rm diag}\le[\gamma(N\!-\!d_2\!-\!1), \dots, \gamma(N-1) \ri] \\
&& \m{\underline \b_j}^{N} :={\rm
diag}\le(\b_j(N-1),\b_j(N),\dots \b_j(N\!+\!d_2\!-\!1)\ri)\ ,\ \ j=0,\dots,d_1\ .
\label{betabarjN}\\
&&{\m{\underline \b}^{N}}\!_{-1} :={\rm
diag}\le(\gamma(N-1),\gamma(N),\dots,\gamma(N\!+\!d_2\!-\!1)\ri)\ .
\eea
\bea
 &&
 \m{D_2}^N(y) := 
 \stackrel{N}{\a}_{-1}\m{\B}_{N-1}\!^{-1} +
\stackrel{N}{\a_0} + \sum_{j=1}^{d_2}\stackrel{N}{\a_j}\B_{N+j-1}
\B_{N+j-2}\cdots
\B_N\ \in\ gl_{d_1+1}[y]  \ . \label{defD2}\\
&&\m{\underline D_2}^N(y)  := \ds{\m{\un\B}^{N}\!^{-1}
{\m{\underline \a}^{N}}\!_{-1}  + {\m{\underline\a_0}^{N}}
+\sum_{j=1}^{d_2}
\m{\un\B}^{N-1}\m{\un\B}^{N-2}\cdots
\m{\un\B}^{N-j}
{\m{\underline\a_j}^{N}} } \ \in\ gl_{d_1+1}[y]\ ,\\
&& \stackrel{N}\a_j := 
{\rm diag}\le[\a_j(N\!+\!j\!-\!d_1),\a_j(N\!+\!j\!-\!
d_1\!+\!1),\dots,\a_j(N\!+\!j) \ri] \ ,\ 
j=0, \dots d_2\ , \label{betajN}\\
&&\stackrel{N}{\a}_{-1} :=
{\rm diag}\le[\gamma(N\!-\!d_1\!-\!1), \dots, \gamma(N-1) \ri] \\
&& \m{\underline \a_j}^{N} :={\rm
diag}\le(\a_j(N-1),\a_j(N),\dots \a_j(N\!+\!d_1\!-\!1)\ri)\ ,\ \ j=0,\dots,d_2\ .
\label{alphabarjN}\\
&&{\m{\underline \a}^{N}}\!_{-1} :={\rm
diag}\le(\gamma(N-1),\gamma(N),\dots,\gamma(N\!+\!d_1\!-\!1)\ri)\ .
\eea
Moreover the systems (\ref{ODEs1Psi})--(\ref{ODEs2}) are compatible
with the ladder relations (\ref{ladder1}), (\ref{ladder2}) since 
\bea
&& \m{D_1}^{N+1}(x) = \A_N(x)\m{D_1}^N(x){\A_N}^{-1}(x) - \hbar
\A_N\!'(x){\A_N}^{-1}(x)  \label{shiftgaugex}\\
&&  \m{\un D_1}^{N+1}(x) = \m{\un\A}^N(x)\m{\un D_1}^N(x)\m{\un\A}^N\!^{-1}(x) - \hbar
\m{\un\A}^N\!'(x)\m{\un\A}^N\!^{-1}(x)  \label{shiftgaugexinv} \ .
\eea
\bea
&& \m{D_2}^{N+1}(y) = \B_N(y)\m{D_2}^N(y){\B_N}^{-1}(y) - \hbar
\B_N\!'(y){\B_N}^{-1}(y)  \label{shiftgaugey}\\
&&  \m{\un D_2}^{N+1}(y) = \m{\un\B}^N(y)\m{\un D_2}^N(y)\m{\un\B}^N\!^{-1}(y) - \hbar
\m{\un\B}^N\!'(y)\m{\un\B}^N\!^{-1}(y)  \label{shiftgaugeyinv} \ .
\eea
\el
\br
\label{lads}
We stress that the windows taken from any wave-vector solution to eqs.
(\ref{matrixreca}), (\ref{matrixrecb}) and eqs. (\ref{matrixrecduala}),
(\ref{matrixrecdualb})  will automatically satisfy the systems specified in
Lemma \ref{D1s} (differential and difference equations).
\er

In the following, we will only consider the fundamental systems of
solutions to the sequence of equations (\ref{ODEs1Psi}),  (\ref{ODEs1}),
(\ref{shiftgaugex}), (\ref{shiftgaugexinv}), since the
corresponding dual sequence of equations for the quantities
$\ds{\m{\Phi}_N(y),\ \m{\un\Psi}^N(y)}$  may be treated analogously by just
making the appropriate interchange  of notations $x \leftrightarrow y$, $\Psi
\leftrightarrow \Phi$, $d_1 \leftrightarrow d_2$, etc., and the
interchange of integration contours in the $x$ and $y$ planes.

In  \cite{BEH} it was shown  that there exists a natural nondegenerate
{\em Christoffel--Darboux} pairing between any pair of solutions
$\ds{\m{\widetilde\Psi}_N(x)}$ and $\ds{\m{\un{\widetilde\Phi}}^N(x)}$ to eqs.
(\ref{ladder1}) and (\ref{ODEs1}) such that
\be
\le(\m{\un{\widetilde \Phi}}^N,\m{\widetilde\Psi}_N\ri)_N:=
\m{\un{\widetilde \Phi}}^N(x)\Amat^N\m{\widetilde\Psi}_N(x)\ ,\label{pairing}
\ee
is {\em constant} both in $x$ {\em and } $N$, where the invertible matrix
$\ds{\Amat^N}$ defining the pairing is
\bea
 \Amat^N := \le[
\begin{array}{cccc|c}
0&0&0&0&\!\!-\!\gamma(\!N\!\!-\!1\!)\!\!\cr\hline
\a_{d_2}\!(\!N\!)& \cdots & \a_{2}(\!N\!)& \a_1(\!N\!)& 0\cr
0& \a_{d_2}\!(\!N\!\!+\!1\!) & \cdots & \a_2(\!N\!+\!1\!)& 0\cr
0&0&\a_{d_2}\!(\!N\!\!+\!2\!) &\cdots & 0\cr
0&0&0&\a_{d_2}\!(\!N\!\!+\!d_2\!-\!1\!)&0
\end{array}
\ri]\ .
\label{CDAmatrix}
\eea
This follows from the fact that the matrices $\ds{\m{D_1}^N}$ and
$\ds{\m{\un D_1}^N}$ are conjugate to each other by means of the matrix
$\ds{\Amat^N}$ (Theorem 4.1 in \cite{BEH})
\bea
\Amat^N \m{D_1}^N(x) = \m{\un D_1}^N(x)\Amat^N\ ,
\label{D1intertwining}
\eea
together with the shift relation (\ref{shiftgaugex}).
The $(d_2+1)\times (d_2+1) $ matrix $\ds{\Amat^N}$ is the only nonzero
block in the commutator $\ds{\le[\m{\Pi}^{N-1}_0,Q\ri]}$, 
where $\ds{\m{\Pi}^{N-1}_0}$ denotes the projector onto the first  $N$
basis elements (a ``canonical'' projector).
In the following it is convenient to use the same notation
$\ds{\Amat^N}$ both for the finite matrix and the semi-infinite one.

It was also proved in \cite{BEH} (Proposition 3.3) that one can choose the
windows $\ds{\m{\widetilde \Psi}_N}$ and $\ds{\m{\un{\widetilde
\Phi}}^N}$  as joint solutions of the PDE's following from the infinitesimal
changes of the coefficients $\{u_J\}$ and $\{v_K\}$ of the potentials
(deformation equations).  For this choice,  the pairing also becomes independent
of these deformation parameters.

\subsection{Fundamental solutions of the $D_1$ and $\un D_1$ systems}
\label{fundamental}

In this section we explicitly construct solutions of the pairs of
dual ODE's defined by the matrices $\ds{\m{D_1}^N(x)}$ and $\ds{\m{\un
D_1}^N(x)}$.  In fact these solutions will simultaneously satisfy the
deformation equations and the ladder recursion relations in $N$. It will be left
to the reader to formulate the corresponding statements for the other pairs
($\ds{\m{D_2}^N(y)}, \ds{\m{\un D_2}^N(y)}$), which are essentially the same,
{\em mutatis mutandis}.

As mentioned in Remark \ref{lads}, taking windows within any
wave-vector solution to eqs. (\ref{matrixreca}), (\ref{matrixrecduala})
one obtains solutions to the difference-differential
equations in Lemma \ref{D1s}. Therefore we could try to construct $d_2+1$
such wave-vector solutions in order to obtain a fundamental system for 
 eqs. (\ref{ODEs1}). On the other hand it will be shown in Proposition
\ref{unmodrex} below  that the relations (\ref{matrixrecduala}) have precisely
$d_2$ linearly independent wave-vector solutions (given by the $d_2$
Fourier--Laplace transforms (\ref{FLT})) while relations (\ref{matrixreca})
have only one solution (given by the quasipolynomials). Therefore we will look
for solutions to slightly modified recursion relations differing from
(\ref{matrixreca}), (\ref{matrixrecdualb}) by changing the initial
terms\footnote{This is not new in the context of orthogonal polynomials, where
there exist, besides the orthogonal polynomials, solutions of the second kind.
(See e.g. \cite{chihara, Szego}).}. 
 The solutions to these modified recursion relations $\{\tilde
\psi_n(x)\}_{n\geq 0}$ or $\{\tilde {\un\psi_n} (x)\}_{n\geq 0} $   
will still satisfy the {\em unmodified} recursions relations, but only for
$n$ large enough. This will provide us  with the required solutions of the
difference-differential equations of Lemma \ref{D1s}.

We start by proving that the Fourier--Laplace transforms and the
quasipolynomials are the only wave-vector solutions to eqs.
(\ref{matrixreca}), (\ref{matrixrecduala}).

\noindent
 {\bf Notational remark}. Throughout the remainder of this section, 
$\ds{\m{\Psi}_\infty(x)}$ and $\ds{\m{\un\Phi}_\infty (x)}$ will be used
generally to denote {\em arbitrary} wave-vectors consisting of
arbitrary solutions and not just the previously defined quasipolynomials and
their Fourier--Laplace transforms.
\bp
\label{unmodrex}
The semi-infinite systems
\bea
&& \le\{
\begin{array}{l}
\ds{x\m{\Psi}_\infty(x) = Q\m{\Psi}_\infty(x)}\\
\ds{\hbar\pa_x\m{\Psi}_\infty(x) = -P\m{\Psi}_\infty(x)}
\end{array}
\ri. \label{unmodifa} \ , \\[10pt]
&& \le\{
\begin{array}{l}
\ds{x\m{\un\Phi}_\infty(x) = \m{\un\Phi}_\infty(x)Q}\\
\ds{\hbar\pa_x\m{\un\Phi}_\infty(x) = \m{\un\Phi}_\infty(x)P}
\end{array}\ri.  \label{unmodifb}
\eea
have $1$ and $d_2$ linearly independent solutions, respectively, 
for the wave-vectors $\ds{\m{\Psi}}_\infty(x) $
and $\ds{\m{\un\Phi}}_\infty(x) $.
(A similar statement holds for $\ds{\m{\Phi}_\infty,\ \m{\un\Psi}_{\infty}}$,
as solutions to eqs. (\ref{matrixrecb}) and (\ref{matrixrecdualb}) with
$d_2\rightarrow d_1$).
\ep
{\bf Proof.} 
The compatibility is guaranteed by the string equation (\ref{comm}).
Recalling  \cite{BEH} that
\be
\bigg(P-V'_1(Q)\bigg)_{\geq 0} = \bigg(Q-V'_2(P)\bigg)_{\leq 0} = 0 \ ,
\ee
we have
\bea
 &&0= \le[\le(P-V'_1(Q)\ri)\m{\Psi}_\infty\ri]_0 = -\hbar \psi_0'(x) -
V'_1(x)\psi_0(x)\\
&&0= \le[\m{\un\Phi}_\infty\le(Q-V'_2(P)\ri)\ri]_0 = x\un\phi_0(x) -
V'_2(\hbar\pa_x)\un\phi_0(x)\ .\label{dualgrnd}
\eea
Thus we have only one solution of the first equation (up to normalization) and
$d_2$ independent solutions for the second. Using the $x$ recursion relations
for the $\psi$ sequence, we can build the rest of the sequence by starting
from the first term $\psi_0(x)$,  since the matrix $Q$ has
only one nonzero diagonal above the principal one. On the other hand, using the
$\pa_x$ recursion relations we can build the rest of the $\un\phi_n$ sequence
starting from any given solutions $\un\phi_0$ of (\ref{dualgrnd}) because the
matrix $P$ has only one diagonal below the main one. Q.~E.~D.

The solutions of Propostion \ref{unmodrex} are, up to multiplicative constants,
exactly the solutions given by the quasipolynomials for the wave-vector
$\ds{\m{\Psi}_\infty}$ and the different possible $\ds{\m{\un\Phi}_\infty}$'s
corresponding to the different solutions of eq. (\ref{dualgrnd}), which are
just the $d_2$ different Fourier--Laplace transforms of the quasipolynomials
$\phi_n(y)$. Indeed, the solutions $\un\phi_0(x)$ of eq. (\ref{dualgrnd}) can be
expressed by
\be
\un\phi_0(x) \propto \int_\Gamma\!{\rm d}y\,{\rm e}^{\frac 1 \hbar
(xy-V_2(y))}\ ,
\ee
 where $\Gamma$ is any of the contours $\Gamma_k^{(y)}$ or a linear combination
of them.

As indicated in the introduction to this section, in order to
find a fundamental system of solutions to the Difference-Differential
equations in Lemma \ref{D1s}, we  consider systems for the
wave-vectors modified by the addition of terms that only change the
equations involving the first few entries of the wave-vectors.
The recursion relations will therefore remain unchanged for $n$ sufficiently
large.
\bp
\label{modrex}
The semi-infinite systems:
\bea
&& \le\{
\begin{array}{l}
\ds{x\m{\Psi}_\infty(x) = Q\m{\Psi}_\infty(x) - W_2(-\hbar\pa_x) F(x)}\\
\ds{\hbar\pa_x\m{\Psi}_\infty(x) = -P\m{\Psi}_\infty(x) + F(x)}
\end{array}
\ri.\label{modifrec1},\\[10pt]
&&  \le\{
\begin{array}{l}
\ds{x\m{\un\Phi}_\infty(x) = \m{\un\Phi}_\infty(x)Q + U(x)}\\
\ds{\hbar\pa_x\m{\un\Phi}_\infty(x) = \m{\un\Phi}_\infty(x)P +U(x) W_1(x)}
\end{array}\ri.\label{modifrec2},
\eea
where
\bea
&& W_1(x):= \frac {V'_1(Q)-V'_1(x)}{Q-x}\ ,\qquad
W_2(\hbar\pa_x):= \frac {V'_2(P)-V'_2(\hbar\pa_x)}{P-\hbar\pa_x}\
,\\ 
&& F(x):= [f(x),0,0,\cdots]^t\ ,\qquad U(x):= [u(x),0,0,\cdots] \ ,
\eea
both have $d_2+1$ linearly independent solutions for the extended unknown
wave-vectors $\le(f(x), \
\ds{\m{\Psi}_\infty(x)}\ri)$ and $\le(u(x), 
\ds{\m{\un\Phi}_\infty(x)}\ri)$.
\ep
\br
The  terms in the RHS of eqs. (\ref{modifrec1}), (\ref{modifrec2}) which
are not present in eqs. (\ref{unmodifa}), (\ref{unmodifb}) are
semi-infinite vectors with at most  $d_2$ nonzero entries for
(\ref{modifrec1}) or $d_1$ for (\ref{modifrec2}). Indeed the
matrices $W_1$ and $W_2$ (which are polynomials in $Q$ and $P$ of
degrees $d_1-1$ and $d_2-1$ respectively) have finite band sizes, and
hence when acting on the vectors $U(x)$ and $F(x)$, which have only a
first nonzero entry, produce a finite number of nonzero entries. As a result,
windows taken from any solution of the modified wave-vector systems
(\ref{modifrec1}), (\ref{modifrec2}) will provide solutions of  the
Difference-Differential systems of Lemma \ref{D1s}, but only beyond a minimal
$N$ value ($d_1$ or $d_2$ respectively).
\er
{\bf Proof.}
The compatibility of these systems is not obvious. We have:
\bea
&& \hbar\m{\Psi}_\infty = [\hbar\pa_x,x]\m{\Psi}_{\infty} =
 \hbar\pa_x\le(Q\m{\Psi}_\infty - W_2(-\hbar\pa_x)F\ri) -
x\le(-P\m{\Psi}_\infty + F\ri) \\
&&=Q\le(-P\m{\Psi}_\infty + F\ri)  - \pa_x W_2(-\hbar\pa_x)F
+P\le(Q\m{\Psi}_\infty - W_2(-\hbar\pa_x)F \ri) + xF \\
&&= [P,Q]\m{\Psi}_\infty + QF - (\hbar\pa_x+P) \frac
{V'_2(P)-V'_2(-\hbar\pa_x)}{P+\hbar\pa_x} F(x) + xF \\ 
&&= \hbar \m{\Psi}_\infty +\bigg(Q-V'_2(P)\bigg) F  + 
\bigg(x-V'_2(-\hbar\pa_x)\bigg)F . 
\eea
Since only the first entry of $F$ is nonzero and $Q-V'_2(P)$
is strictly upper-triangular \cite{BEH}, the second term vanishes. 
The last term gives the following ODE for the first entry of $F(x)$
\be
\label{dualground}
V_2'(-\hbar\pa_x) f(x) = xf(x).
\ee
The solutions of eq. (\ref{dualground}) are easily written as
Fourier--Laplace integrals, and give a compatible system (see
eq. (\ref{intpsik}), where we also fix the most convenient
normalization for them). We denote them by  $f^{(\alpha)}$ with
$\alpha=0,\dots,d_2$, with $f^{(0)}\equiv 0$ the trivial solution, 
corresponding to the unmodified system (\ref{unmodifa}).

Fixing any such solution $f(x)$, we can now solve for $\Psi$. First of all, one
can prove by induction that 
\be
Q^k\m{\Psi}_\infty = x^k\m{\Psi}_\infty + \frac {Q^k-x^k}{Q-x}
 W_2(-\hbar\pa_x)F.
\ee
Next we compute as for the previous proposition
\bea
&& 0= \le[\le(P-V'_1(Q)\ri)\m{\Psi}_\infty\ri]_0 = 
\le[-\hbar\pa_x\m{\Psi}_\infty + F -
V'_1(x)\m{\Psi}_\infty  - \frac {V'_1(Q)-V'_1(x)}{Q-x} W_2(-\hbar\pa_x)F\ri]_0 
\\
&&=
 - \hbar \psi_0'(x) -
V'_1(x)\psi_0(x) +\le[\1 - W_1(x)W_2(-\hbar\pa_x)\ri]_{00} f(x). 
\label{bootstrap}
\eea
Thus,  $\psi_0(x)$ must solve this first order system of ODE's. Since
there are $d_2+1$ choices for the function $f=f^{(\alpha)}$, we
correspondingly obtain $d_2+1$ independent solutions $\Psi^{(\alpha)}(x)$ to
the system.\par Now consider the second system (\ref{modifrec2}). The
compatibility gives
\bea
&& \hbar\m{\un\Phi}_\infty = [ \hbar\pa_x,x]\m{\un\Phi}_\infty 
 =  \hbar\pa_x\le(\m{\un\Phi}_\infty Q  + U \ri) -
x\le(\m{\un\Phi}_\infty P + U W_1(x)\ri) \\
&&=\le(\m{\un\Phi}_\infty P +  U W_1(x)\ri)Q  +  \hbar \pa_x U
-\le(\m{\un\Phi}_\infty Q + U \ri)P - U W_1(x)x \\
&&=\m{\un\Phi}_\infty[P,Q] - U P +U  \frac
{V'_1(Q)-V'_1(x)}{Q-x} (Q-x) + \hbar \pa_x U \\ 
&&=\hbar \m{\un\Phi}_\infty -U\bigg(P-V'_1(Q)\bigg)  + \bigg( \hbar\pa_x-V'_1(x)\bigg)U(x) 
\eea
Since the first entry of $U$ is nonzero and  $P-V'_1(Q)$
is strictly lower-triangular, the second term vanishes. The last term
gives the following ODE for the first entry of $U(x)$:
\be
 \hbar u'(x) = V_1'(x)u(x) \ \ \Rightarrow\ u(x) = c{\rm e}^{\frac 1
\hbar\,V_1(x)} \hbox{ or } u(x)\equiv 0\ .
\ee
We next consider the solutions $\ds{\m{\un\Phi}_\infty}$. By a  computation 
similar to the previous one, we have 
\bea
&&0= \le[\m{\un\Phi}_\infty \bigg(Q-V'_2(P)\bigg)\ri] _0 =   \le[
x\m{\un\Phi}_\infty -U -
V'_2( \hbar\pa_x)\m{\un\Phi}_\infty  + UW_1(x)\frac{V'_2(P)-V_2'(\hbar \overleftarrow
 \pa_x)}{P-
 \hbar \overleftarrow \pa_x} \ri]_0 \\
&&= \bigg(x-V'_2( \hbar \pa_x)\bigg)\phi_0(x) - u(x)\le[\1 - W_1(x)W_2(
\hbar\overleftarrow \pa_x)\ri]_{00} \label{comps}
\eea
This is a $d_2$-th order inhomogeneous ODE for the function
 $\un\phi_0(x)$. Choosing $u(x)\equiv 0$ as solution to
(\ref{comps}) we  have $d_2$ independent solutions corresponding to the
``unmodified'' system (\ref{unmodifb}). The choice 
\bea
u(x)=c{\rm
e}^\frac 1\hbar V_1(x)
\eea 
leads to one more independent solution of the inhomogeneous equation. Q.~E.~D.
 
We denote the solutions to system (\ref{modifrec2}) in general as
$\ds{\m{\un\Phi}_\infty\!^{(\alpha)}}$, $\alpha=0,\dots d_2$, where
$\alpha=0$ corresponds to the inhomogeneous solutions and 
$\alpha=1,\dots d_2$ correspond to the independent solutions to the
homogeneous case ($U(x)\equiv 0$). We will give explicit integral
representations for the $d_2+1$ solutions in the next section.
With the solutions $\ds{\m{\Psi}_\infty\!^{(\alpha)}, \
\m{\un\Phi}_\infty\!^{(\beta)}}$, $\alpha,\beta=0,\dots,d_2$, we can construct
the $(d_2+1)\times (d_2+1)$ modified kernels
\be
\m{K}^{N}\!^{(\alpha,\beta)}_{11}(x,x') := \sum_{n=0}^{N-1}
\un\phi_n^{(\alpha)}(x) \psi_{n}^{(\beta)}(x') =
\m{\un\Phi}_\infty\!^{(\alpha)}(x)
\m{\Pi}^{N-1}_{0} \m{\Psi}_\infty\!^{(\beta)}(x')\ .
\ee 
and  also obtain the following modified Christoffel--Darboux formulae
(cf. \cite{BEH}).
\bp[Christoffel--Darboux Kernels]
\bea
&& \hspace{-1.5cm}(x-x')\m{K}^N\!^{(\alpha,\beta)}_{11}(x,x') \\
&& \hspace{-1cm}=\le(\m{\un\Phi}_\infty\!^{(\alpha)}(x)Q+\delta_{\alpha 0} U(x)
\ri)\!\m{\Pi}^{N-1}_{0} \m{\Psi}_\infty\!^{(\beta)}(x') -
\m{\un\Phi}_\infty\!^{(\alpha)}(x)
\m{\Pi}^{N-1}_{0}\le( Q\m{\Psi}_\infty\!^{(\beta)}(x') - W_2(- \hbar
\pa_x')F^{(\beta)}(x')\ri) \\ 
&& \hspace{-1cm}=  -\m{\un\Phi}_\infty\!^{(\alpha)}(x)
\le[\m{\Pi}^{N-1}_{0},Q\ri] \m{\Psi}_\infty\!^{(\beta)}(x')  + \delta_{\alpha 0}
u(x) \psi_0^{(\beta)}(x') + \m{\un\Phi}_\infty\!^{(\alpha)}(x)\m{\Pi}^{N-1}_{0}
W_2(- \hbar \pa_x')F^{(\beta)}(x') \ .
\eea
\ep
Recall that the commutator of the finite band  matrix $Q$ with
the projector $\ds{\m{\Pi}^{N\!-\!1}_0}$ gives a finite-rank semi-infinite
matrix that corresponds to the Christoffel--Darboux kernel matrix
$\ds{\Amat^N}$. As a corollary, we obtain the pairing between the solutions of
the systems (\ref{ODEs1}) by setting $x=x'$
\bc
\bea
\bigg(\m{\un\Phi}^N\!^{(\alpha)}, \m{\Psi}_N\!^{(\beta)}\bigg)_N:= 
\m{\un\Phi}^N\!^{(\alpha)}(x)\Amat^N \m{\Psi}_N\!^{(\beta)}(x) = 
\m{\un\Phi}_\infty\!^{(\alpha)}(x)
\le[\m{\Pi}^{N-1}_{0},Q\ri] \m{\Psi}_\infty\!^{(\beta)}(x) =\cr
=   \delta_{\alpha 0}
u(x) \psi_0^{(\beta)}(x) + \m{\un\Phi}_\infty\!^{(\alpha)}(x)\m{\Pi}^{N-1}_{0}
W_2(- \hbar \pa_x)F^{(\beta)}(x)\ .
\eea
\ec
As shown in \cite{BEH}, this is  a constant (in $x$) if $N>d_2$. But
if $N>d_2$  the projector in the second term is irrelevant because the vector 
$W_2(- \hbar \pa_x)F^{(\beta)}(x)$ has only its first $d_2+1$ entries
nonvanishing, and hence we have, for $N>d_2$,
\be
\bigg(\m{\un\Phi}^N\!^{(\alpha)}, \m{\Psi}_N\!^{(\beta)}\bigg)_N
 =   \delta_{\alpha 0}
u(x) \psi_0^{(\beta)}(x) +\m{ \un\Phi}_\infty\!^{(\alpha)}(x)
\frac{V_2'(P)-V'_2(- \hbar \pa_x)}{P+\hbar\pa_x} F^{(\beta)}(x) \ .
\ee
If $\alpha\neq 0$ then 
\bea
\ds{\m{\un\Phi}_\infty\!^{(\alpha)}(x) P =
 \hbar \pa_x\m{\un\Phi}_\infty\!^{(\alpha)}(x)} \ ,
\eea
 and hence we have, for $ N>d_2, \quad \alpha\neq 0$,
\bea
 \m{\un\Phi}_\infty\!^{(\alpha)}(x)
\Amat^{N} \m{\Psi}_\infty\!^{(\beta)}(x) &=&  \m{\un\Phi}_\infty\!^{(\alpha)}(x)
\frac{V_2'( \hbar\overleftarrow
\pa_x)-V'_2(- \hbar\overrightarrow\pa_x)} { \hbar \overleftarrow\pa_x+
\hbar \overrightarrow
\pa_x} F^{(\beta)}(x) \cr
&=&\le.\frac{V_2'(
 \hbar\pa_{x'})-V'_2(- \hbar\pa_x)}{ \hbar\pa_{x'} + \hbar \pa_x} 
\un\phi_0^{(\alpha)}(x')f^{(\beta)}(x)\ri|_{x'=x} \ .
\label{acca}
\eea
In this expression $\un\phi_0$ and $f$ are kernel solutions of the  pair
of adjoint differential equations
\be
(V'_2( \hbar\pa_x)-x)\un\phi_0^{(\alpha)}(x) = 0\ ,\ \ \
(V'_2(- \hbar \pa_x)-x)f^{(\beta)}(x) = 0\ ,
\ee
and the last expression in eq. (\ref{acca}) is just the bilinear
concomitant of the pair (which is a constant in our case).
For $\beta=0$  we have 
\be
 \m{\un\Phi}_\infty\!^{(\alpha)}(x)
\Amat^{N} \m{\Psi}_\infty\!^{(0)}(x) = \delta_{\alpha 0} 
u(x)\psi_0^{(0)}(x) = \delta_{\alpha 0} c {\rm
e}^{\frac 1 \hbar V_1(x)}\frac 1{\sqrt{h_0}} {\rm e}^{-\frac 1 \hbar 
V_1(x)} =  \delta_{\alpha 0}\frac
c{\sqrt{h_0}}\ .\label{A0m} 
\ee

 For completeness we recall that the matrices $P, Q$  satisfy the following
deformation equations 
\bea
\hbar \pa_{u_K}Q =-\bigg [Q,{\mathrm U}^K \bigg]  \ ,&& \hbar \pa_{v_J}Q
=\bigg[Q, {\mathrm V}^J\bigg]\label{defQ} \ , \\
\hbar \pa_{u_K}P =-\bigg [P,{\mathrm U}^K \bigg] \ , &&\hbar \pa_{v_J}P
=\bigg[P, {\mathrm V}^J\bigg]\label{defP} \ ,
\eea
where 
\be
 {\mathrm U}^{K} :=-\frac 1 K \le\{ \le[Q^K\ri]_{>0} + \frac 1 2
\le[Q^K\ri]_0\ri\}\ ,\ \  
{\mathrm V}^{J} :=-\frac 1 J \le\{ \le[P^J\ri]_{<0} + \frac 1 2 
\le[P^J\ri]_0\ri\}\ .
\label{UKVJexpr}
\ee
(Note that in \cite{BEH} the matrix that here is denoted by $P$ was denoted
$-P$.) Here the subscripts $<0$ (resp. $>0$) mean the part of the matrix below
(resp., above)  the principal diagonal and the subscript $0$ denotes the
diagonal.
\bp
\label{rotturadica}
The equations in Proposition \ref{modrex} are compatible with the deformation
equations
\bea
&& \le\{
\begin{array}{l}
\ds{\hbar  \pa_{u_K}\m{\Psi}_\infty(x) = {\mathrm U}^K
\m{\Psi}_\infty(x)}\\[9pt]
\ds{\hbar\pa_{u_K}F(x) = \mathrm U^K F(x)}
\end{array}\ri. 
\label{defupsi}\\[4pt]
&&\le\{\begin{array}{l}
\ds{\hbar  \pa_{v_J}\m{\Psi}_\infty(x) = -{\mathrm V}^J
\m{\Psi}_\infty(x) - \frac 1 J \frac{P^J-(-\hbar
\pa_x)^J}{P+\hbar \pa_x} F(x)} \\  
\ds{\hbar  \pa_{v_J}F(x) = \le(\frac {(-\hbar \pa_x)^J}{J} + \mathrm
V^J_{00}\ri)F(x)}
\end{array}
\ri. \label{defvpsi}\\[4pt]
&&\le\{\begin{array}{l}
\ds{ \hbar\pa_{u_K}  \m{\un\Phi}_\infty(x) = - \m{\un\Phi}_\infty(x)
\mathrm U^K + \frac 1 K U(x)\frac{Q^K-x^K}{Q-x}}\\
\ds{ \hbar\pa_{u_K} U(x) = \le(\frac {x^K}K+\mathrm U^K_{00} \ri)U(x)}
\end{array}\ri.
 \label{defuphi}\\[4pt]
&&\le\{\begin{array}{l}
\ds{\hbar\pa_{v_J} \m{\un\Phi}_\infty(x) =  \m{\un\Phi}_\infty(x) \mathrm
V^J}\\
\ds{\hbar\pa_{v_J} U(x) = U(x)\mathrm V^J \ . }
\end{array}\ri.
\label{defvphi}
\eea
\ep
{\bf Proof}.
We only sketch the proof, which is quite straightforward but rather long. 

Compatibility of the deformation equations amongst themselves
follows from the zero curvature equations 
\be
[\hbar\pa_{u_K}\!\!+\!\mathrm U^K, \hbar\pa_{u_{K'}}\!\!+\!\mathrm U^{K'}] =
[\hbar\pa_{v_J}\!\!-\!\mathrm V^J, \hbar\pa_{v_{J'}}\!\!-\!\mathrm V^{J'}]
=[\hbar\pa_{u_K}\!\!+\!\mathrm U^K, \hbar\pa_{v_{J}}\!\!-\!\mathrm V^{J}] =0
\ee
(which are established in  the general theory \cite{UT,BEH}),
together with the fact that both vectors $U(x)$ and $F(x)$ have, by
assumption,  nonzero entries only in the first position.

Compatibility with eqs. (\ref{modifrec1}), (\ref{modifrec2})
requires some more computation. We first prove the compatibility between eq.
(\ref{modifrec2}) and the two eqs. (\ref{defuphi}), (\ref{defvphi}).
Compatibility between the equations involving multiplication by $x$ and
application of $\pa_{v_J}$:
\bea
&&\hbar x\pa_{v_J}\m{\un\Phi}_\infty(x) =\m{\un\Phi}_\infty(x) Q\mathrm V^J+
U(x) \mathrm V^J\label{d1e1}\\
&& \hbar \pa_{v_J}x \m{\un\Phi}_\infty(x) =\m{\un\Phi}_\infty(x) \mathrm V^J
Q+ \m{\un\Phi}_\infty(x)[Q, \mathrm V^J]+U(x)  \mathrm V^J  =
\hbox{ RHS of (\ref{d1e1})}\ .
\eea
Compatibility between the equation involving multiplication by $x$ and
application of $\pa_{u_K}$:
\bea
&&x\hbar \pa_{u_K}\m{\un\Phi}_\infty(x) =-\m{\un\Phi}_\infty(x)Q\mathrm
U^K-U(x)\mathrm U^K +  \frac 1 K U(x) \frac{Q^K-x^K}{Q-x}x \label{r13r}\\
&&\hbar \pa_{u_K}x\m{\un\Phi}_\infty(x) =\hbar \pa_{u_K} \m{\un\Phi}_\infty(x) Q-
\m{\un\Phi}_\infty(x) [Q, \mathrm U^K] + \hbar\pa_{u_K}U(x) \\
&&\ \ \ = - \m{\un\Phi}_\infty(x) \mathrm U^KQ +  \frac 1 K U(x)
\frac{Q^K-x^K}{Q-x}Q - \m{\un\Phi}_\infty(x) [Q,\mathrm U^K] +
\le(\frac {x^K} K+ \mathrm U^K_{00} \ri)U(x) \\
&&\ \ \ =- \m{\un\Phi}_\infty(x)Q \mathrm U^K+\frac 1 K U(x)
(Q^K-x^K)  + \frac 1 K U(x)
\frac{Q^K-x^K}{Q-x}x + \le(\frac {x^K} K+ \mathrm U^K_{00} \ri)U(x) 
\\
&&\ \ \ \m{=}^{(\star)} \hbox{ RHS of (\ref{r13r})}, 
\eea
where in the step $(\star)$, we have used the fact that (since only the
first entry of $U(x)$ is nonzero) 
\be
\frac 1 K U(x)Q^K + \mathrm U^K_{00} U(x) =- U(x)\mathrm U^K\label{cool} \ .
\ee

Compatibility between the $\pa_x$ and  $\pa_{u_K}$ equations is a bit more
involved:
\bea
&&\hbar^2\pa_x\pa_{v_J}\m{\un\Phi}_\infty(x) = \hbar\pa_x \m{\un\Phi}_\infty(x)
\mathrm V^J  =  \m{\un\Phi}_\infty(x)P\mathrm V^J +U(x)W_1(x)\mathrm
V^J\label{1231}\\
&&\hbar^2\pa_{v_J}\pa_x\m{\un\Phi}_\infty(x) = \hbar\pa_{v_J} \le(\m{\un\Phi}_\infty(x)P+U(x)W_1(x)\ri)
=\cr
&&\ \ \ =  \m{\un\Phi}_\infty(x)\mathrm V^J P +
\m{\un\Phi}_\infty(x)[P,\mathrm V^J ]+  U(x)\mathrm V^J W_1(x)+
U(x)[W_1(x),\mathrm V^J ] = \hbox{ RHS of (\ref{1231})}\ .
\eea
Finally we have  the $\pa_x, \pa_{v_J}$ compatibility:
\bea
&&\hbar^2\pa_x\pa_{v_J}\m{\un\Phi}_\infty(x)
=\hbar\pa_x\le(-\m{\un\Phi}_\infty(x)\mathrm U^K + \frac 1 K U(x)\frac
{Q^K-x^K}{Q-x} \ri) \cr
&&= -\le(\m{\un\Phi}_\infty(x) P+ U(x)W_1(x)
\ri)\mathrm U^K  + \frac 1 K \hbar \pa_x U(x)  \frac {Q^K-x^K}{Q-x} + \frac
1 K U(x) \hbar\pa_x\frac {Q^K-x^K}{Q-x} \cr
&&= -\le(\m{\un\Phi}_\infty(x) P+ U(x)W_1(x)
\ri)\mathrm U^K  + \frac 1 K( \hbar \pa_x-V_1'(x)) U(x)  
\frac {Q^K-x^K}{Q-x} \cr
&&\hspace{3cm}+ \frac 1 K U(x)
\le(V_1'(x)-V_1'(Q)+\m{V_1'(Q)-P}^{\hbox{lower tri.}}\ri)
\frac {Q^K-x^K}{Q-x} \cr
&&\hspace{3cm}+ \frac 1 K U(x) P \frac {Q^K-x^K}{Q-x} 
+ \frac 1 K U(x)
\hbar\pa_x\frac {Q^K-x^K}{Q-x} \cr 
&&\m{=}^{(\star \star)}-\le(\m{\un\Phi}_\infty(x)
P+ U(x)W_1(x)
\ri)\mathrm U^K  -\frac 1 K U(x)W_1(x)(Q^K-x^K) + \frac 1 K U(x) 
\frac {Q^K-x^K}{Q-x}P \cr
&&\hspace{3cm}+ \frac 1 K( \hbar \pa_x-V_1'(x)) U(x)  
\frac {Q^K-x^K}{Q-x}+ \frac 1 K U(x) (\hbar\pa_x+\hbar\pa_Q)
\frac {Q^K-x^K}{Q-x} \cr
 &&= -\m{\un\Phi}_\infty(x) P -U(x)W_1(x)\le(\frac 1 K
Q^K+\mathrm U^K\ri) +\frac {x^K}K U(x)W_1(x) + \frac 1 K U(x) 
\frac {Q^K-x^K}{Q-x}P  \cr
&&\hspace{3cm} + \frac 1 K( \hbar \pa_x-V_1'(x)) U(x)  \frac {Q^K-x^K}{Q-x} 
+\hbar U(x) \frac {Q^{K-1}-x^{K-1}}{Q-x}  \cr
&&=  -\m{\un\Phi}_\infty(x) P -U(x)\le(\frac 1 K Q^K+\mathrm
U^K\ri) W_1(x)+\frac {x^K}K U(x)W_1(x) + \frac 1 K U(x) 
\frac {Q^K-x^K}{Q-x}P  \cr
&&\hspace{3cm} + \frac 1 K( \hbar \pa_x-V_1'(x)) U(x)  \frac {Q^K-x^K}{Q-x} 
+\hbar U(x) \frac {Q^{K-1}-x^{K-1}}{Q-x}-U(x)[W_1(x),\mathrm U^K] \cr
 &&\m{=}^{(\star \star \star)}-\m{\un\Phi}_\infty(x) P +\le(\frac {x^K}K+
\mathrm U^K_{00}\ri) U(x) W_1(x) + \frac 1 K U(x) 
\frac {Q^K-x^K}{Q-x}P  \cr
&&\hspace{1.5cm} + \frac 1 K( \hbar \pa_x-V_1'(x)) U(x)  \frac {Q^K-x^K}{Q-x}
 + \hbar U(x) \frac {Q^{K-1}-x^{K-1}}{Q-x}-U(x)[W_1(x),\mathrm
U^K]\label{a2sw}\ .
\eea
In step $(\star\star)$ we have used the fact that the commutator of $P$
with a function of $Q$ is equivalent to the (formal) derivative $\hbar
\pa_Q$, while  in step $(\star \star \star)$ we have used eq. (\ref{cool}).
Computing the derivatives in the opposite order we obtain
\bea
&&\hbar^2\pa_{v_J}\pa_x\m{\un\Phi}_\infty(x)=
\hbar\pa_{v_J}\le(\m{\un\Phi}_\infty(x)P + U(x)W_1(x)\ri) \cr
&& =\le(-\m{\un\Phi}_\infty(x)\mathrm U^K + \frac 1 K U(x)\frac {Q^K-x^K}{Q-x}
\ri)P   + \le(\frac {x^K}K+\mathrm U^K_{00} \ri)U(x)W_1(x) \cr
&&\qquad  + \hbar  U(x) \frac{x^{K-1}-Q^{K-1}}{x-Q} 
- U(x)[W_1(x),\mathrm U^K]\label{1112}.
\eea
Subtracting eq. (\ref{a2sw}) from eq. (\ref{1112}), we are left with
\bea
&&
0= \hbox{ RHS of  (\ref{a2sw}) -  RHS of (\ref{1112})} =  \frac 1 K(
\hbar \pa_x-V_1'(x)) U(x)  \frac {Q^K-x^K}{Q-x}
\cr
&&\Longrightarrow \ \big(\hbar
\pa_x-V_1'(x)\big) U(x) = 0\ ,
\eea
where the implication follows from the fact that only the first entry
of $U(x)$ is nonzero. This is precisely the same compatibility
condition that we found in Prop. \ref{modrex}.
The other compatibility checks are also rather long but routine and are
left to the reader to fill in. Q.~E.~D.\par\vskip 4pt

\subsection{Explicit Integral representations for the dual wave-vectors }
\label{explicit}
\bp
\label{intexp}
 
The $d_2+1$ semi-infinite wave-vectors $\ds{\m{\un
\Phi}_\infty\!^{(\alpha)}}$
and  functions $u^{(\alpha)} (x)$, $\alpha=0,\dots d_2$ defined by 
\bea
&& \m{\un \Phi}_\infty\!^{(0)}(x) = 
{\rm e}^{\frac 1 \hbar V_1(x)}
\int_{\varkappa\Gamma} \!\!\!\!\!{\rm d}s\,{\rm d}y \, \frac {{\rm
e}^{-\frac 1 \hbar (V_1(s)-sy)}}{x-s}{\m{\Phi}}^t(y)\ ,\ \  u^{(0)}(x):= \sqrt{h_0}
    {\rm e}^{\frac 1 \hbar V_1(x)} \label{intphi0}\\[10pt]
&&\m{\un\Phi}_\infty\!^{(k)}(x)
 = \int_{\Gamma_k^{(y)}} {\rm d}y \,{\rm e}^{\frac
{xy}\hbar } {\m{\Phi}}^t(y)\ ,\ \ u^{(k)}(x)\equiv 0\label{intphik},\qquad
k=1,\dots ,d_2,
\eea
are independent solutions of the modified system (\ref{modifrec2}).
\ep
{\bf Proof.}
There is nothing to prove for $\ds{\m{\un\Phi}_\infty\!^{(k)}}$, since for
$k>0$  these are just the Fourier-Laplace transforms, which satisfy the
corresponding unmodified system (\ref{unmodifa}), which is the same as the
modified system with $u(x)\equiv 0$. Let us therefore consider
$\ds{\m{\un\Phi}_\infty\!^{(0)}}$. We first check the $x$ recurrence relations.
\beas
&& x\m{\un \Phi}_\infty\!^{(0)}(x)= 
{\rm e}^{\frac 1 \hbar V_1(x)}
\int_{\varkappa\Gamma} \!\!\!\!\!{\rm d}s\,{\rm d}y \, \le[\frac {(x-s){\rm
e}^{-\frac 1 \hbar(V_1(s)-sy)}}{x-s}{\m{\Phi}}^t(y) +\frac {{\rm
e}^{-\frac 1 \hbar(V_1(s)-sy)}}{x-s}s{\m{\Phi}}^t(y)\ri] \\
&&= \le[\sqrt{h_0} {\rm
e}^{\frac 1 \hbar V_1(x)},0,\dots\ri]  - {\rm e}^{\frac 1 \hbar
V_1(x)}\int_{\varkappa\Gamma} \!\!\!\!\!{\rm 
d}s\,{\rm d}y \,  \frac {{\rm
e}^{-\frac 1 \hbar(V_1(s)+sy)}}{x-s}\hbar \pa_y {\m{\Phi}}^t(y) \\
&&\m{=}^{(\star)} U^{(0)}(x)+ \m{\un \Phi}_\infty\!^{(0)}(x)Q\ ,
\eeas 
where in $(\star)$ we have used eq. (\ref{matrixrecb}) and defined  the
semi-infinite vector 
\be
U^{(0)}(x) :=  \le[\sqrt{h_0} {\rm
e}^{\frac 1 \hbar V_1(x)},0,\dots\ri].\label{defU}
\ee
It is easy to prove by induction in $k$,  starting from
the first part of eq. (\ref{modifrec2}), that 
\be
x^k \m{\un\Phi}_\infty\!^{(0)}(x) = U^{(0)}(x)
\frac{x^k-Q^k}{x-Q} + \m{\un\Phi}_\infty\!^{(0)}(x)Q^k\label{ww}\ .
\ee
We now consider the $\pa_x$ differential recursion. After shifting
the derivative in $x$ to one in $s$ inside the integral, and integrating by
parts, using eq. (\ref{ww}),  we get 
\bea
&& \hbar\pa_x \m{\un\Phi}_\infty\!^{(0)}(x) = 
U^{(0)}(x)\frac{V_1'(x)-V_1'(Q)}{x-Q} + \m{\un\Phi}_\infty\!^{(0)}(x)P \label{diffrhoeq}\ .
\eea
This proves that the given integral expression is indeed the additional
solution to eq. (\ref{modifrec2}).
{Q.~E.~D.}
\br
We would also have solutions for any other choice of
admissible contours (or linear combinations of them) in eq. (\ref{intphik}).
This arbitrariness will be used later.
\er
\br
The functions $\un \phi_n^{(0)}(x)$ are piecewise analytic functions
in each connected component of $\C_x\setminus \bigcup_{j=1}^{d_1}
 \Gamma_j^{(x)}$, but can be analytically continued from each such
 connected component to entire functions by deforming the contours of
integration.
\er
With regard to the deformation equations studied in Propostion 
\ref{rotturadica}, we have:
\bl
\label{uvPSiPhidefs}
The wave-vectors $\ds{\m{\Psi}_\infty\!^{(0)}(x)}$ (i.e. the
quasipolynomials) and $\ds{\m{\un\Phi}_\infty\!^{(k)}(x)}$,
$k=1,\dots,d_2$ (i.e. the Fourier--Laplace transform of the
quasipolynomials  $\phi_n(y)$) satisfy the following deformation equations
\bea
\hbar \pa_{u_K} \m{\Psi}_\infty\!^{(0)}(x) &=& {\mathrm U}^K
\m{\Psi}_\infty\!^{(0)}(x)\ , \label{defUK}\\ 
\hbar \pa_{v_J} \m{\Psi}_\infty\!^{(0)}(x)  &=& -{\mathrm V}^J \m{\Psi}_\infty\!^{(0)}(x)  \
,\label{defVJt}\\
 \hbar \pa_{u_K} \m{\un\Phi}_\infty \!^{(k)}(x)&=&  -\m{\un\Phi}_\infty\!^{(k)}(x) {\mathrm U}^K
\ ,\label{defUKt}\\ 
\hbar \pa_{v_J} \m{\un\Phi}_\infty\!^{(k)}(x)&=& \m{\un\Phi}_\infty \!^{(k)}(x){\mathrm V}^J\ ,
\label{defVJ} \eea
\el
{\bf Proof.}
This follows straightforwardly from the  two equations (\ref{defQ}),
(\ref{defP}) together with the fact that taking a Fourier--Laplace transform
commutes with differentiation with respect to the deformation parameters
 Q.~E.~D. \par\vskip 4pt

We also remark that the given integral expressions satisfy the modified
deformation equations given by the following proposition.
\bp
\label{defwave}
The dual wave-vectors defined in Proposition \ref{intexp} satisfy
the following deformation equations
\bea
&& \hbar \pa_{u_K}\m{\un\Phi}_\infty\!^{(0)}(x) = -
\m{\un\Phi}_\infty\!^{(0)}(x){\mathrm U}^K+\frac 1 K U^{(0)}(x) \frac {x^K-Q^K}{x-Q} \\
&& \hbar \pa_{v_J}\m{\un\Phi}_\infty\!^{(0)}(x) =
\m{\un\Phi}_\infty\!^{(0)}(x){\mathrm V}^J\ ,
\eea
where the semi-infinite vector $U^{(0)}(x)$ is defined in eq. (\ref{defU}).
\ep
{\bf Proof}. 
As for the \{$v_J$\} deformation equations for 
$\ds{\m{\un\Phi}_\infty\!^{(0)}}$, the proof follows from Lemma
\ref{uvPSiPhidefs} and from the fact that 
the integral transform defining this wave-vector does not depend on
the 
$v_J$'s. On the other hand, the $u_K$ deformations give (again  using Lemma
\ref{uvPSiPhidefs}):
\be
\hbar \pa_{u_K}\m{\un\Phi}_\infty\!^{(0)}(x) = -
\m{\un\Phi}_\infty\!^{(0)}(x){\mathrm U}^K+\frac 1
K \m{\un\Phi}_\infty\!^{(0)}(x) \le(x^K-Q^K\ri)\ ,
\ee
from which the proof follows using eq. (\ref{ww}).
Q.~E.~D.

\br
Note that from the definition of the vector $U(x)$, one can verify directly
that it satisfies the deformation equations (\ref{defuphi}), (\ref{defvphi}) by
using the equations for the normalization constant $h_0$: 
\bea && h_0 =
\int_{\varkappa\Gamma}{\rm d}x{\rm d}y\, {\rm e}^{-\frac 1 \hbar
(V_1(x)+V_2(y)-xy)},\\ 
 && \frac \hbar 2 \pa_{u_K}\ln(h_0) = \mathrm U^K_{00}\ ,\ \ \frac \hbar 2
\pa_{v_J}\ln(h_0) = \mathrm V^J_{00}\ .
\eea
\er

\br
\label{defwaverem}
In particular, the sequences of functions defined in Proposition \ref{intexp}
satisfy identical deformation equations for $N$ large enough ($N>d_1$) 
 and hence the relevant windows satisfy the full DDD equations
specified in \cite{BEH} for the barred (dual) quantities.
\er

We conclude with a discussion regarding the construction of an integral
representation for the wave-vectors $\ds{\m{\Psi}_\infty}$.
For the unmodified system (\ref{unmodifa}) the solution is explicitly
given by the quasipolynomials $\ds{\m{\Psi}_\infty\!^{(0)}(x)}$. The other
solutions of the wave-equations (\ref{modifrec1}) are in $1-1$ correspondence
with the $d_2$ solutions of eq. (\ref{dualground}) via eq. (\ref{bootstrap}). 
We could try to define an integral representation of the form
\be
\m{\Psi}_\infty\!^{(k)}(x) =  \int_{\widetilde
\Gamma_k^{(y)}}\!\! \frac{{\rm d}y}{2i\pi\hbar} \,
{\rm e}^{\frac 1 \hbar (V_2(y)-yx)} \int_{\varkappa\Gamma}\!\!\!\!\!{\rm d}z\,{\rm
d}t \,\frac {{\rm e}^{-\frac 1 \hbar (V_2(t)-zt)}}{t-y}\m{\Psi}(z)\ ,\label{formalpsi}
\ee
corresponding to the following solution of
eq. (\ref{dualground})
\be
 f^{(k)}(x) := \frac{\sqrt{h_0}}{2i\pi\hbar} 
\int_{\widetilde \Gamma_k^{(y)}} \!\!\!\!
    {\rm d}y\, {\rm e}^{\frac 1 \hbar (V_2(y)-xy)}
 \qquad k=1,\dots ,d_2 .\label{intpsik}
\ee
However formula (\ref{formalpsi}) is not well-defined, since the inner integral
defines only a piecewise analytic function with jump discontinuities
along the $\Gamma_k^{(y)}$'s, and each of the anti-wedge
contours $\widetilde\Gamma_k^{(y)}$ crosses at least one of the contours
$\Gamma_j^{(y)}$. One could alternatively take the analytic continuation of the
inner double integral from one connected component of $\C_y\setminus
\bigcup_{k=1}^{d_2} \Gamma_k^{(y)}$ to an entire function,
 which amounts to deforming the
contours in $\varkappa\Gamma$ to avoid any jump discontinuities.
 But then the outer integral is in general divergent, due
to the behaviour of the integrand as $y\rightarrow \infty$.

 Thus, as it stands, eq. (\ref{formalpsi}) may only be viewed as a formal
expression which at best defines a divergent integral.  However, treating
it naively (as though it were uniformly convergent with respect to the
parameters $\{u_J, v_K\}$, as well as $x$) we would find that it
indeed satisfies eqs. (\ref{modifrec1})  (\ref{defupsi}) and (\ref{defvpsi}).
Nonetheless we  know that there are solutions
$\ds{\m{\Psi}_\infty\!^{(k)}}$ of eq. (\ref{modifrec1})
(\ref{defupsi})  corresponding to the $f^{(k)}(x)$ defined in
(\ref{intpsik}), determined 
recursively from the solutions of eq. (\ref{bootstrap}). Such solutions are
uniquely defined by the choice of a particular solution to eq.
(\ref{dualground}), up to the addition of a solution of the homogeneous
equation (that is, the quasipolynomials) forming
$\ds{\m{\Psi}_\infty\!^{(0)}}$.
 
There does exist a way to arrive at convergent integral representations,
however, by forming suitable linear combinations of Fourier-Laplace transforms
of a piecewise analytic function defined along appropriate contours. This was
done for the case of cubic potentials by Kapaev in \cite{kapa}. In Appendix A,
it will be indicated how the procedure used there may be extended to polynomial
potentials of higher degree. For  present purposes, however, we may view the
solutions as just defined by the recursion relations (\ref{modifrec1}),
together with eq. (\ref{bootstrap}). It follows from compatibility that these
satisfy the deformation equations as well.

As a further remark, we note that the solutions $\ds{\m{\Psi}_\infty\!^{(k)}}$
($\ds{\mod \C\m{\Psi}_\infty\!^{(0)}}$) associated to the $f^{(k)}$s defined 
by eq. (\ref{intpsik}), may be associated to any linear combination
of the anti-wedge contours, used as contours of integration in
(\ref{intpsik}). This freedom will be used in the next section.

\subsection{Diagonalization of the Christoffel--Darboux pairing}
\label{diagonalization}

We know from Proposition \ref{defwave} that the windows constructed from the
integral representations in Proposition \ref{intexp} provide solutions to the
DDD equations for the barred system (for
$N>d_1$).
On the other hand, we are guaranteed by  Theorem 4.1 and Corollary
4.1 of \cite{BEH} that the Christoffel--Darboux pairing between such
solutions and any solution of the unbarred DDD system does not depend on
 $x$, $N$, or the deformation parameters determining the two potentials. It is
of interest therefore to compute this pairing for the explicit solutions at
hand.

\begin{figure}[!ht]
\centerline{
\epsfxsize 10cm
\epsfysize 10cm
\epsffile{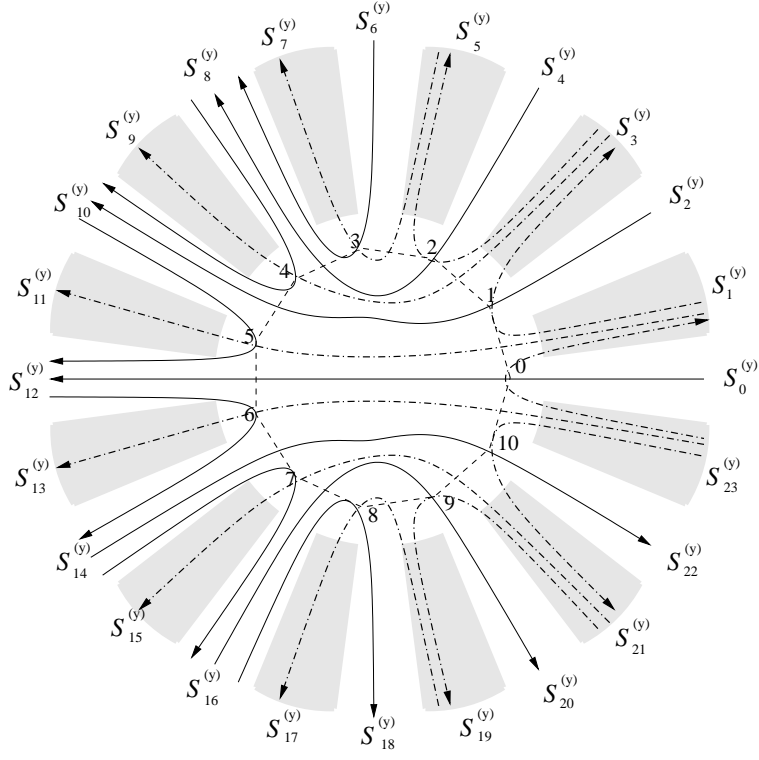}}
\caption{Example of the asymptotic (as $|x|\to \infty$) 
SDC's $\gamma$ (solid) and SAC's $\tilde \gamma$
(line-dot-line) in the $y$-plane for a  
potential of degree $d_2+1=12$ with positive leading coefficient 
($v_{d_2+1}>0$) and for the
non-Stokes line $\arg(x)=0$. The numbers at the vertices of the
endecagon label the $d_2=11$ critical points and the pair of dual SDC and
SAC passing through each them.}\label{fig1}
\end{figure}

The solutions of the unbarred DDD equations are obtained by taking
suitable windows in the wave-vector solutions of eq. (\ref{modifrec1}),
consisting of the quasipolynomials $\ds{\m{\Psi}_\infty\!^{(0)}(x)}$
and the solutions $\ds{\m{\Psi}_\infty\!^{(k)}(x)}$ associated to the
$f^{(k)}$ (eq. \ref{intpsik}). It should be clear that the diagonalization of
the pairing depends on a careful choice  of the contours of integration
in eqs. (\ref{intphik}), (\ref{intpsik}). Here we prove that the choice of
contours that diagonalizes the pairing (\ref{pairing}) is linked to the notion
of dual steepest descent-ascent contours, whose definition is  given here and
will be needed again in the following section.
\bd
\label{defSDC}
The steepest descent  contours (SDC's) and the dual steepest ascent contours
(SAC's) for integrals of the form  
\be
I(x):= \int_\Gamma\!\!\!{\rm d}y\, {\rm e}^{-\frac 1 \hbar (V_2(y)-xy)} H(y)\ ,\qquad
\widetilde I(x) := 
\int_{\widetilde \Gamma}\!\!\!{\rm d}y\, {\rm e}^{\frac 1 \hbar (V_2(y)-xy)} H(y)\ , 
\ee
respectively, passing through any saddle point $y_k(x)$, $k=0 \dots d_2-1$
with $H(y)$ at most of exponential type, are the contours $\gamma_k$
and $\tilde \gamma_k$, respectively,  
uniquely defined by
\bea
&& \gamma_k:=\le\{y\in \C;\ \Im(V_2(y)-xy) =
\Im\le(V_2(y_k(x))-xy_k(x)\ri)\  , \Re
(V_2(y))\m{\longrightarrow}_{
\shortstack{
\scriptsize $y\to\infty$\\
\scriptsize $y\in \gamma_k$}} +\infty\  \ri\} \\
 && \widetilde\gamma_k:=\le\{y\in \C;\ \Im(V_2(y)-xy) =
\Im\le(V_2(y_k(x))-xy_k(x)\ri)\  , \Re
(V_2(y))\m{\longrightarrow}_{
\shortstack{
\scriptsize $y\to\infty$\\
\scriptsize $y\in \widetilde \gamma_k$}} -\infty\ \ri\}\ . 
\eea
Here  $y_k(x)$ denotes  one of the $d_2$ branches of the solution to the
algebraic equation
\be
V'_2(y)=x\ ,
\ee
which behaves like 
\be
y_k(x)\m{\sim}_{x \to \infty} (v_{d_2+1})^{-\frac 1{d_2}}\omega^k 
 x^{\frac 1{d_2}}\ ,\ \ \ \omega:= {\rm e}^{\frac {2i\pi}{d_2}}\ ,
\ee 
where $x^{\frac 1{d_2}}$ denotes the principal $d_2$-th root of $x$.
(Note that the homology class of the SDC's and SAC's becomes a constant for
$|x|$ sufficiently large along a generic ray 
(see Fig. \ref{fig1}). It will
be proved in Section \ref{stokesmats} that the homology class is also  {\em
locally constant} with respect to the angle of the ray.
\ed

For the following,  we also define the sectors
\bea
&&\mathcal S_k:=\le\{x\in \C\ ,\ {\rm arg}(x)\in
\le(-\vartheta_y+\frac{(2k-1)\pi}{2(d_2+1)}, -\vartheta_y+\frac{(2k+1)\pi}{2(d_2+1)} \ri)\ri\},\
\hbox{ for $d_2$ odd}\nonumber \\
&&\mathcal S_k:=\le\{x\in \C\ ,\ {\rm arg}(x)\in
\le(-\vartheta_y+\frac{k\pi}{(d_2+1)},-\vartheta_y+ \frac{(k+1)\pi}{(d_2+1)} \ri)\ri\},\
\hbox{ for $d_2$ even}\nonumber \\
&&k=0,\dots,
2d_2+1\ ,\label{Stokessects}
\eea
(where $\vartheta_y$ was defined in eq. (\ref{secty})),
and denote by $\mathcal R_k$, $k=0,\dots 2d_2+1$  the 
rays separating them, counting counterclockwise, starting from $\mathcal S_0$.
These will be shown in the following section to be interpretable as Stokes' rays,
defining a part of the Stokes' sectors in our Riemann--Hilbert
problem. It will be shown in  
Sect. \ref{asymptotics} that if $x$ approaches $\infty$ along a ray
distinct from the $\mathcal R_k$'s, then the homology class of the
SDC's and SAC's in 
Def. \ref{defSDC} is well defined. 
We then have 
\bp
Fix any (non-Stokes) ray $\{ \arg(x)=\alpha= {\rm const.}\}\neq
{\mathcal R}_k $ and the corresponding
limiting (for large $|x|$) homology class of the 
dual steepest descent-ascent contours
$\gamma_k$  and  $\tilde \gamma_k$, $k=1,\dots d_2$ of Def. \ref{defSDC}.
With this choice of the contours of integration, use the SDC's to redefine the
Fourier-Laplace transform (\ref{intphik}), and the SAC's to redefine the
$f^{(k)}$s  of eq. (\ref{intpsik}),  and the associated wave-vectors 
$\ds{\m{\Psi}_\infty\!^{(k)} \mod
\C\m{\Psi}_\infty\!^{(0)}}$.
Then 
\bea
C^{\alpha\beta}:= 
\m{\un\Phi}^{N}\!^{(\alpha)} (x)\Amat^{N} \m{\Psi}_{N}\!^{(\beta)}(x) =
\le[
\begin{array}{c|c}
1&0\\
\hline
0 & \begin{array}{ccc}
&&\\
&\1_{d_2}&\\
&&
\end{array}
\end{array}
\ri] = \1_{d_2+1}\ ,\qquad N>d_2\ ,
\eea
up to the addition of a  suitable multiple of $\ds{\m{\Psi}_\infty\!^{(0)}(x)}$ to
the solutions
$\ds{\m{\Psi}_\infty\!^{(k)}(x)}$.
\ep
{\bf Proof.}\\
For $\beta=0$  the statement follows  from eq. (\ref{A0m}) (where the
constant $c$ equals  $\sqrt{h_0}$).
For $\alpha=0\neq \beta$ we use the fact that the solutions
$\ds{\m{\Psi}_\infty\!^{(k)}(x)}$ associated to $f^{(k)}$ are defined only
up to the addition of a multiple of $\ds{\m{\Psi}_\infty\!^{(0)}(x)}$.
Indeed, we know from the general theory that  the pairing of the two solutions
is constant, and we may use this freedom to normalize the constant to zero
by adding a suitable multiple of  the homogeneous solution of eq.
(\ref{modifrec1}) $\ds{\m{\Psi}_\infty\!^{(0)}(x)}$ (i.e., the
quasipolynomials), which is ``orthogonal'' to the solutions
$\ds{\m{\un\Phi}_\infty\!^{(\alpha)}(x),\
\alpha\neq 0}$, as follows again from eq. (\ref{A0m}).\par  Finally for
$k=\alpha\neq 0 \neq \beta=j$, $C^{kj}$ equals the bilinear concomitant of
the corresponding functions (eq. \ref{acca}). Let us consider $x$ belonging to
a fixed ray, and choose a basis of steepest descent contours  $\gamma_k$ and
steepest ascent contours $\widetilde
\gamma_k$ (whose homology does not change as $x\to\infty$ on the fixed
ray for $|x|$ big enough). We use
formula (\ref{acca}) with 
\be
f^{(j)}(x):= \frac {\sqrt{h_0} }{2i\pi\hbar} 
\int_{\widetilde\gamma_j}\!\!\!\!\!{\rm d}y\, {\rm
e}^{\frac 1 \hbar (V_2(y)-xy)}\ ,\ \ \un\varphi_0^{(k)}(x):=\frac 1{\sqrt{h_0}} 
\int_{\gamma_k}\!\!\!\!{\rm d}y\,{\rm e}^{-\frac 1 \hbar (V_2(y)-xy)}\ .
\ee 
The respective asymptotic behaviors for large $|x|$, computed by the saddle--point
method, are
\bea
&& f^{(j)}(x)\simeq 
 \frac {\sqrt{h_0} }{2i\pi\hbar} 
{\rm e}^{\frac 1 \hbar (V_2(y_j(x))-xy_j(x))} \sqrt{\frac
{-2\pi\hbar }{V_2''(y_j(x))}}\big(1+\mathcal O(\lambda^{-1}) \big)\label{aa}\\
&&\un\varphi_0^{(k)}(x) \simeq \frac 1{\sqrt{h_0}} {\rm
e}^{-\frac 1 \hbar (V_2(y_k(x))-xy_k(x))} \sqrt{\frac
{2\pi\hbar }{V_2''(y_k(x))}} \big(1+\mathcal O(\lambda^{-1})
\big)\label{bb}\ ,
\eea
where $y_k(x)$ are as in Def. \ref{defSDC}. The bilinear
 concomitant is a constant, so it must vanish for $j\neq k$ since
the exponential parts of the asymptotic forms in eqs. (\ref{aa}), (\ref{bb})
cannot give a nonzero constant when multiplied together. 

For $j=k$, the bilinear concomitant is given by the integral
\bea
&& \le.\frac{V_2'(\hbar
\pa_{x'})-V'_2(-\hbar\pa_x)}{\hbar\pa_{x'}+\hbar \pa_x} 
\un\varphi_0^{(k)}(x')f^{(k)}(x)\ri|_{x'=x}\cr
&&\qquad =\frac 1{2i\pi\hbar} 
\int_{\widetilde \gamma_k\times \gamma_k} \!\!\!\!\!\!\! {\rm
d}y\,{\rm d}y'\, {\rm e}^{\frac 1 \hbar (V_2(y)-V_2(y')-x(y-y'))} \frac
{V_2'(y)-V_2(y')}{y-y'}  \cr
&& \qquad   
\simeq \frac 1{2i\pi\hbar}{\rm e}^{\frac 1 \hbar (V_2(y_k(x))-xy_k(x))} \sqrt{\frac
{2\pi\hbar }{V_2''(y_k(x))}}   \cr
&& \qquad\qquad\times {\rm
e}^{-\frac 1 \hbar (V_2(y_k(x))-xy_k(x))} \sqrt{\frac
{-2\pi\hbar}{V_2''(y_k(x))}}\big(1+\mathcal O(\lambda^{-1})
\big)V_2''(y_k(x)) = 1\ .
\eea
This concludes the proof. Q.~E.~D.\par\vskip 3pt
Had we chosen the contours as $\widetilde \Gamma_k^{(y)}$ and
$\Gamma_k^{(y)}$ rather than the steepest descent contours, we
would have had a constant matrix for
$C^{\alpha\beta}$. Notice that the pairing of these integrals is also
independent of the deformation parameters determining $V_1,\ V_2$, as well
as of the choice of the integer $N$ defining the window. As shown in
\cite{BEH}, this can always be accomplished through a suitable choice of basis.
But here we have explicitly shown how this occurs for the particular
normalizations chosen in the integrals.

\section{Asymptotic behavior at infinity and Riemann--Hilbert problem}
\label{asymptotic}

\subsection{Stokes sectors and sectorial asymptotics}
Given the duality between the ODE's involving  $\m{D_1}^N$ and $\m{\un D_1}^N$
implied by eq. (\ref{D1intertwining}), it is clear that the Stokes matrices
around the irregular singularity at $x=\infty$ for systems of the form 
(\ref{ODEs1Psi}) and (\ref{ODEs1}) are related. Using the explicit
integral representations of the fundamental solutions, we can determine the
asymptotic behavior  from these integral representations by saddle point
methods.  The solutions for the $\ds{\m{\un D_1}^N}$ system are simpler to
analyze and do not involve the problems of divergent integrals discussed in
the section \ref{explicit} (and the Appendix). But in principle one could
consider the asymptotic behavior of the solutions to the
system (\ref{ODEs1Psi}) directly, by taking  suitable windows in the
wave-vector solutions of eq. (\ref{modifrec1}).

\br
\label{duality}
We have seen in Sect. \ref{diagonalization} that one can choose the
wave-vector solutions to eqs. (\ref{modifrec1}), (\ref{modifrec2}) 
 so that the Christoffel-Darboux pairing between the dual windows of
fundamental solutions implies the identity
\be 
\m{\mathbf{\un\Phi}}^{N} (x)\Amat^{N} \m{\mathbf \Psi}_{N} (x)=\1_{d_2+1}
 ,\qquad N>d_2\ .
\ee
Therefore, if we formulate a Riemann--Hilbert problem for
$\ds{\m{\mathbf{\un\Phi}}^{N} (x)}$, we can immediately derive the
corresponding one for $\ds{\m{\mathbf \Psi}_{N} (x)}$. Since the asymptotic
forms are mutual inverses, the Stokes (and jump) matrices for the one must
be the inverses of those for the other. (In our conventions, the Stokes
matrices for $\ds{\m{\mathbf{\un\Phi}}^{N} (x)}$ act on the left, while for
$\ds{\m{\mathbf \Psi}_{N} (x)}$ they act on the right.)  The formulation for
the dual wave-functions is considerably easier, so this is what we analyze here
in detail. 
\er

From the general theory of ODE's, since the matrix $\ds{\m{\un D_1}^N(x)}$ is of
degree $d_1$, one would expect $d_1+1$ Stokes sectors. However, this is true
only if the leading term of the matrix has a nondegenerate spectrum. In the case
at hand, however, we have 
\be
\m{\un D_1}^{N}(x) \sim  x^{d_1}\le[\begin{array}{c|c}
1& 0\\
\hline
0&\begin{array}{ccc}
&&\\
&0&\\
&&
\end{array}
\end{array} 
\ri]+ \mathcal O(x^{d_1-1})\ .
\ee
Since the spectrum of the leading term has a $d_2$-fold degeneracy, we
have more complicated asymptotic behavior and the occurrence of more
Stokes sectors.

A few more preparatory remarks are required. From the discussion in sections
\ref{fundamental} and \ref{explicit} we obtain that the fundamental system for
the differential-difference equations specified in Lemma \ref{D1s} is provided
by 
\bea
\m{\mathbf {\un\Phi}}^N(x) := \le[
\begin{array}{c}
\ds{\m{\un\Phi}^N\!^{(0)}(x)}\\
\ds{\m{\un\Phi}^N\!^{(1)}(x)}\\
\vdots\\
\ds{{\m{\un\Phi}^N\!^{(d_2)}}(x)}
\end{array}
\ri]\ ,\qquad N\geq d_1+1,\label{piecewise}
\eea
where $\{\ds{\m{\un\Phi}^N\!^{(k)}(x)}\}_{k=0,\dots d_2}$
are windows constructed from the wave-vectors defined in Proposition
\ref{intexp}.  Given these integral representations of the solutions, the
asymptotic behavior and the Riemann--Hilbert problem can be determined by means
of the steepest descent method.

\begin{figure}[!ht]
\centerline{
\epsfxsize 14cm
\epsffile{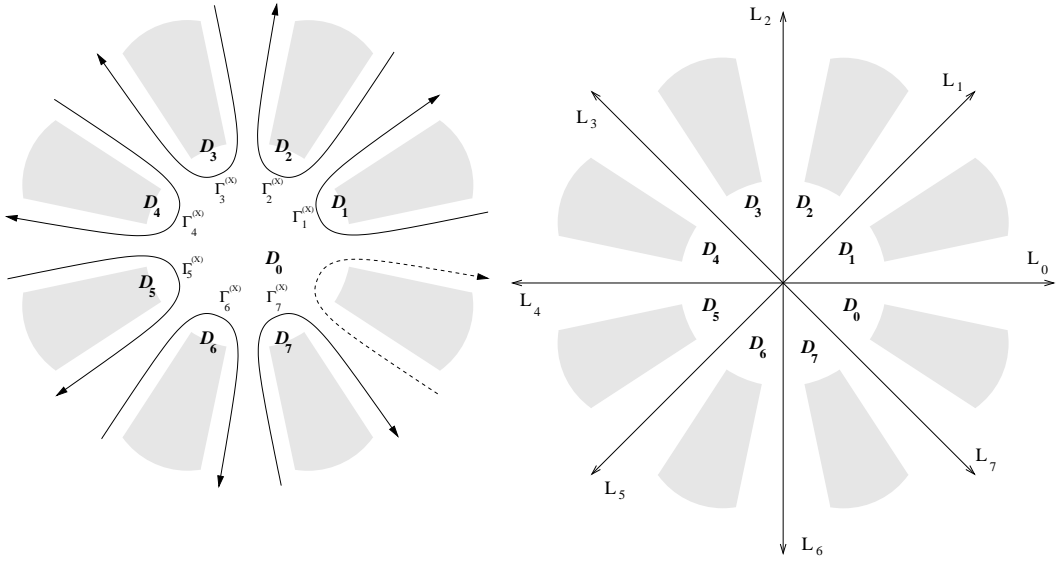}}
\caption{Example of the two possible 
choices for the domains of definition of $\un\phi_n^{(0)}$ (in the $x$-plane)
 in the case $d_1=7$}
\label{fig3}
\end{figure}

 While  $\{\ds{\m{\un\Phi}^N\!^{(k)}(x)}\}_{k=1,\dots d_2}$ are entire
functions, the integral defining the window
$\ds{\m{\un\Phi}^N\!^{(0)}(x)}$ defines in fact a piecewise analytic
vector function. Its domains of analyticity are the $d_1+1$ connected 
components in which the $x$-plane is partitioned by the contours
$\Gamma_j^{(x)}$, $j=1,\dots d_1$.
We denote by  $\mathcal D_j$ the connected
domain to the right of the contours $\Gamma_j^{(x)}$ for
$j=1,\dots,d_1$ and by 
$\mathcal D_{0}$ the domain to the left of all contours.
(See figure \ref{fig3} for the case $d_1=7$.)
The fundamental piecewise-analytic solution $\ds{\m{\mathbf{\un\Phi}^N}(x)}$
defined in eq. (\ref{piecewise}) then satisfies the jump equations
\bea
{\m{\bf \un \Phi}^N}_{+} (x) = \le[\begin{array}{ccccc}
1 &  2i\pi\varkappa^{\mu,1}& 2i\pi\varkappa^{\mu,2}& \cdots &
2i\pi\varkappa^{\mu,d_2}\\
0&1&0&\cdots &0\\
0&0&1&\cdots &0\\
0&0&0&\ddots& \\
0&0&0&\cdots&1
\end{array}\ri]{\m{\bf \un \Phi}^N}_{-} (x)\ ,\ \ \
x\in\Gamma_\mu^{(x)},\ \mu=1,\dots,d_1\ ,
\eea
where the subscripts $_+$, $_-$ denote the limiting values from the
right or the left, respectively, with respect to  the orientation of the
contour. Since we can arbitrarily deform  the contours in the finite part of
the $x$-plane, we can,  by retracting the $d_1$ contours to the origin arrange
that the $d_1+1$ regions all become wedge-shaped sectors. Using this
freedom in the choice of the  contours $\Gamma_j^{(x)}$,
 denote by  $L_\mu$ $\mu=0,\dots,d_1$ the oriented rays starting from the origin
and going to infinity in the sectors $\mathcal S^{(x)}_{2\mu}$ defined
in (\ref{sectx}). From this point on  we choose the contours $\Gamma_j^{(x)}$ as
follows
\be
\Gamma_j^{(x)} := L_{j}-L_{j-1},\ j=1,\dots,d_1.
\ee
In this way all the regions, including $\mathcal D_0$, become 
wedge-shaped sectors (see Fig. \ref{fig3}). The corresponding (equivalent) jump
discontinuities are 
\bea
&& {\m{\bf \un \Phi}^N}_+ (x)\bigg|_{x\in L_\mu}  = G_\mu  {\m{\bf \un \Phi}^N}_{-} (x):=
   \le[\begin{array}{ccccc}
1 &  2i\pi J_{\mu,1} &
2i\pi J_{\mu,2} & \cdots &
2i\pi J_{\mu,d_2}\\
0&1&0&\cdots &0\\
0&0&1&\cdots &0\\
0&0&0&\ddots& \\
0&0&0&\cdots&1
\end{array}\ri]{\m{\bf \un \Phi}^{N}}_{-} (x)  \cr
&& J_{\mu,j}:= \varkappa^{\mu,j}-\varkappa^{\mu+1,j}\ ,
\mu=0,\dots,d_1,\ j=1,\dots, d_2 , \
\varkappa^{0,j} :=  \varkappa^{d_1+1,j}:=0\label{jumpStokes}
\eea

In order to formulate the complete RH problem, we need to supplement this
discontinuity data with the sectorial asymptotics around the irregular
singularity at $x=\infty$ and the Stokes matrices.
 In doing so one should be
careful that the lines $L_\mu$ for which the discontinuities are
defined do not coincide with any of the Stokes' lines. We can always
arrange this by perturbing the rays $L_\mu$ within the same
sector $\mathcal S_{2\mu}^{(x)}$.

It should be clear that each of the piecewise analytic functions
$\ds{\m{\bf \un\Phi}^{N}(x)}$ can  alternatively be
analytically continued to entire functions, since the contours
$\Gamma_j^{(x)}$ can be deformed arbitrarily in the finite part of
the $x$-plane. Therefore the ``discontinuities'' in the definition of
the Hilbert integral are just apparent and have an intrinsic meaning
only when studying the asymptotic behavior at infinity.
Indeed in the final formulation of our Riemann--Hilbert problem we
prefer to let  the lines $L_\mu$ also  play the roles of Stokes' lines.
\bp[Sectorial Asymptotics]
\label{asymptotics}
In each of the sectors around $x=\infty$ with boundaries given by the 
 Stokes' lines
$L_\mu$, $\mu=0,\dots, d_1$ and $\mathcal R_k$, $k=0,\dots 2d_2+1$
(defined right after  (\ref{Stokessects})) 
the system 
\be
\hbar \frac {d}{dx} \m{\mathbf {\un\Phi}}^N(x) =   \m{\mathbf{
\un\Phi}}^N (x)\m{\un D_1}^{N}(x)\ 
\ee
possesses a solution whose  leading  asymptotic form at $x=\infty$
coincides within this sector with the following formal asymptotic
expansion:
\be
{\mathbf {\un\Phi}}^N\!_{form}(x)\sim {\rm e}^{\frac 1 \hbar T(x)} W\,x^{G}  Y(x^{\frac 1 {d_2}})
\ee
where $ Y =Y_0+\mathcal O(x^{-1/{d_2}})$ is a matrix--valued function
analytic at infinity, $Y_0$ is a diagonal invertible matrix (specified
in the proof) and 
\bea
&& T(x):= \sum_{j=0}^{d_2} \frac {d_2 t_j}{d_2-j+1} x^{\frac {d_2+1-j}{d_2}}
\Omega^{d_2+1-j} + V_1(x) E\\
&& W:=\le[\begin{array}{c|cccc}
1&0&\cdots &0&0\\
\hline
0&\omega &\omega^2&\cdots&\omega^{d_2}\\
0&\omega^2&\omega^4&\cdots&\omega^{2d_2}\\
\vdots & & & \vdots\\
0&\omega^{d_2}&\omega^{2d_2}&\cdots &\omega^{{d_2}^2}
\end{array}
\ri] \label{defmatrixW}\\
&& G:= {\rm diag}\le(-N,
\frac {N+\frac 12-\frac {d_2}2}{d_2},\frac{N+\frac 32-\frac {d_2}2}
{d_2}, \dots,\frac {N-\frac 12+\frac {d_2}2}{d_2}\ri)\\
&& \Omega:={\rm diag}(0,\omega,\omega^2,\dots,\omega^{d_2-1},\omega^{d_2})\ ,\ \
\omega:=  {\rm e}^{\frac{2i\pi}{d_2}}\\
&& E:= {\rm diag}(1,0,\dots,0)\\
&& t_0:= (v_{d_2+1})^{-\frac 1{d_2}}\ ,\quad
 t_1:= -\frac 1{d_2} \frac {v_{d_2}}{v_{d_2+1}} \ ,\\
 &&t_j:= \frac 1{j-1} \res{y=\infty}\big(V_2'(y)\big)^{\frac
{j-1}{d_2}} {\rm d}y  \qquad j=2,\dots,d_2. 
\eea
\ep
{\bf Proof}.
In any given sector $\mathcal S_k$ bounded by the lines $\mathcal
R_{k-1}$ and $\mathcal R_k$ (eq. \ref{Stokessects}) 
 we can choose a basis of steepest descent contours
$\gamma_k$, $k=1,\dots,d_2$. The reason why the $\mathcal S_k$'s 
 are Stokes sectors  and the proper construction of the steepest descent
contours is delayed to the discussion of the Stokes' matrices in Sect.
\ref{stokesmats}. We Fourier-Laplace transform the
quasipolynomials $\phi_n(y)$ along these contours in order to obtain
the functions $\un\phi_n(x)$. Notice that they are not necessarily
the same as the previously introduced $\un\phi_n^{(k)}(x)$'s since
the steepest descent contours do not necessarily coincide with the contours
$\Gamma_k^{(y)}$ defined previously. However they are suitable linear
combinations with integer coefficients of such $\un\phi_n^{(k)}(x)$'s
since the choice of the steepest descent contours is just a different
basis in the homology space of the $y$-plane. Now consider the asymptotic
expansions for
\be
\un\varphi_n^{(k)}(x) := \int_{\gamma_k}\!\!\!{\rm d}y\, {\rm
e}^{-\hbar^{-1}(V_2(y)-xy)} \phi_n(y)\ .
\ee
Here we use the notation $\varphi$ rather than $\phi$ to stress that
these are Fourier--Laplace transforms along contours of a
homology class equivalent to SDC's.
The leading asymptotic term in the sector $\mathcal S_k$ is given at the critical point of
the exponent $V_2(y)-xy$ corresponding to the steepest descent contour
$\gamma_k$. That is, we must compute $V_2(y)-xy$ near a solution to:
\be\label{Vxy}
V'_2(y)-x=0
\ee
asymptotically as $x\to \infty$ within the specified sector.
Let us solve eq.(\ref{Vxy}) in a series expansion in the
local parameter at $\infty$ given by one determination, $\lambda$, of the
$d_2$-th root of $x$:
\bea
&& v_{d_2+1}y^{d_2} + v_{d_2}y^{d_2-1} + \dots=V'_2(y) = x :=\lambda^{d_2}\\
&& y(\lambda) = \lambda \sum_{j=0}^{\infty} t_j\lambda^{-j}\ .\label{yx}
\eea
We then have the formulae (recalling $\lambda = (V_2'(y))^{\frac 1{d_2}}$)
\be
\begin{array}{l}
\ds{t_0 = (v_{d_2\!+\!1})^{-\frac 1{d_2}}\ ,}\\[5pt]
\ds{t_1 = \res{\lambda=\infty} y(\lambda)\frac{{\rm d}\lambda}\lambda  = 
\frac 1{d_2} \res{y=\infty} y\frac {V''_2(y)}{V'_2(y)}{\rm d}y =
-\frac 1{d_2} \frac{v_{d_2}}{v_{d_2\!+\!1}}}\\[5pt]
\ds{t_j = \res{\lambda=\infty} \frac{\lambda^{j-1}}{j-1} y'(\lambda){\rm
d}\lambda = \frac 1{j-1} \res{y=\infty}
\big(V_2'(y)\big)^{\frac{j-1}{d_2}} {\rm d}y} \, , \, j=2\dots \infty\ .
\end{array}\label{yxt}
\ee
As before, denote by $\{y_k(x)\}_{k=0, \dots d_2+1}$ the $d_2$ solutions of 
the equation $V'_2(y)=x$, which are solved  in(\ref{yxt}) by a series in the
$d_2$-th root of $x$. We then have in a neighborhood of
$x=\infty$
\be
V_2(y_k(x)) -xy_k(x) = -\int^x y_k(x'){\rm d}x' =
-d_2\int^{\lambda_k}\!\!\!\!
y(\lambda')\lambda'^{d_2-1}{\rm d}\lambda' = - \sum_{j=0}^\infty
\frac {d_2t_j}{d_2\!+\!1-j} \lambda_k^{d_2\!+\!1-j} -c_k\ ,\label{critvalseries}
\ee
where $\lambda_k := \omega^k \lambda$ and 
 $c_k$ is a constant depending only on the coefficients of $V_2$
and the branch of the solution $y_k(x)$.
This formula is proved by taking the derivative (with respect to  $x$) of both
sides and using the defining equation for $y_k(x)$.
Notice that there is no logarithmic contribution since $t_{d_2\!+\!1}=0$ as 
follows immediately from the residue formula (\ref{yxt}).
The different saddle  points are computed by replacing $\lambda$ with
$\lambda_k:= \omega^k\lambda$.
Substituting into the integral representation of the functions
$\un\varphi_n^{(k)}(x)$, we get
\bea
&&\un\varphi_n^{(k)}(x) := \frac 1{\sqrt{h_n}} \int_{\gamma_k} \!\!\!\!{\rm d}y
\, {\rm e}^{-\frac 1\hbar(V_2(y)-xy)}\sigma_n(y) \\
&&\simeq \frac{ {\rm e}^{\frac {c_k}\hbar}} {\sqrt{h_n}} {\rm
e}^{\frac 1{\hbar}\int^x
y_k(x'){\rm d}x'} \sigma_n(y(\lambda_k)) \int_\R {\rm e}^{-\frac 1 {2\hbar}
V_2''(y(\lambda_k))t^2}{\rm d}t   \\
&&
=\frac { {\rm e}^{\frac {c_k}\hbar}}{\sqrt{h_n}} {\rm e}^{\int^x
y_k(x'){\rm d}x'} \sigma_n(y(\lambda_k))\sqrt{\frac {2\pi\hbar}{V_2''(y(\lambda_k))}}
\eea
Differentiating $V_2'(y(\lambda_k))={\lambda_k}^{d_2}$ implicitly we obtain the
relation
\be
V_2''(y(\lambda_k)) = \frac{d_2{\lambda_k}^{d_2-1}}{y'(\lambda_k)} \ ,
\ee
where $y'(\lambda)$ means differentiation with respect to  $\lambda$.
Therefore we obtain 
\bea
&& \un\varphi_n^{(k)} (x) \sim
{\rm e}^{-\hbar^{-1}(V_2(y_k(x))-xy_k(x))} 
\overbrace{\frac
{\sigma_n(y_k(x))}{\sqrt{h_n}}\sqrt{\frac
{2\pi\hbar}{V_2''(y_k(x))}}}^{
\ds{\simeq \sqrt{\frac {2\pi\hbar}{h_n}}{\lambda_k}^{n-\frac {d_2-1}2}
v_{d_2\!+\!1}^{-\frac {2n-1}{2d_2}}}} \nonumber \\ 
&&\ \ \sim \sqrt{\frac {2\pi\hbar}{d_2 h_n}}
 {\rm e}^{-\hbar^{-1}(V_2(y_k(x))-xy_k(x))}
{\lambda_k}^{n-\frac {d_2-1}2}
v_{d_2\!+\!1}^{-\frac {2n-1}{2d_2}} \label{asympdir}\\
&&\ \ =
 \frac 1{\sqrt{h_n}} {\rm e}^{\frac 1 \hbar \int^x
y_k(x'){\rm d}x'} \sigma_n(y(\lambda_k))\sqrt{\frac {2\pi\hbar
y'(\lambda_k)}{d_2{\lambda_k}^{d_2-1}}} \label{ciccio}\\
&&\ \ = 
  C_k \frac {{v_{d_2\!+\!1}}^{-\frac {n+1}{d_2}}}{\sqrt{h_n}}
{\lambda_k}^{n+\frac {1-d_2}2} {\exp}
\le[\frac 1 \hbar
\sum_{j=0}^{d_2}
\frac {d_2t_j}{d_2\!+\!1-j} {\lambda_k}^{d_2\!+\!1-j}
\ri] \le(1+\mathcal O\le(\lambda^{-1}\ri)\ri) \ ,\label{balocco}
\eea
where
\bea
 C_k:=  {\rm e}^{\frac {c_k}\hbar}\sqrt{\frac{2\pi\hbar}
{{v_{d_2\!+\!1}}^{\frac 1 {d_2}}{d_2}}}.
\eea
Note that the full series (\ref{critvalseries}) which should appear in the 
exponent of (\ref{balocco}) has been truncated to $j\leq d_2$ because the terms
corresponding to $j>d_2+1$  contribute to negative powers in the exponential
and hence  give a $\big(1+\mathcal O(\lambda^{-1})\big)$  term and, as remarked
above, there is no term for $j=d_2+1$ since $t_{d_2+1}=0$.

On the other hand  the functions $\phi_n^{(0)}(x)$ have the following asymptotic
expansion as $x\to\infty$ within the $\mathcal D_\mu$ sectors:
\bea
&&\phi_n^{(0)}(x) := {\rm e}^{\frac 1 \hbar V_1(x)}\int_{\varkappa
\Gamma}\!\!\!\!\!\!{\rm d}s\, {\rm d}y\, \frac{{\rm
e}^{-\hbar^{-1}(V_1(s)-sy)}\phi_n (y)} {(x-s)} \cr
&&\simeq  {\rm
e}^{\frac 1 \hbar V_1(x)}\sum_{k=0}^{\infty}
x^{-k-1}\int_{\varkappa\Gamma}\!\!\!\!\!\! {\rm d} s\,{\rm d}y\,
s^k{\rm e}^{-\hbar^{-1}V_1(s)}\un\phi_n(y)  \cr
&&= \sqrt{h_n}
x^{-n-1}{\rm e}^{\hbar^{-1}V_1(x)}\le(1+\mathcal O \le(x^{-1}\ri)\ri)
\label{hilbertasymp}\ .
\eea
Therefore the matrix of leading terms of  $\ds{\m{\bf \un\Phi}^N(x)}$
as defined  in eq. (\ref{piecewise}) is given by
\bea
&& {\rm diag}\le(1,C_1,\cdots C_{d_2}\ri) {\rm e}^{\frac 1 \hbar T(x)} \le[
\begin{array}{ccccc}
 h_{N\!-\!1}v_{d_2\!+\!1}^{\frac {N}{d_2}}x^{-N}& &\cdots & &
h_{N\!+\!d_2\!-\!1}v_{d_2\!+\!1}^{\frac {N+d_2}{d_2}}x^{-N-d_2}\\ 
{\lambda_1}^{N-\frac  12-\frac {d_2}2} & &\cdots & & {\lambda_1}^ {N-\frac  12+\frac
{d_2}2}\\
{\lambda_2}^{N-\frac  12-\frac {d_2}2} & &\cdots & & {\lambda_2}^ {N-\frac  12+\frac
{d_2}2}\\
\vdots &&\cdots&&\vdots\\
{\lambda_{d_2}}^{N-\frac  12-\frac {d_2}2} & &\cdots & &
{\lambda_{d_2}}^ {N-\frac  12+\frac
{d_2}2}
\end{array}
\ri]\cdot \cr 
&&\hspace{3cm} \cdot {\rm diag}\le(\frac { v_{d_2\!+\!1}^{-\frac {N}{d_2}}}
{ \sqrt{h_{N-1}}},\dots,\frac { v_{d_2\!+\!1}^{-\frac
{N\!+\!d_2}{d_2}}} 
{ \sqrt{h_{N\!+\!d_2\!-\!1}}}\ri)\ .
\eea

The determinant of the Vandermonde--like matrix is
very simple: by computing it along the first row one realizes
that only the first and last minors are not zero. Indeed for all other
minors the corresponding submatrix has the first and last column
proportional. The first minor is a constant in $x$ while the last is
of order $x^{-d_2}$.
Therefore we can write 
\bea
&& \le[ 
\begin{array}{ccccc}
h_{N\!-\!1}{v_{d_2\!+\!1}}^{\frac {N}{d_2}}x^{-N}\!\!\!\!\!& &\cdots & & 
\!\!\!\!\!h_{N\!+\!d_2\!-\!1}{v_{d_2\!+\!1}}^{\frac {N+d_2}{d_2}}x^{-N-d_2}\\
{\lambda_1}^{N-\frac  12-\frac {d_2}2} & &\cdots & & {\lambda_1}^ {N-\frac  12+\frac
{d_2}2}\\
{\lambda_2}^{N -\frac  12-\frac {d_2}2} & &\cdots & & {\lambda_2}^ {N-\frac  12+\frac
{d_2}2}\\
\vdots &&\cdots&&\vdots\\
{\lambda_{d_2}}^{N-\frac  12-\frac {d_2}2} & &\cdots & &
{\lambda_{d_2}}^ {N-\frac  12+\frac
{d_2}2}
\end{array}
\ri]  \cr
&&=
\le[ 
\begin{array}{ccccc}
h_{N\!-\!1}{v_{d_2\!+\!1}}^{\frac {N}{d_2}}x^{-N}\!\!\!\!\!\!\!\!\!\!
&0 &\cdots & & 0 \\
0 & {\lambda_1}^{N+\frac  12-\frac {d_2}2}\!\!\!\!\! 
 &\cdots & &\!\!\!\!\! {\lambda_1}^ {N-\frac  12+\frac {d_2}2}\\
 0&{\lambda_2}^{N -\frac  12-\frac {d_2}2} \!\!\!\!\!
&\cdots & &\!\!\!\!\! {\lambda_2}^ {N-\frac  12+\frac{d_2}2}\\
\vdots &&\cdots&&\vdots\\
 0& \!\!\!\!\!{\lambda_{d_2}}^{N-\frac  12-\frac {d_2}2}\!\!\!\!\!
&\cdots & &\!\!\!\!\! {\lambda_{d_2}}^ {N-\frac  12+\frac {d_2}2}
\end{array}
\ri] (1+\mathcal O(x^{-1}))\\
&&=\le[ 
\begin{array}{ccccc}
h_{N\!-\!1}{v_{d_2\!+\!1}}^{\frac {N}{d_2}}x^{-N}&0 &\cdots & & 0 \\
0 & {\lambda_1}^{N+\frac  12-\frac {d_2}2} &\cdots & & {\lambda_1}^ {N-\frac  12+\frac
{d_2}2}\\
 0&{\lambda_2}^{N -\frac  12-\frac {d_2}2} &\cdots & & {\lambda_2}^ {N-\frac  12+\frac
{d_2}2}\\
\vdots &&\cdots&&\vdots\\
 0& {\lambda_{d_2}}^{N-\frac  12-\frac {d_2}2}&\cdots & &
{\lambda_{d_2}}^ {N-\frac  12+\frac
{d_2}2}
\end{array}
\ri] (1+\mathcal O(x^{-1})) \\
&&=
\le[ 
\begin{array}{ccccc}
1&0 &\cdots & & 0 \\
0 & {\omega}^{N+\frac  12-\frac {d_2}2} &\cdots & & {\omega}^ {N-\frac  12+\frac
{d_2}2}\\
 0&{(\omega^2)}^{N +\frac  12-\frac {d_2}2} &\cdots & & {(\omega^2)}^ {N-\frac  12+\frac
{d_2}2}\\
\vdots &&\cdots&&\vdots\\
 0& 1&\cdots & & 1
\end{array}
\ri] \cr
&& \qquad \times {\rm diag}\le(h_{N\!-\!1}{v_{d_2\!+\!1}}^{\frac {N}{d_2}}x^{-N}
,
\lambda^{N+\frac  12-\frac {d_2}2},\dots,\lambda^{N-\frac  12+\frac
{d_2}2} \ri)(1+\mathcal O(x^{-1}))  
\eea
When inserting this into the asymptotic form, we see that, up to factoring
the  constant invertible (diagonal) matrix on the left 
\be
{\rm diag}(1, C_1\omega^{N+\frac 1 2 -\frac {d_2} 2},
C_2\omega^{2(N+1/2 - d_2/2)},\cdots , C_{d_2}) \,\, ,
\ee
which is irrelevant for the asymptotics and depends on $N$ in a rather trivial
manner, we obtain a solution with the asymptotic form
\bea
&&\m{\bf\un \Phi}^N\!_{form}(x) \simeq \exp\frac 1 \hbar \le( E V_1(x)
 + \sum_{j=0}^{d_2} \frac {d_2 t_j \lambda^{{d_2\!+\!1-j}}}{d_2-j+1}
\Omega^{d_2\!+\!1-j} \ri) W   \cr
&&\times{\rm diag}\le(h_{N-1}{v_{d_2\!+\!1}}^{\frac N{d_2}}x^{-N} ,
\lambda^{N+\frac  12-\frac {d_2}2},\dots,\lambda^{N-\frac  12+\frac
{d_2}2} \ri)
{\rm diag}\le(\frac {{v_{d_2\!+\!1}}^{-\frac {N}{d_2}}}
{ \sqrt{h_{N-1}}},\dots,\frac {{v_{d_2\!+\!1}}^{-\frac
{N\!+\!d_2}{d_2}}} 
{ \sqrt{h_{N\!+\!d_2\!-\!1}}}\ri) (1+\mathcal O(\lambda^{-1})) \cr
&&=
{\rm e}^{\frac 1 \hbar T(x)} W \ {\rm diag}\le(\sqrt{h_{N-1}} x^{-N} ,
\frac {{v_{d_2\!+\!1}}^{-\frac {N+1}{d_2}}\lambda^{N+\frac 12-\frac {d_2}2}}
{ \sqrt{h_{N}}}
,\dots,\frac {{v_{d_2\!+\!1}}^{-\frac
{N\!+\!d_2}{d_2}}\lambda^{N-\frac  12+\frac
{d_2}2} } 
{ \sqrt{h_{N\!+\!d_2\!-\!1}}} \ri)(1+\mathcal O(\lambda^{-1}))
 \cr
&&=\exp\le(\frac 1 \hbar  WT(x)W^{-1} \ri) x^G {\rm diag} \le(\sqrt{h_{N-1}}, \frac
{{v_{d_2\!+\!1}}^{-\frac {N+1}{d_2}}}{ \sqrt{h_{N}}}, \dots,
 \frac {{v_{d_2\!+\!1}}^{-\frac {N\!+\!d_2}{d_2}}} { \sqrt{h_{N\!+\!d_2\!-\!1}}} 
\ri)(1+\mathcal O(\lambda^{-1})) \ , 
\eea
where $W$ is the matrix defined in eq. (\ref{defmatrixW}).
Note that $W^{-1}\Omega W$ is just the permutation matrix (in
the subblock). Q.~E.~D.

\subsection{Stokes Matrices for the Fourier--Laplace transforms}
\label{stokesmats}

The fundamental solution of the system $\un D_1$ is formed from $d_2$
Fourier-Laplace transforms $\ds{\m{\un\Phi}^N\!^{(k)}(x)}$, $k=1,\dots d_2$, and
one Hilbert-Fourier-Laplace transform $\ds{\m{\un\Phi}^N\!^{(0)}(x)}$. The
asymptotic behavior of the $d_2$ FL transforms is analyzed by means of the
steepest descent method in each of the sectors $\mathcal S_k$, $k=0,\dots
2d_2+1$ (\ref{Stokessects}), while the behavior of the (piecewise) analytic
function $\ds{\m{\un\Phi}^N\!^{(0)}(x)}$ is obtained (in each $\mathcal D_\mu$)
from eq. (\ref{hilbertasymp}). The computation is achieved by expressing the
change of homology basis from the $\Gamma_1^{(y)},\dots,\Gamma_{d_2}^{(y)}$
contours to the steepest descent contours. In order to simplify the analysis of
the Stokes matrices we point out that there is no essential loss of generality
in assuming $V_2(y)= y^{d_2+1}\frac 1 {d_2+1}$. Indeed, we are concerned with
just the homology classes of the SDC's, and as $x\to\infty$ the $d_2$ solutions
of the equation $ V_2'(y)=x$ entering Def. \ref{defSDC} are distinct and
asymptotic to the $d_2$ roots of $x$ (up to a nonzero factor).  The particular
choice of the leading coefficient is also essentially irrelevant. If we
choose a different coefficient, we must just appropriately rotate by
$\vartheta = \arg(v_{d_2+1})/(d_2+1)$ counterclockwise in the pictures to
follow, but without any essential difference. Therefore, we proceed with
$v_{d_2+1}$ set equal to unity.

With these simplifications, the Stokes phenomenon can be studied directly on
the  integrals
\be
\int {\rm d}y\, {\rm e}^{-\frac 1 \hbar \le(\frac
{y^{d+1}}{d+1}-xy\ri) } = |x|^{\frac 1 d} \int {\rm d}z\,
\exp\le[
-\Lambda\le(\frac{z^{d+1}}{d+1} -{\rm e}^{i\alpha} z\ri)\ri]\ , 
\ee
where
\be
\Lambda:= \frac 1 \hbar |x|^{\frac{d+1}d} \ ,\quad \alpha:={\rm arg}(x)\ ,  
\label{integrs}
\ee
and, in order to avoid too many subscripts in the formulae to follow, we
have here, and for the remainder of this section, have set $d=d_2$. 
Disregarding the positive factor $|x|^{\frac 1 d}$, which is inessential for
these considerations,  the integrals in (\ref{integrs}) can be written as 
\be
\int {\rm d}z\,
\exp\le[
-\Lambda\le(\frac{z^{d+1}}{d+1} -{\rm e}^{i\alpha} z\ri)\ri] =
\int{\rm d}s\, {\rm e}^{-\Lambda s} \frac {{\rm d}z}{{\rm d}s},
\ee
where $z=Z(s)$ is the $D+1$-valued inverse to 
\be
s=S(z):= \frac {z^{d+1}}{d+1} - {\rm e}^{i\alpha} z\ ,\ \ 
z=Z(s).
\ee
It defines a $(d+1)$-fold covering of the $s$-plane branching around the points
$(z,s)$ whose projection on the $s$-plane are the $d$ critical values
\be
s_{cr}^{(j)} = s_{cr}^{(j)}(\alpha) = -\frac d{d+1}{\rm e}^{i\frac{(d+1)\alpha} d}\omega^j \ ,\ \
\omega:={\rm e}^{\frac {2i\pi}{d}}\ ,\   \ \
j=0,\dots, d-1.
\ee
In realizing this $d+1$-fold covering, we take the branch cuts on the
$s$-plane to  be the  rays
$\Im(s)=\Im(s_{cr}^{(j)})=$const, extending to $\Re(s)=+\infty$.  
As $\Lambda\to +\infty$ the integrals (\ref{integrs})
 have leading asymptotic behavior that
depends only on  the critical values of the map $s(z)$ and on the
homology class of the contour.
 
We now return to the computation of the Stokes' lines.
By the definition of the SDC's $\gamma_k$
(Def. \ref{defSDC}, with  $V_2(y)$ now taken as just $y^{d+1}/(d+1)$),
 their image in the $s$-plane consists of contours which come from
$\Re(s)=+\infty$ on one side of the branch-cut (and on the appropriate sheet)
and go back to $\Re(s)=+\infty$ on the other side of the branch-cut, on the
same sheet. (For the SAC's, we choose cuts extending to $\Re(s)=-\infty$.) The
cuts on the $s$-plane may overlap only for those values of
$\alpha=\arg(x)$ for which the imaginary parts of two different critical values
$s_{cr}^{(i)}(\alpha)$ and
$s_{cr}^{(j)}(\alpha)$ coincide. A straightforward computation, with our  
simplifying assumptions on $V_2(y)$, yields the lines separating the sectors
$\mathcal S_k$ to be those defined in eqs. (\ref{Stokessects}) 
(with $d_2$ replaced by $d$ and $\vartheta=0$).

We also need the following. 
\bd
For a given sector $\mathcal S$, of width $A<\pi$, centered around a ray
${\rm arg}(x)=\alpha_0$,  the {\bf dual sector} $\mathcal
S^\vee$ is the sector centered around the ray ${\rm arg}(x)=
-\alpha_0+\pi$, with width $\pi-A$.
\ed
For $x\to\infty$  in each sector the SDC's are  constant integral
linear combinations of the contours $\Gamma_k^{(y)}$'s. When $x$ crosses the
Stokes line between two adjacent sectors, the homology of the SDC's
changes discontinuously. We denote the SDC's relative to the sector $\mathcal
S_k$ by $\gamma_j^{(k)}$, $j=0\dots d-1$, and denote the column vector with
these as entries $\vec \gamma^{(k)}$, and similarly, let $\vec \Gamma$
denote the column vector with entries $\{\Gamma_j^{(y)}\}_{j=1\dots d}$. 
Denoting the matrix of change of basis by $C_k$, we have 
\be
\vec \gamma^{(k)} = C_k \vec \Gamma\ , C_k\in GL(d,\Z)\ .
\ee

Our first objective is to compute these matrices $C_k$.
For each fixed {\em generic}  $\alpha$ (i.e. away from the Stokes' lines),
 we can construct a diagram (essentially a
Hurwitz diagram) which describes the sheet structure of the inverse
map $z=Z(s)$. 
We draw $d+1$ identical ordered $d$-gons each representing a copy of
the $s$-plane and whose (labeled) vertices represent the projections of the $d$
critical values $s_{cr}^{(j)}$. Two vertices with the same label  of two different $d$-gons
are joined by a segment if the two sheets are glued together along a
horizontal branch-cut originating at the corresponding critical value
and going to $\Re(s)=+\infty$. Since all the branch-points of the
inverse map are of order $2$, there are at most two sheets glued along
each cut.
Furthermore we give an orientation to the segments (represented by an
arrow) with the understanding that this gives an orientation to the
corresponding SDC. The convention is that an arrow going from sheet
$j$ to sheet $k$ means that the SDC runs on sheet $j$ coming from
$\Re(s)=+\infty$ below the cut and goes back above the same cut (or,
what is ``homologically'' the same, the contour runs on sheet $k$
coming from $+\infty$ above the cut and returns to $+\infty$ below it).

\begin{figure}[!ht]
\centerline{
\epsfxsize 15cm
\epsfysize 13cm
\epsffile{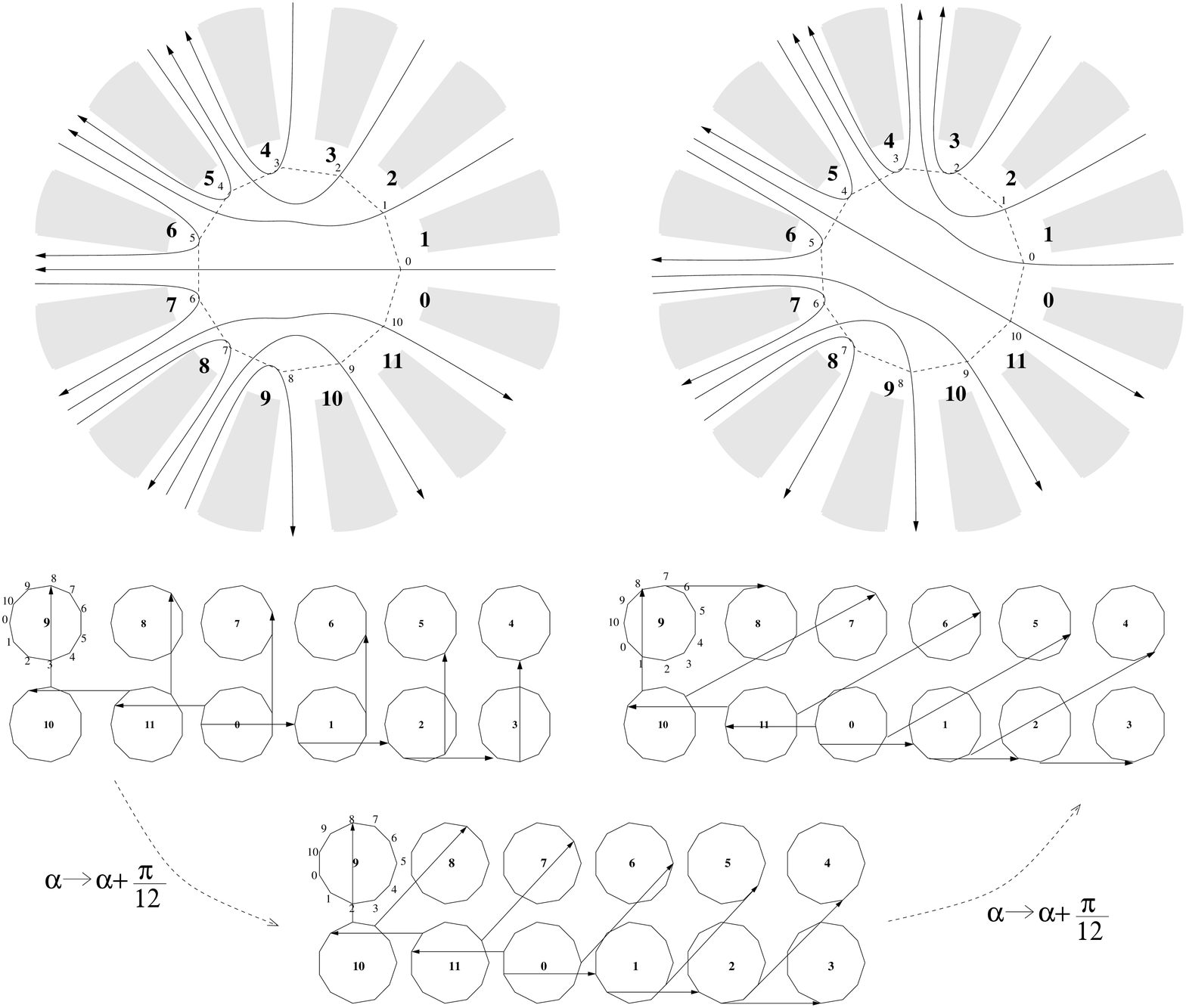}}
\caption{Example of the sheet structure for the case $d=11$, $\alpha=0$.
The contours depicted in the figure are the SDC's in the $z$ plane. The
second image represents the contours after incrementing $\alpha$ by $\pi/{6}$,
i.e., after crossing {\em two} Stokes lines. The labeling of the contours
(smaller numbers in the top figures) is by the subscripts of the $\gamma_k$'s
and $\widetilde \gamma_k$'s passing through the critical points
$z_{cr}^{(k)}:= {\rm e}^{i\frac \alpha d} \omega^k$. This also  corresponds
to the numbering of the vertices of the endecagons representing the critical
values $s_{cr}^{(k)}:= -\frac d{d+1} {\rm e}^{\frac {d+1}{d}\alpha} \omega^k$.
 The different sheets of the $s$-plane are mapped onto the
connected components in which the $z$-plane is cut by the SDC's. These
are the images in the $z$-plane of the cuts in the $s$-plane.  The labeling of
each sheet is given by the boldface numbers and corresponds to the labels inside
 the endecagons appearing in the extended Hurwitz diagrams below the
corresponding $z$-planes. The intermediate Hurwitz diagram represents the
gluing of the sheets after crossing the intermediate Stokes line, i.e., after
incrementing $\alpha$ by $\pi/12$.}
\label{fig2}
\end{figure}

The diagram can be uniquely associated to a matrix $Q_k$ of size
$d\times(d+1)$, in which each row corresponds to a  SDC and each column
to a sheet. The matrix element $(Q_k)_{ij}$ is taken to equal:\\
\indent  $-1$ if the $i$-th SDC points to the $j$-th sheet,\\
\indent  $\phantom{-}1$  if the $i$-th SDC originates on the $j$-th sheet, 
\\
\indent  $\phantom{-}0$ otherwise. \\
Hence each row of the matrix has exactly one $+1$ and one $-1$ entry.

In Fig. \ref{fig2}, we indicate, for the case $d=11$, $\alpha=0$, the SDS
contours and sheet structure represented by the extended Hurwitz diagram. 
 The matrix corresponding to this diagram is: 
\be
Q_0:=\m{
\left [\begin {array}{cccccccccccc} 1&-1&0&0&0&0&0&0&0&0&0&0
\\ 0&1&\!\!\!\!\!-1&0&0&0&0&0&0&0&0&0\\
0&0&1&\!\!\!\!\!-1&0&0&0&0&0&0&0&0\\
0&0&0&1&\!\!\!\!\!-1&0&0&0&0&0&0&0\\
0&0&1&0&0&\!\!\!\!\!-1&0&0&0&0&0&0\\
0&1&0&0&0&0&\!\!\!\!\!-1&0&0&0&0&0\\
1&0&0&0&0&0&0&\!\!\!\!\!-1&0&0&0&0\\
0&0&0&0&0&0&0&0&\!\!\!\!\!-1&0&0&1\\
0&0&0&0&0&0&0&0&0&\!\!\!\!\!-1&1&0\\
0&0&0&0&0&0&0&0&0&0&\!\!\!\!\!-1&1\\ 
1&0&0&0&0&0&0&0&0&0&0&\!\!\!\!\!-1
\end {array}\right]
}^{\ds{\begin{array}{cccccccccccc} \,\,\bf 0& \bf 1& \bf  2
& \bf 3& \bf  4& \bf 5& \!\bf 6& \!\bf
 7& \!\bf 8& \!\bf 9 & \!\bf  
10 &\bf 11
\end{array}
}} \matrix{0\cr 1\cr \,2\cr 3\cr 4\cr 5\cr 6\cr 7\cr 8\cr 9\cr
10}\ ,
\ee
where the boldface numbers labeling the columns are the corresponding
labels for the sheets, while the numbers labeling the rows are the
labels of the SDC's.
Since  $\alpha={\rm arg}(x)$ ranges within a fixed sector $\mathcal S_k$,
the diagram does not change topology 
and the corresponding matrix $Q_k$ remains unchanged.\par
\begin{figure}[!ht]
\centerline{
\epsfxsize 13cm
\epsffile{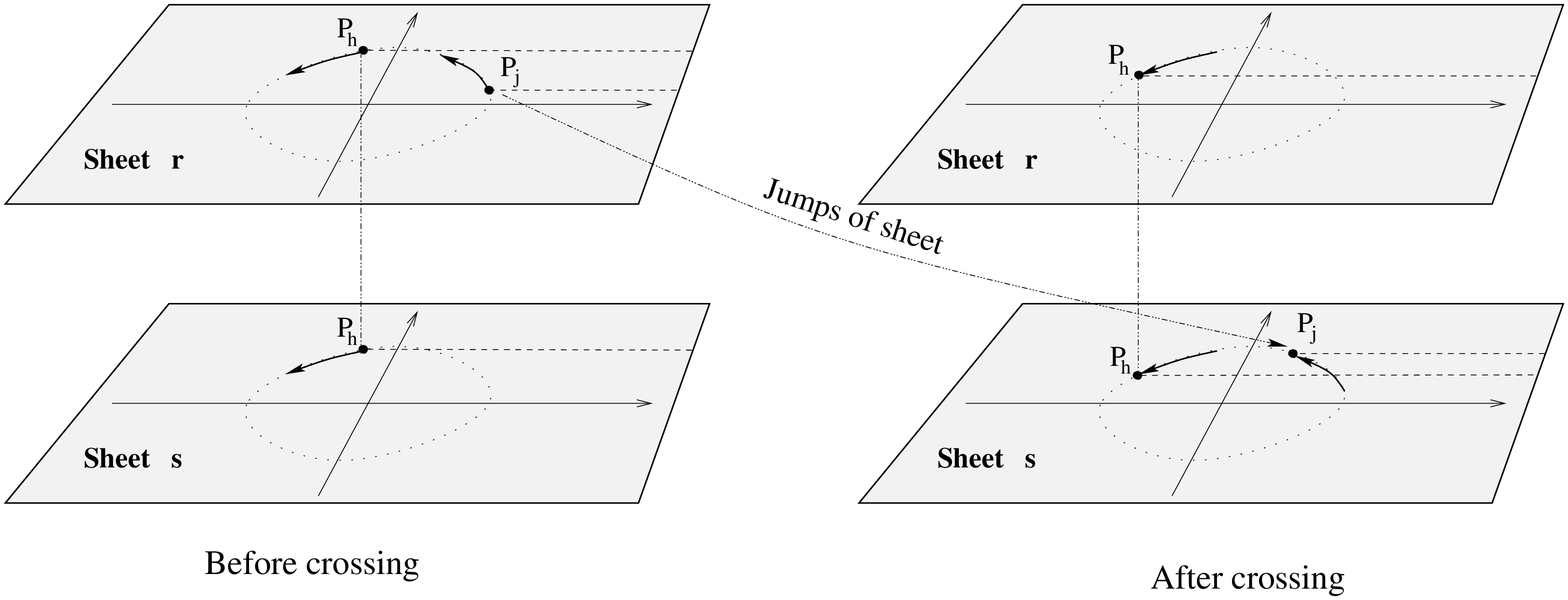}}
\caption{The crossing of two critical values and the corresponding cuts.}
\label{fig2a}
\end{figure}

We can now describe how a given diagram changes when $\alpha=\arg(x)$
crosses the line between two adjacent sectors $\mathcal S_k$ and
$\mathcal S_{k+1}$ (counterclockwise). These lines
correspond precisely to the values of $\alpha$ for which two
distinct critical values  have the same imaginary parts (so that the cuts may
overlap if they are on the same sheet). We leave to the reader to check
that these lines are precisely the boundaries of the Stokes sectors
$\mathcal S_k$ defined in eq. (\ref{Stokessects}).
As $\alpha$ {\em increases} by $\pi/(d+1)$ from $\mathcal S_k$ to
 $\mathcal S_{k+1}$. the $d$-gons rotate by $\pi/d$. In this process the
connections between the sheets change according to the following rule:
if the branch-point $P_j$ on sheet $r$ crosses the cut originating
from a different branch-point $P_{h}$ on the same sheet (on the left of $P_j$
on sheet $r$, and hence $P_j$ crosses the cut from below as it
moves upwards), then the $P_j$ (and its cut) jumps to the sheet
$s$ which is glued to sheet $r$ along the cut originating at $P_h$
(see Fig. \ref{fig2a}). Diagrammatically, the tip (or the tail) of the
corresponding arrow  moves from one $d$-gon to another one connected along the
vertex $h$. In terms of the matrix $Q_k$, the $j$-th row  reflects along the
hyperplane orthogonal to the $h$-th row. The homology class of the SDC
$\gamma_h^{(k)}$ will then change because the branch-cut attached to $P_j$
which ``emerges'' from the branch cut attached to $P_h$ ``extracts'' a
contribution proportional to $\gamma^{(k)}_j$. The proportionality factor is 
$\pm 1$ depending on the relative orientations.
The corresponding SDC's $\gamma_j^{(k)}$ and $\gamma_h^{(k)}$ are
related to the SDC's $\gamma_j^{(k+1)}$ and $\gamma_h^{(k+1)}$ by the
relations
\be
\begin{array}{l}
\gamma_j^{(k+1)}=\gamma_j^{(k)}  \\[10pt]
\gamma_h^{(k+1)}= \gamma_h^{(k)} + \epsilon_{hj} \gamma_j^{(k)}
\end{array}
\label{Mmatr}
\ee
where the {\em incidence number} $\epsilon_{hj} = \epsilon_{jh}$ is
$1$ if the SDC's $\gamma_j^{(k)},\gamma_h^{(k)}$ have the opposite
orientation and $-1$ if they have the same orientation.
Alternatively, the incidence number is just the (standard) inner product of the
corresponding rows $h,j$ of the matrix $Q_k$.

We denote by $M_k$ the $d\times d$ matrix which expresses the 
changes in the SDC's described by eq. (\ref{Mmatr})  through the relation
\be
\vec\gamma^{(k+1)}  = M_k \vec\gamma^{(k)}\ .
\ee
These will be seen presently to be precisely the Stokes matrices for the
passage between these sectors. One can  check that the matrices $Q_k$ and
$Q_{k+1}$ are related by  
\be
Q_{k+1} = {M_k^t}^{-1} Q_k\ .
\ee
Therefore we can reconstruct the matrices $M_k$ once we have an
initial diagram representing the sheet structure.

Before constructing the initial diagram, we  note that when
$\alpha$ increases by $2\pi/(d+1)$ (i.e. when we cross {\em two} Stokes
lines),  the steepest descent (unoriented) contours 
are like the original ones, but rotated by the same amount {\em
clockwise}.
 That is, the {\em unoriented} diagram (i.e. forgetting the
orientation of the SDC's) is the same, up to cyclically permuting
 the labels of the $(d+1)$ sheets and the $d$ cuts. Note also that the critical
points  rotate by $\frac {2\pi}{d(d+1)}$ and the critical values by
$\frac{2\pi}{d}$ {\em counterclockwise}. Indeed we have
\be
s\le(z;\alpha+{2 \pi \over {d+1}}\ri) = \frac {z^{d+1}}{d+1}-{\rm e}^{i\alpha}
{\rm e}^{i{2\pi\over {d+1}}} z =s({\rm e}^{i{2\pi\over {d+1}}}z;\alpha)\ .  
\ee
As for the orientation of the SDC's relative to the new labeling,
 the $d$-th SDC passing through $z_{cr}^{(d-1)}$  reverses its
orientation relative to the (oriented) SDC's obtained by just rotating
the initial SDC's (see Figure \ref{fig2}).
This implies that the matrices $Q_k$ representing the diagrams in the
various sectors  and the Stokes' matrices $M_k$ satisfy the recursion relations
\bea
&& Q_{k+2} = p^{-1}Q_{k}\mathfrak S_{d+1}\ ,\qquad M_{k+2} = p\cdot
M_k\cdot p^{-1} \label{stokesrec}\\
&& \mathfrak S_{d+1} = \le[\begin{array}{ccccc}
0&0&\cdots&0&1\cr
1&0&\cdots&0&0\cr
0&1&\cdots&0&0\cr
0&0&\ddots&0&\vdots\cr
0&0&\cdots&1&0
\end{array}
\ri]\in\  {\rm GL}(d+1,\Z)\ ,\ \ \ 
p:=\le[\begin{array}{ccccc}
0&0&\cdots&0&\!\!\!\!-\!1\cr
1&0&\cdots&0&0\cr
0&1&\cdots&0&0\cr
0&0&\ddots&0&0\cr
0&0&\cdots&1&0
\end{array}\ri] \in \ {\rm GL}(d,\Z)
\nonumber
\eea
It is therefore only necessary to compute $Q_0,Q_1$ and $M_0, M_1$.

Notice also that the wedge contours $\Gamma_j^{(y)} ,\  j=0,\dots,d$ are in
$1-1$ correspondence with the sheets of the map $Z(s)$. Indeed, each
connected component to the right of the wedge contours contains only
one sector $S_{2k+1}^{(y)}$ in the $z$-plane which corresponds to the
sector $\Re(s)<0$ in the $s$-plane. 
Thus, the same argument used to arrive at the recursion (\ref{stokesrec})
proves that 
\bea
&& C_{k+2} = p\,C_k\, P\ ,\label{stokesrec2}\\
&& P:= \le[
\begin{array}{ccccc}
\!\!\!\!-1&\!\!\!\!-1&\cdots&\!\!\!\!-1&\!\!\!\!-1\cr
1&0&\cdots&0&0\cr
0&1&\cdots&0&0\cr
0&0&\ddots&0&0\cr
0&0&\cdots&1&0
\end{array}
\ri]\in {\rm GL}(d,\Z) \ ,\nonumber
\eea
where both the matrix $\mathfrak S_{d+1}$ in eq. (\ref{stokesrec})
and $P$ in eq. (\ref{stokesrec2}) generate a representation of the
cyclic group $\Z_{d+1}$ (although $P$ is of size $d\times d$). 
The matrix $P$ is just the generator of $\Z_{d+1}$ on the ``wedge'' contours;
i.e. with the additional constraint $\sum_{i=0}^{d} \Gamma_i^{(y)} =0$.

We now describe how to derive the initial diagram - for example, how Figure
\ref{fig2} is obtained. We start by noting that the ``wedge'' contours
$\Gamma_k^{(y)}$ enclose the sector $\mathcal S_{2k-1}$ of width $\pi/(d+1)$.
It is to see that the corresponding integrals are (more than) exponentially
decreasing in the  dual sector, since
\be
\le|\int_{\Gamma_k^{(y)}} {\rm e}^{-\frac 1 \hbar \le(y^{d+1}/(d+1)-xy\ri)} {\rm
d}y
\ri| \leq
\exp\le(|x| M \ri) \int_{\Gamma_k^{(y)}}\le|{\rm e}^{-\frac 1 \hbar
y^{d+1}/(d+1) }
\ri|{\rm d} y  \ ,
\ee
where $M$ is the supremum of $\frac 1{|x|}\Re(xy)$ as $y$ goes along
$\Gamma_k^{(y)}$. Now the contour $\Gamma_k^{(y)}$ can be deformed so as to
approach the sector $\mathcal S_{2k-1}^{(y)}$ as closely as we wish. Then the
constant
$M$ is finite and negative if $x$ lies within the dual sector ${\mathcal
S_k^{(y)}}^\vee$.

We now consider the case of odd $d$, leaving the easy generalization to even
$d$'s to the reader.  (The only difference is that for odd $d$, 
$\alpha={\rm arg}(x)=0$ is not a Stokes' line. To study the case even $d$
case, one  should choose a convenient initial value of $\alpha$
(e.g. $\alpha=\epsilon\ll 1$ or $\alpha = \pi/2(d+1)$, which corresponds
to an anti--Stokes line).\par
Let us start with the (non-Stokes) value $\alpha=0$ and 
focus on any of the SDC's attached
to a critical value lying in the right $s$-plane for this value of
$\alpha$ (remembering that $d$ is assumed to be odd).  Since the real part
of such a critical value is positive, the corresponding integrals
decrease as $\exp(-\frac d {d+1}\omega^j\Lambda)$,
($\Lambda:=\frac 1 \hbar |x|^{\frac {d+1}d}$) i.e., they are (more
than) exponentially suppressed on the line $x\in \R_+$.
As we increase $\alpha$,  the $d$-gons rotate by $\frac {d+1}d \alpha$
counterclockwise.  It should be clear (from the previous description of the
change of homology of the SDC's) that the SDC's attached to such critical
values do not change homology class while they remain in the right half
of the  $s$-plane. This is so because there are no branch points to their
right which can pass through the branch-cut through the given saddle
point (which extends to the right) as $\alpha$ is increased. Therefore the
corresponding integrals are exponentially suppressed as long as $\alpha$ ranges
in a corresponding sector of width $\frac d{d+1}\pi=
\pi-\frac 1{d+1}\pi$. This is precisely the width of the dual sector
to a  sector of width $\frac \pi{d+1}$; i.e., the width of any of the
$S_{2k+1}^{(y)}$'s. By careful inspection of the anti-Stokes lines for
these integrals\footnote{The anti-Stokes' line for exponential integrals of the
type of Def. \ref{defSDC} are the lines along which the integrals are most
rapidly decreasing as $|x|\to\infty$.}, which correspond to the value of
$\alpha$ for which the corresponding critical value is {\em real} and positive, 
 one concludes that they coincide with an appropriate
$\Gamma_k^{(y)}$ in the left half of the  $z$-plane.
This argument proves that all ``wedge'' contours $\Gamma_k^{(y)}$
lying in the left plane are homological to SDC's.
(These are the SDC's in the left $z$-plane in Figure \ref{fig2}.)

We then take the first critical value lying in the {\em left} half of the 
$s$-plane (in Figure \ref{fig2}, the SDC number $9$). As we increase $\alpha$
so as to move this critical value to the right half of the $s$-plane, the
corresponding SDC can acquire a contribution only from the first SDC in the
left half-plane (number
$8$ in our example). As a consequence the corresponding integral is
exponentially suppressed in a sector of width $\pi- 2\frac \pi{d+1}$.
Considering its anti-Stokes line and its linear independence from the previously
identified SDC's, we conclude that it must enclose two odd-numbered
sectors $S_{2k+1}$ and $S_{2k+3}$ (in our example this is contour
$9$). Proceeding this way we can easily identify the homology classes
of all SDC's for $\alpha=0$. 

The labeling of the sheets is largely arbitrary. (The
choice we have made in the fig. 4 is just for ``aesthetic'' reasons). 
The fixed basis of contours $\Gamma_i^{(y)}$, $i=1,\dots d$
and the bases $\gamma_j^{(k)}$, $j=0,\dots d-1\ k=0,\dots 2d+1$ are related by 
\bea
&& \vec \gamma^{(k)} = C_k \vec \Gamma =  M_{k-1}\cdots M_0 C_0\vec \Gamma\\ 
&& \vec \gamma^{(0)} = C_0\vec \Gamma\ .
\eea
Therefore only the matrices $M_k$ and the first  change of basis  matrix 
$C_0$ are needed. The matrix $C_0$  can easily be constructed from the 
initial sheet structure. For our present purposes it is not actually  
necessary to have its  general form, since the objects of primary interest are   
the Stokes matrices, which will be seen to be just the $M_k$'s.
But to illustrate by example the form that $C_0$ takes, within the basis we 
have chosen, for the case $d=11$, it is:
\be 
C_0:= \m{\le[
\begin{array}{ccccccccccc}
1\!&1\!&1\!&1\!&1\!&1\!&0\!&0\!&0&\!0&\!0\\
0\!&1\!&1\!&1\!&1\!&0\!&0\!&0\!&0&\!0&\!0\\
0\!&0\!&1\!&1\!&0\!&0\!&0\!&0\!&0&\!0&\!0\\
0\!&0\!&0\!&1\!&0\!&0\!&0\!&0\!&0&\!0&\!0\\
0\!&0\!&0\!&0\!&1\!&0\!&0\!&0\!&0&\!0&\!0\\
0\!&0\!&0\!&0\!&0\!&1\!&0\!&0\!&0&\!0&\!0\\
0\!&0\!&0\!&0\!&0\!&0\!&1\!&0\!&0&\!0&\!0\\
0\!&0\!&0\!&0\!&0\!&0\!&0\!&1\!&0&\!0&\!0\\
0\!&0\!&0\!&0\!&0\!&0\!&0\!&0\!&1&\!0&\!0\\
0\!&0\!&0\!&0\!&0\!&0\!&0\!&0\!&1&\!1&\!0\\
0\!&0\!&0\!&0\!&0\!&0\!&0\!&1\!&1&\!1&\!1\\
\end{array}
\ri]}
\ee
It is not difficult to give an explicit description of the matrices
$Q_0,Q_1$ and $M_0,M_1$ but, for the sake of brevity,  we will not give it
here. It consists of a lengthy but straightforward calculation, which leads
to the matrices in Table \ref{tableone} (listed for the cases $d=2,\dots,11$),
where one can clearly extrapolate to the correct rule. 

\vskip 4pt
We now turn our attention to the Stokes matrices. Consider the subblock of the
fundamental system corresponding to solutions of the $\un D_1$ ODE given by
the Fourier-Laplace transforms. If we denote by $Y(x)$ the $d_2\times d_2$
such block  whose rows are the integrals of $\ds{\m{\varphi}^N}(y)$ on the
contours $\Gamma_k^{(y)}$, and by $Y_k(x)$ the analogous matrix obtained by
integrating over the SDC's, we have
\be
Y_k(x)= C_k Y(x).
\ee
The Stokes matrices are then given by 
\be
S_k:=Y_{k+1}{Y_k}^{-1}=
 C_{k+1}{C_k}^{-1} = M_k\ .
\ee
These are just the matrices expressing the relative change of homology basis
of the SDC's corresponding to two consecutive Stokes' sectors.
\begin{table}
{\tiny
\bea
&&M_0= \left [\begin {array}{cc} 1&0\\ 0&1\end {array}
\right ] \qquad \qquad \qquad \qquad \qquad \qquad \qquad \qquad \qquad \qquad 
M_1= \left [\begin {array}{cc} 1&0\\ 1&1\end {array}
\right ] \qquad \qquad \qquad \qquad \qquad \qquad \qquad \qquad \qquad
\qquad  d=2\cr
 && M_{{0}}=\left [\begin {array}{ccc} 1&\!\!\!\!-1&0\\ 0&1&0
\\ 0&0&1\end {array}\right ]
\qquad \qquad \qquad \qquad \qquad \qquad \qquad \qquad  \qquad 
M_{{1}}=\left [\begin {array}{ccc} 1&0&0\\ 0&1&0
\\ 0&1&1\end {array}\right ]
\qquad \qquad \qquad \qquad \qquad \qquad \qquad \qquad \qquad  d=3\cr
&& M_{{0}}=\left [\begin {array}{cccc} 1&\!\!\!\!-1&0&0\\ 0&1&0&0
\\ 0&0&1&0\\ 0&0&1&1\end {array}
\right ]
\qquad \qquad \qquad \qquad \qquad \qquad \qquad  \qquad 
M_{{1}}=\left [\begin {array}{cccc} 1&0&0&0\\ 0&1&0&0
\\ 0&0&1&0\\ 0&1&0&1\end {array}
\right ]
\qquad \qquad \qquad \qquad \qquad \qquad \qquad  \qquad  d=4\cr
&&M_{{0}}=\left [\begin {array}{ccccc} 1&0&\!\!\!\!-1&0&0\\ 0&1
&0&0&0\\ 0&0&1&0&0\\ 0&0&0&1&0
\\ 0&0&0&1&1\end {array}\right ]
\qquad \qquad \qquad \qquad \qquad \qquad \qquad 
M_{{1}}=\left [\begin {array}{ccccc} 1&\!\!\!\!-1&0&0&0\\ 0&1
&0&0&0\\ 0&0&1&0&0\\ 0&0&0&1&0
\\ 0&0&1&0&1\end {array}\right ]
\qquad \qquad \qquad \qquad \qquad \qquad \qquad   d=5\cr
&&M_{{0}}=\left [\begin {array}{cccccc} 1&0&\!\!\!\!-1&0&0&0\\ 0
&1&0&0&0&0\\ 0&0&1&0&0&0\\ 0&0&0&1&0
&0\\ 0&0&0&0&1&0\\ 0&0&0&1&0&1
\end {array}\right ]
\qquad \qquad \qquad \qquad \qquad \qquad 
M_{{1}}=\left [\begin {array}{cccccc} 1&\!\!\!\!-1&0&0&0&0\\ 0
&1&0&0&0&0\\ 0&0&1&0&0&0\\ 0&0&0&1&0
&0\\ 0&0&0&1&1&0\\ 0&0&1&0&0&1
\end {array}\right ]
\qquad \qquad \qquad \qquad \qquad \qquad  d=6\cr
&&M_{{0}}=\left [\begin {array}{ccccccc} 1&0&0&\!\!\!\!-1&0&0&0
\\ 0&1&\!\!\!\!-1&0&0&0&0\\ 0&0&1&0&0&0&0
\\ 0&0&0&1&0&0&0\\ 0&0&0&0&1&0&0
\\ 0&0&0&0&0&1&0\\ 0&0&0&0&1&0&1
\end {array}\right ]
\qquad \qquad \qquad \qquad \qquad 
M_{{1}}=\left [\begin {array}{ccccccc} 1&0&\!\!\!\!-1&0&0&0&0
\\ 0&1&0&0&0&0&0\\ 0&0&1&0&0&0&0
\\ 0&0&0&1&0&0&0\\ 0&0&0&0&1&0&0
\\ 0&0&0&0&1&1&0\\ 0&0&0&1&0&0&1
\end {array}\right ]
\qquad \qquad \qquad \qquad \qquad d=7\cr
&&M_{{0}}=\left [\begin {array}{cccccccc} 1&0&0&\!\!\!\!-1&0&0&0&0
\\ 0&1&\!\!\!\!-1&0&0&0&0&0\\ 0&0&1&0&0&0&0&0
\\ 0&0&0&1&0&0&0&0\\ 0&0&0&0&1&0&0&0
\\ 0&0&0&0&0&1&0&0\\ 0&0&0&0&0&1&1&0
\\ 0&0&0&0&1&0&0&1\end {array}\right ]
\qquad \qquad \qquad \qquad 
M_{{1}}=\left [\begin {array}{cccccccc} 1&0&\!\!\!\!-1&0&0&0&0&0
\\ 0&1&0&0&0&0&0&0\\ 0&0&1&0&0&0&0&0
\\ 0&0&0&1&0&0&0&0\\ 0&0&0&0&1&0&0&0
\\ 0&0&0&0&0&1&0&0\\ 0&0&0&0&1&0&1&0
\\ 0&0&0&1&0&0&0&1\end {array}\right ]
\qquad \qquad \qquad \qquad  d=8\cr
&&M_{{0}}=\left [\begin {array}{ccccccccc} 1&0&0&0&\!\!\!\!-1&0&0&0&0
\\ 0&1&0&\!\!\!\!-1&0&0&0&0&0\\ 0&0&1&0&0&0&0
&0&0\\ 0&0&0&1&0&0&0&0&0\\ 0&0&0&0&1
&0&0&0&0\\ 0&0&0&0&0&1&0&0&0\\ 0&0&0
&0&0&0&1&0&0\\ 0&0&0&0&0&0&1&1&0\\ 0
&0&0&0&0&1&0&0&1\end {array}\right ]
\qquad \qquad \qquad 
M_{{1}}=\left [\begin {array}{ccccccccc} 1&0&0&\!\!\!\!-1&0&0&0&0&0
\\ 0&1&\!\!\!\!-1&0&0&0&0&0&0\\ 0&0&1&0&0&0&0
&0&0\\ 0&0&0&1&0&0&0&0&0\\ 0&0&0&0&1
&0&0&0&0\\ 0&0&0&0&0&1&0&0&0\\ 0&0&0
&0&0&0&1&0&0\\ 0&0&0&0&0&1&0&1&0\\ 0
&0&0&0&1&0&0&0&1\end {array}\right ]
\qquad \qquad \qquad  d=9\cr
&&M_{{0}}=\left [\begin {array}{cccccccccc} 1&0&0&0&\!\!\!\!-1&0&0&0&0&0
\\ 0&1&0&\!\!\!\!-1&0&0&0&0&0&0\\ 0&0&1&0&0&0
&0&0&0&0\\ 0&0&0&1&0&0&0&0&0&0\\ 0&0
&0&0&1&0&0&0&0&0\\ 0&0&0&0&0&1&0&0&0&0
\\ 0&0&0&0&0&0&1&0&0&0\\ 0&0&0&0&0&0
&0&1&0&0\\ 0&0&0&0&0&0&1&0&1&0\\ 0&0
&0&0&0&1&0&0&0&1\end {array}\right ]
\qquad \qquad 
M_{{1}}=\left [\begin {array}{cccccccccc} 1&0&0&\!\!\!\!-1&0&0&0&0&0&0
\\ 0&1&\!\!\!\!-1&0&0&0&0&0&0&0\\ 0&0&1&0&0&0
&0&0&0&0\\ 0&0&0&1&0&0&0&0&0&0\\ 0&0
&0&0&1&0&0&0&0&0\\ 0&0&0&0&0&1&0&0&0&0
\\ 0&0&0&0&0&0&1&0&0&0\\ 0&0&0&0&0&0
&1&1&0&0\\ 0&0&0&0&0&1&0&0&1&0\\ 0&0
&0&0&1&0&0&0&0&1\end {array}\right ]
\qquad \qquad  d=10\cr
&&M_{{0}}=\left [\begin {array}{ccccccccccc} 1&0&0&0&0&\!\!\!\!-1&0&0&0&0&0
\\ 0&1&0&0&\!\!\!\!-1&0&0&0&0&0&0\\ 0&0&1&\!\!\!\!-1
&0&0&0&0&0&0&0\\ 0&0&0&1&0&0&0&0&0&0&0
\\ 0&0&0&0&1&0&0&0&0&0&0\\ 0&0&0&0&0
&1&0&0&0&0&0\\ 0&0&0&0&0&0&1&0&0&0&0
\\ 0&0&0&0&0&0&0&1&0&0&0\\ 0&0&0&0&0
&0&0&0&1&0&0\\ 0&0&0&0&0&0&0&1&0&1&0
\\ 0&0&0&0&0&0&1&0&0&0&1\end {array}\right ]
\qquad 
M_{{1}}=\left [\begin {array}{ccccccccccc} 1&0&0&0&\!\!\!\!-1&0&0&0&0&0&0
\\ 0&1&0&\!\!\!\!-1&0&0&0&0&0&0&0\\ 0&0&1&0&0
&0&0&0&0&0&0\\ 0&0&0&1&0&0&0&0&0&0&0
\\ 0&0&0&0&1&0&0&0&0&0&0\\ 0&0&0&0&0
&1&0&0&0&0&0\\ 0&0&0&0&0&0&1&0&0&0&0
\\ 0&0&0&0&0&0&0&1&0&0&0\\ 0&0&0&0&0
&0&0&1&1&0&0\\ 0&0&0&0&0&0&1&0&0&1&0
\\ 0&0&0&0&0&1&0&0&0&0&1\end {array}\right ]
\qquad  d=11\ . \nonumber
\eea
}
\caption{The first two Stokes matrices for degrees $d_2=2 \dots 11$.
The remaining ones are obtained using eq. (\ref{stokesrec}).}
\label{tableone}
\end{table}

In order to complete the description of the Riemann--Hilbert problem,
we need to also consider the extra solution given by a
Hilbert--Fourier-Laplace transform. In doing so, we extend the
previously computed Stokes matrices $M_k$ to the full fundamental
system of $d_2+1$ solutions by means of 
\be
\widehat M_k:=\le[\begin{array}{c|c}
1&0\\
\hline 0&M_k
\end{array}
\ri]\ .\label{FLTStokes}
\ee
Summarizing the whole discussion, we have proved the following theorem.
\bt[Riemann--Hilbert Problem]
\label{RHpr}
There exists a fundamental system of solutions, $\ds{\m{\bf\un\Phi}^{N}(x)}$
which is analytic where defined, with the properties:
\begin{enumerate}
\item In each of the sectors around $x=\infty$ defined by removing the 
lines $L_\mu$, $\mu=0,\dots d_1$ and $\mathcal R_k$, $k=0,\dots,2d_2+1$,  
 $\ds{\m{\bf\un\Phi}^{N}(x)}$  exists, is analytic and invertible, and it can be
normalized on the left by a constant matrix to have the same  asymptotic
behavior  as  $\ds{\m{\bf\un\Phi}^{N}\!_{form}(x)}$ defined in Proposition
\ref{asymptotics}.
\item Crossing the lines $L_\mu$, the corresponding Stokes matrix is
given by $G_\mu$ as defined in eq. (\ref{jumpStokes}).
\item Crossing the lines $\mathcal R_k$ the corresponding Stokes
matrix is given by $\widehat M_k$ in eq. (\ref{FLTStokes}),
constructed according to the algorithm described in section \ref{stokesmats}. 
\end{enumerate}
\et
\begin{figure}[!ht]
\epsfxsize 10cm
\epsfysize 10cm
\centerline{\epsffile{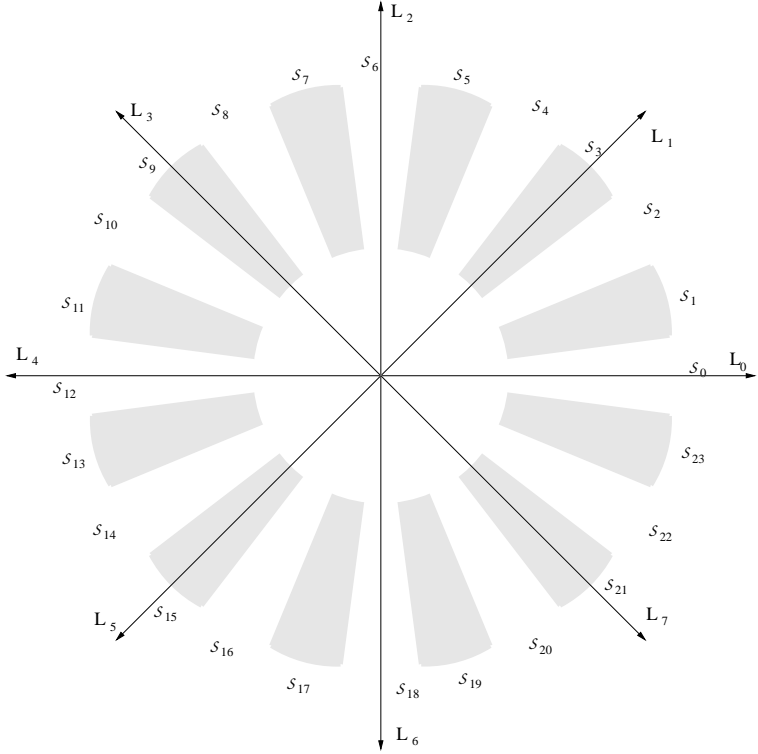}}
\caption{Example of the  structure of Stokes sectors and discontinuity
lines $L_\mu$ (which can equivalently be viewed as Stokes lines) in the
$x$-plane for the case $d_1=7$ and $d_2=11$ and both leading coefficients of
the potentials real and positive.}
\label{fig4}
\end{figure}

\noindent For an example of all Stokes' sectors and lines see Fig. \ref{fig4}
and for examples of explicit Stokes' matrices $M_0, M_1$ from which
all the others are computed easily, see Table \ref{tableone}.

We conclude this section with the remark that in the formulation of
Theorem \ref{RHpr} the lines $L_\mu$ were viewed as Stokes' lines. However, we
could alternatively have formulated  an equivalent Riemann--Hilbert 
problem, in which the lines $L_\mu$ define discontinuities, according to eq.
(\ref{jumpStokes}), by retaining a fixed choice of the contours of
integration along the boundaries of the wedge sectors, thereby giving rise to
genuine jump discontinuities across the boundaries. 
%
\section{Summary and comments on large $N$ asymptotics in multi-matrix models}
 
A complete formulation of the Riemann--Hilbert problem characterizing
fundamental systems of solutions to the differential - recursion equations
satisfied by biorthogonal polynomials associated to $2$-matrix models with 
polynomial potentials is provided in Theorem \ref{RHpr}. The approach derived
here  can also be extended quite straightforwardly to the case of
biorthogonal polynomials  associated to a finite chain of  coupled matrices
with polynomial potentials,  following  the lines indicated in the appendix of
\cite{BEH}.
  
  The Riemann-Hilbert data, consisting essentially of the Stokes matrices at 
$\infty$, are independent of both the integer parameter $N$ corresponding to 
the matrix size and the deformation parameters determining the
 potentials; that is, the fundamental solutions constructed are
 solutions simultaneously of a differential-difference generalized
 isomonodromic deformation problem. This provides the first main step towards a 
rigorous analysis of the $N\to \infty,\ \hbar\,N=\mathcal O(1)$ limit 
of the partition function and the Fredholm kernels determining the spectral
statistics of coupled random matrices (essential, e.g., to the question
of  universality in $2$-matrix models). Such an analysis should follow similar
lines to those previously  successfully applied to ordinary orthogonal
polynomials in the $1$-matrix case
\cite{BlIt,FIK,isomIts,stringIts,dkmvz,dkmvz2}.   The main difference in the
$2$-(or more) matrix case is that in the double--scaling limit the  functional
dependence  of the free energy on the eigenvalue distributions is not as
explicit as in the  $1$-matrix models \cite{matytsin, Guionnet}. 
It is also clear that the hyperelliptic spectral curve that arises in the
solution of the one-matrix model must be replaced by a  more general algebraic
curve, which arises naturally in the spectral duality of the  spectral curves
of \cite{BEH} (cf. \cite{eynard}).

In order determine the large $N$ asymptotics with the help of the data
defining the Riemann-Hilbert problem,  one should begin with an ansatz that can
be checked {\em a posteriori} against the  given case. In the $1$-matrix case
\cite{dkmvz,dkmvz2}, this was provided by means  of hyperelliptic
$\Theta$-functions. The physical heuristics and the basic tools for  generating
such ansatz were also given in \cite{bonnet, eynard, matytsin, Guionnet}. Much
of of these heuristics can be extended to the $2$-matrix case \cite{AMSE}, and
this will be the subject of a subsequent work \cite{ansatz}. 
\appendix
\section{Appendix: Convergent integral representations of
$\ds{\m{\Psi}_\infty\!(x)}$}
\label{chissa}
\renewcommand{\theequation}{\Alph{section}-\arabic{equation}}
\setcounter{equation}{0}

In this appendix, we indicate how to overcome the problem in the
formal definition of eq. (\ref{formalpsi}), illustrated through an example.
The idea follows that used in \cite{kapa} for the case of cubic
potentials\footnote{We wish to thank A. Kapaev for pointing out the problem of
divergent integrals appearing in an earlier version of this paper, and for
helpful discussions. The approach outlined here, for potentials of arbitrary
degrees, will be the subject of further work, and the
results detailed in a subsequent, joint publication}.

 As discussed in  section $\ds{\m{\Psi}_\infty\!^{(0)}(x)}$, a natural
approach to finding solutions of eq. (\ref{modifrec1}) would be to take the
inverse Fourier-Laplace transform of the function
\be
\m{\un\Psi}_\infty\!^{(0)} (y):= {\rm e}^{\frac 1 \hbar V_2(y)}
\int_{\varkappa\Gamma} \!\!\!\!\!{\rm d}x\,{\rm d}t \, \frac {{\rm
e}^{-\frac 1 \hbar (V_1(x)-xt)}}{y-t}{\m{\Psi}}^t(x)\ ,\
\label{auno}
\ee
along the anti-wedge contours $\widetilde \Gamma_k^{(y)}$.
Eq. (\ref{auno}) defines a  piecewise analytic function 
in the domains 
\be
\C_y \setminus \bigcup_{k=1}^{d_2} \Gamma_k^{(y)} = \bigcup_{\nu
=0}^{d_2} \mathcal D_\nu ,
\ee
with jumps across the $\Gamma_k^{(y)}$ given by
\be
\m{\un\Psi}_\infty\!^{(0)} (y_+) - \m{\un\Psi}_\infty\!^{(0)} (y_-) =
2i\pi \sum_{j=1}^{d_1} \varkappa^{j,k} \m{\un\Psi}_\infty\!^{(j)} (y)\
,\qquad y\in \Gamma_k^{(y)}.\label{ajump}
\ee
However, either retaining this definition or modifying it through analytic
continuation leads to a divergent integral when the Fourier-Laplace transform
is applied.  In order to resolve this difficulty, the idea is to consider a
suitable contour approaching $y=\infty$ within the odd-numbered sectors
$\mathcal S_y^{(2k+1)}$ and define a locally analytic function in a
neighborhood of this contour so as to render the Fourier--Laplace transform
convergent.

 A general approach resolving this problem is not needed for the results
presented in this paper since, as pointed out, the Riemann-Hilbert problem  for
$\ds{\m{\mathbf \Psi}_N(x)}$ is more simply obtained by means of duality 
(Remark \ref{duality}). Therefore, we just give an indication of how to proceed
by means of an example.
\subsection{An Example: $d_1=7$, $d_2=4$}

\begin{figure}[!ht]
\begin{center}
\epsfysize 12cm
\epsffile{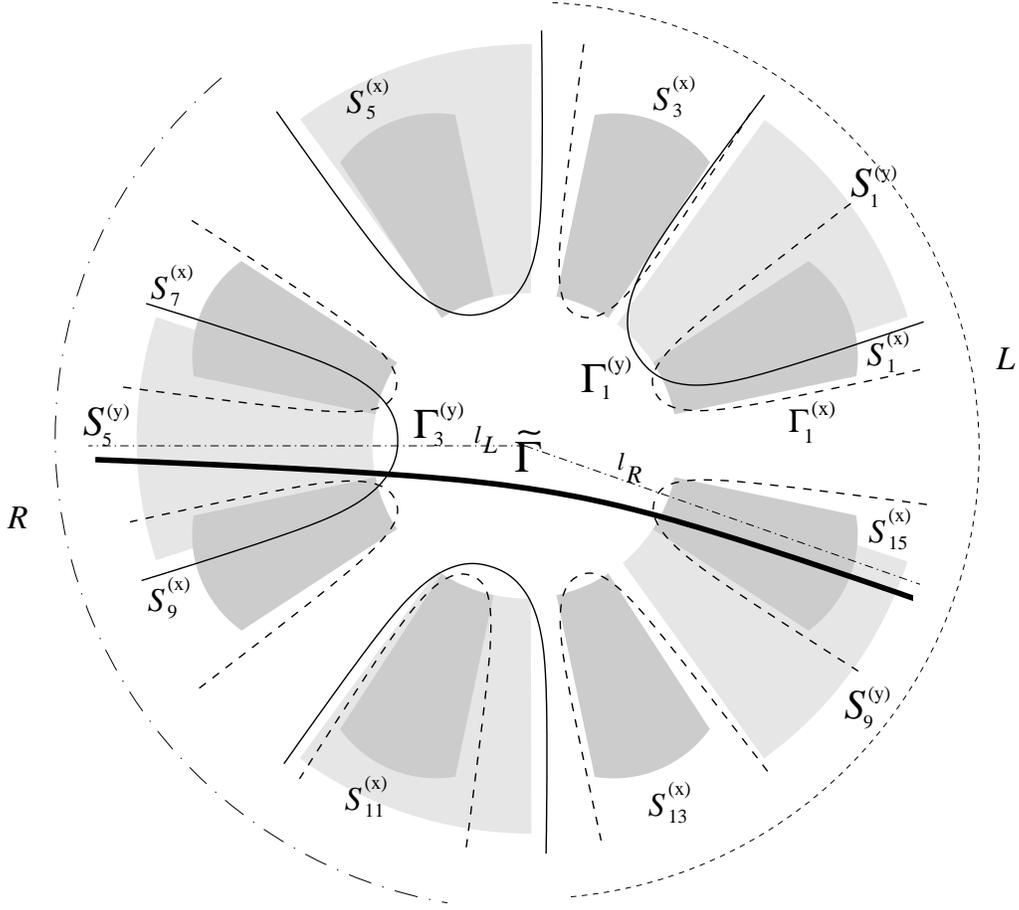}
\end{center}
\caption{The figure illustrating the definition of
$\ds{\m{\Theta}_\infty\!^{(Left,Right)}(y)}$. The contour
$\widetilde \Gamma$ must approach $\infty$ within the sectors
$\mathcal S_{2k+1}^{(y)}$'s (lighter shaded) and asymptotic to the two
rays $\ell_{\mathcal L}, \ell_{\mathcal R}$. The terms added to
$\ds{\m{\un\Psi}_\infty\!^{(0)}}$ in the definition of
$\ds{\m{\Theta}_\infty\!^{(Left)}}$
($\ds{\m{\Theta}_\infty\!^{(Right)}}$)  must correspond to sectors
$\mathcal S_{2j+1}^{(x)}$ which fall in the dual sector $\mathcal L$
($\mathcal R$) indicated by the  dashed (dash-dotted) arc.}
\label{examp}
\end{figure}

\renewcommand{\theequation}{\Alph{section}-\arabic{subsection}-\arabic{equation}}
\setcounter{equation}{0}
The example we consider consists of two potentials $V_1(x)$ and $V_2(y)$ of
degrees $8$ and $5$ respectively, both with positive real leading
coefficients. We  refer to Fig. \ref{examp} as the ``the figure''  throughout
this section.
The $\mathcal S_y^{(k)}$ sectors are the light-shaded ones while the
$\mathcal S_x^{(j)}$ are the darker-shaded ones.
Fix an asymptotic  ray $\ell_{\mathcal L}$ with $\arg(y)= \alpha_0 (=\pi)$ within an
odd-numbered sector, in our example $\mathcal S_y^{(5)}$,  and an
$\epsilon$
sector around it. Let $\mathcal L$ be the dual sector, i.e., the
sector centered around $\pi-\alpha_0 (=0)$ with width $\pi-\epsilon$.
Any of the functions $\ds{\m{\un\Psi}_\infty\!^{(j)} (y)}$ whose wedge
contours $\Gamma_j^{(x)}$ lie in the sector $\mathcal L$ decay
faster than any exponential ${\rm e}^{-My}$ on the given ray
$\ell_{\mathcal L}$.
These are the sectors $\mathcal S_x^{(1,3,13,15)}$ in Fig. \ref{examp}.
Fix another ray $\ell_{\mathcal R}$ within a different odd-numbered
sector $\mathcal S_y^{(2k+1)}$  in such a way that the dual sector to the
$\epsilon$ sector around $\ell_{\mathcal R}$ (denoted by $\mathcal R$)
contains  all but one of the remaining sectors $\mathcal S_x^{(2j+1)}$. 
In the choice shown in the figure $\ell_{\mathcal R}$ lies in
$\mathcal S_9^{(y)}$ and the sector $\mathcal R$ contains 
the sectors $\mathcal S_x^{(7,9,11)}$.
Finally, fix a contour $\widetilde \Gamma$ which goes off to infinity
asymptotic to the two rays $\ell_{\mathcal L},\ \ell_{\mathcal R}$ (as
in figure).

With these choices, redefine the duality pairing according
to the following rule:\\
 (i) In the $x$-plane use as basis only the contours $\Gamma_j^{(x)}$ which
fall within one or the other of the sectors $\mathcal L,\ \mathcal R$. In our
example this leaves out contour $\Gamma_3^{(x)}$, which has not been drawn in
the figure. \\
(ii) In the $y$-plane use a basis in which the contours $\widetilde\Gamma$
intersect only one of the $\Gamma_k^{(y)}$s. In our example this would leave
out either $\Gamma_3^{(y)}$ or $\Gamma_0^{(y)}$ and we have chosen the latter.

Note that we can accomplish an equivalent duality pairing using such
bases by exploiting the homological equations
\be
\sum_{j=0}^{d_1} \Gamma_j^{(x)} = \sum_{k=0}^{d_2} \Gamma_k^{(y)} = 0.
\label{homologyconstr}
\ee
Indeed, 
using the homological equation (\ref{homologyconstr}), one can avoid one contour
and redefine the $\varkappa$'s:
\be
\varkappa\Gamma = \sum_{i\neq 0,j\neq 0} \varkappa^{i,j}
\Gamma_i^{(x)}\times
\Gamma_j^{(y)}
= \sum_{i\neq 3,j\neq 0} \widetilde\varkappa^{i,j} \Gamma_i^{(x)} \times
\Gamma_j^{(y)} ,
\ee
where $\widetilde\varkappa^{i,j}$ in this example would be
\be
\widetilde\varkappa^{i,j} :=\varkappa^{i,j}-\varkappa^{3,j} \ ,\ \ i\neq
3,0; \ \ 
\widetilde\varkappa^{0, j} :=-\varkappa^{3,j}\ .
\ee
Now define the wave-vector
\be
\m{\mathbf \Theta}_{\infty}\!^{Left} (y) := \m{\un\Psi}_\infty\!^{(0),t}
(y) 
-2i\pi\!\!\!\!\sum_{i= 0,1,2,7}\!\!\! \widetilde\varkappa^{i,3}
\m{\un\Psi}_\infty\!^{(i),t} (y)=:
 \m{\un\Psi}_\infty\!^{(0),t} (y)
-2i\pi\sum_{i\in I_{\mathcal L}} \widetilde\varkappa^{i,3}
\m{\un\Psi}_\infty\!^{(i),t} (y)\ ,\label{adue}
\ee
for $y$ belonging to the part of $\widetilde \Gamma$ inside contour
$\Gamma_3^{(y)}$. (Note the transposition of row vectors to  column vectors.)
In eq. (\ref{adue}) in general $I_{\mathcal L}$
 would be the set of indices $j$ corresponding to
the contours $\Gamma_j^{(x)}$ contained in sector $\mathcal L$.
 Note that this function decays on the ray $\ell_{\mathcal L}$ because $\ell_{\mathcal L}$
is in the dual sector of all the $\ds{\m{\Psi}_\infty\!^{(j)}}$ entering
eq. (\ref {adue}). Moreover the piecewise analytic function
$\ds{\m{\Psi}_\infty\!^{(0)}}$   decays exponentially in each
sector $\mathcal S_k^{(y)}$.\par  
As we cross the contour $\Gamma_3^{(y)}$ along $\widetilde \Gamma$, 
by virtue of the jump condition (\ref{ajump}), the function
$\ds{\m{\mathbf
\Theta}_{\infty}\!^{Left} (y)}$ is analytically continued to the function 
\be
\m{\mathbf \Theta}_{\infty}\!^{Right} (y) :=  \m{\un\Psi}_\infty\!^{(0),t}
(y)
+2i \pi\!\!\sum_{j\not\in L} \!\widetilde
\varkappa^{j,3}\m{\un\Psi}_\infty\!^{(j),t} (y) := 
\m{\un\Psi}_\infty\!^{(0),t} (y)
+2i \pi\!\!\!\! \sum_{j=4,5,6}\!\!\! \widetilde
\varkappa^{j,3}\m{\un\Psi}_\infty\!^{(j),t} (y)\ , 
\ee
which, due to the choices made above,  also decays  along the asymptotic
direction $\ell_{\mathcal R}$.  The two functions $\ds{\m{\mathbf
\Theta}_{\infty}\!^{Left,Right} (y)}$ define together a function 
$\ds{\m{\mathbf \Theta}_{\infty}\!^{\widetilde \Gamma} (y)}$ which is
(locally) analytic
in a neighborhood of the contour $\widetilde \Gamma$ and decays
sufficiently fast in either direction to allow us to take an inverse
Fourier-Laplace transform using
\be
\m{\Psi}_\infty\!^{(\widetilde\Gamma)} (x):= \int_{\widetilde \Gamma}
\!\!{\rm d}y \, {\rm e}^{-\frac {xy}\hbar}\m{\mathbf
\Theta}_{\infty}\!^{(\widetilde
\Gamma)} (y)\ .
\ee

This wave-vector is a solution of eqs. (\ref{modifrec2}),
(\ref{defupsi}) and (\ref{defvpsi}) as we now show.
The equations
\bea
&&y \m{\mathbf \Theta}_\infty\!^{(\widetilde\Gamma)} (y) =
P\m{\mathbf \Theta}_\infty\!^{(\widetilde\Gamma)} (y)  + \le[\sqrt{h_0}
{\rm
e}^{\frac 1 \hbar V_2(y)},0,\dots\ri]^t\\
&&\hbar \pa_y\m{\mathbf \Theta}_\infty\!^{(\widetilde\Gamma)} (y) =
Q\m{\mathbf \Theta}_\infty\!^{(\widetilde\Gamma)} (y)  +
W_2(y) \le[\sqrt{h_0} {\rm
e}^{\frac 1 \hbar V_2(y)},0,\dots\ri]^t\\
&& \hbar \pa_{u_K} \m{\mathbf \Theta}_\infty\!^{(\widetilde\Gamma)} (y)  =
\mathrm U^K
\m{\mathbf \Theta}_\infty\!^{(\widetilde\Gamma)} (y) \\
&&\hbar  \pa_{v_J} \m{\mathbf \Theta}_\infty\!^{(\widetilde\Gamma)} (y)  =
-\mathrm V^J
\m{\mathbf \Theta}_\infty\!^{(\widetilde\Gamma)} (y) - \frac 1 J \frac
{P^J-y^J}{P-y}  \le[\sqrt{h_0} {\rm
e}^{\frac 1 \hbar V_2(y)},0,\dots\ri]^t\ ,
\eea
are satisfied as a consequence of the similar equations for
$\ds{\m{\un\Psi}_\infty\!^{(0)}(y)}$ and
$\ds{\m{\un\Psi}_\infty\!^{(k)}(y)}$, $k=1,\dots d_1$.
From these, eqs. (\ref{modifrec2}), (\ref{defupsi}) and (\ref{defvpsi}) for
$\ds{\m{\Psi}_\infty\!^{(\widetilde
\Gamma) }(x)}$ follow from standard manipulations of the inverse
Fourier--Laplace transform which are now made rigorous by the
convergence of the integral. 
In particular this solution of (\ref{modifrec2}) is associated with
the function 
\be
f^{(\widetilde \Gamma)}(x) := \int_{\widetilde\Gamma}\!\!{\rm
d}y\,{\rm e}^{\frac 1 \hbar(V_2(y)-xy)}\ .
\ee

One should repeat the scheme outlined here for other contours $\widetilde
\Gamma$ as well, until they span the same homology space spanned by the
antiwedge contours $\{\widetilde \Gamma_j^{(y)}\}$, so as to obtain a basis of
solutions to eq. (\ref{modifrec1}). Just to complete the given example we
indicate how to choose the other contours and corresponding bases:
\begin{enumerate}
\item The contour asymptotic to the rays $\arg(y)=-\pi$ and $\arg (y) =
\pi/10+\epsilon$. The basis on which to re-define the coefficients
$\varkappa$ (which we now denote by $\widetilde \varkappa$) 
is given by $\Gamma_i^{(x)}\times \Gamma_j^{(y)}$, $i\neq 6$,
$j\neq 1$ and the wave vectors are given by
\bea
\m{\mathbf \Theta}_{\infty}\!^{Left} (y) := \m{\un\Psi}_\infty\!^{(0),t}
(y)
-2i\pi\!\!\!\!\sum_{i= 0,1,2,7}\!\!\! \widetilde\varkappa^{i,3}
\m{\un\Psi}_\infty\!^{(i),t} (y) \\
\m{\mathbf \Theta}_{\infty}\!^{Right} (y) :=
\m{\un\Psi}_\infty\!^{(0),t} (y)
+2i\pi\!\!\!\!\sum_{j=3,4,5}\!\! \widetilde
\varkappa^{j,3}\m{\un\Psi}_\infty\!^{(j),t} (y)\ .
\eea
\item The contour asymptotic to the rays $\arg(y)=\pi/2+\epsilon$ and
$\arg (y) = 3\pi/2-\epsilon$. 
The basis on which to re-define the coefficients
$\varkappa$ is given by $\Gamma_i^{(x)}\times \Gamma_j^{(y)}$, $i\neq 3$,
$j\neq 4$.
\bea
\m{\mathbf \Theta}_{\infty}\!^{Left} (y) := \m{\un\Psi}_\infty\!^{(0),t}
(y)
-2i\pi\!\!\!\!\sum_{i= 0,1,4} \!\!\!\widetilde\varkappa^{i,2}
\m{\un\Psi}_\infty\!^{(i),t} (y) \\
\m{\mathbf \Theta}_{\infty}\!^{Right} (y) :=
\m{\un\Psi}_\infty\!^{(0),t} (y)
+2i\pi\!\!\!\!\sum_{j=5,6,7,0}\!\!\! \widetilde \varkappa^{j,2}\m{\un\Psi}_\infty\!^{(j),t} (y)\ .
\eea
\item The contour asymptotic to the rays $\arg(y)=\pi/2+\epsilon$ and
$\arg (y) = 17 \pi/10+\epsilon$. 
 The basis on which to re-define the coefficients
$\varkappa$ is given by $\Gamma_i^{(x)}\times \Gamma_j^{(y)}$, $i\neq 0$,
$j\neq 0$.
\bea
\m{\mathbf \Theta}_{\infty}\!^{Left} (y) := \m{\un\Psi}_\infty\!^{(0),t}
(y)
-2i\pi\!\!\!\!\sum_{i= 1,2,3,4}\!\!\! \widetilde\varkappa^{i,2}
\m{\un\Psi}_\infty\!^{(i),t} (y) \\
\m{\mathbf \Theta}_{\infty}\!^{Right} (y) :=
\m{\un\Psi}_\infty\!^{(0),t} (y)
+2i\pi\!\!\!\!\sum_{j=5,6,7}\!\!\!
 \widetilde \varkappa^{j,2}\m{\un\Psi}_\infty\!^{(j),t} (y)\ .
\eea
\end{enumerate}
 We leave the details of the general case to subsequent work.



\begin{thebibliography}{99} 

\bibitem{ansatz} M. Bertola, B. Eynard, J. Harnad, ``An ansatz for the 
solution of the Riemann--Hilbert problem for biorthogonal 
polynomials'', in preparation. 
 
\bibitem{BEH} M. Bertola, B. Eynard, J. Harnad, ``Duality, 
Biorthogonal Polynomials and Multi--Matrix Models'', 
{\it  Commun. Math. Phys.}, , Commun. Math. Phys. {\bf 229}, 
 1, 73-120 (2002).
 
\bibitem{Needs} M. Bertola, B. Eynard, J. Harnad, ``Duality of 
spectral curves arising in two-matrix models'' 
{\it  Theor. Math. Phys.}  (2003, in press) (NEEDS 2001 proceedings)  
nlin.SI/0112006.
 
\bibitem{AMSE} M. Bertola, B. Eynard, J. Harnad, ``Genus zero large n
asymptotics of bi-orthogonal polynomials involved in the random 2-matrix
model'', Presentation by B.E. at AMS northeastern regional meeting, Montreal,
May 3-5, 2002. 

\bibitem{Marco} M. Bertola, ``Bilinear semi--classical moment 
functionals and their integral representation'', CRM-2842
(2002), math.CA/0205160. 

\bibitem {BlIt} P. Bleher, A. Its, ``Semiclassical asymptotics of 
orthogonal polynomials, Riemann-Hilbert problem,  
and universality in the matrix model'' {\em Ann. of Math.} 
 (2) {\bf 150}, no. 1, 185--266 (1999).  
 
\bibitem{bonnet}  G. Bonnet, F. David, B. Eynard,  
``Breakdown of universality in multi-cut matrix models'', 
{\em  J. Phys. A} {\bf 33} 6739-6768  (2000). 
 
\bibitem{chihara} T. S. Chihara, ``An introduction to orthogonal 
polynomials'',  Mathematics and its Applications, Vol. {\bf 13} 
 Gordon and Breach Science Publishers, New 
York-London-Paris, 1978. 

\bibitem{KazakoVDK} J.M. Daul, V. Kazakov, I.K. Kostov, ``Rational Theories of  
2D Gravity from the Two-Matrix Model'', {\em Nucl. Phys.} {\bf B409}, 311-338  
(1993), hep-th/9303093. 
 
\bibitem{dkmvz} P. Deift, T. Kriecherbauer, K. T. R. McLaughlin, 
S. Venakides, Z. Zhou, ``Uniform asymptotics for polynomials 
orthogonal with respect to varying exponential weights and 
applications to universality questions in random matrix theory'', {\it 
Commun. Pure Appl. Math.} {\bf 52}, 1335--1425 (1999). 
 
\bibitem{dkmvz2} P. Deift, T. Kriecherbauer, K. T. R. McLaughlin, 
S. Venakides, Z. Zhou, ``Strong asymptotics of orthogonal polynomials 
with respect to exponential weights'', {\it Commun. Pure Appl. Math.} 
{\bf 52}, 1491--1552, (1999). 
 
\bibitem{ZJDFG} P. Di Francesco, P. Ginsparg, J. Zinn-Justin, ``2D Gravity and  
Random Matrices'', {\em Phys. Rep.} {\bf 254}, 1 (1995). 
 
\bibitem{McLaughlin} N. M. Ercolani and K. T.-R. McLaughlin ``Asymptotics and  
integrable structures for biorthogonal polynomials associated to a random  
two-matrix model'', {\em Physica D},  {\bf 152-153}, 232-268 (2001). 
 
\bibitem{eynard} B. Eynard, ``Eigenvalue distribution of large random matrices, 
 from one matrix to several coupled matrices'' {\em Nucl. Phys. B }{\bf 506},  
633 (1997), cond-mat/9707005. 
 
\bibitem{eynardchain} B. Eynard, ``Correlation functions of eigenvalues of  
multi-matrix models, and the limit of a time dependent matrix'',  
{\em J. Phys. A: Math. Gen.} {\bf 31}, 8081 (1998), cond-mat/9801075. 

 
\bibitem{eynardmehta} B. Eynard, M.L. Mehta, ``Matrices coupled in a chain:  
eigenvalue correlations'', {\em J. Phys. A: Math. Gen.} {\bf 31}, 4449 (1998), 
cond-mat/9710230.  

\bibitem{FIK} A. Fokas, A. Its, A. Kitaev, ``The isomonodromy approach 
to matrix models in 2D quantum gravity'', {\em Commun. Math. Phys.} 
{\bf 147}, 395--430 (1992).  

\bibitem{Guionnet} A. Guionnet, Zeitouni, ``Large deviations
asymptotics for spherical integrals'', 
{\em J. F. A.} {\bf 188}, 461--515 (2002).

\bibitem{isomIts} A. R. Its, A. V.  Kitaev, A. S. Fokas  ``Matrix
models of two-dimensional quantum gravity and isomonodromy solutions of ``discrete
Painleve equations'' '',{\em  Zap. Nauch. Sem. LOMI}, {\bf  187}, 3-30
(1991) (Russian), translation in {\em  J. Math. Sci.} {\bf 73}, no. 4, 415--429  (1995). 
 
\bibitem{stringIts}  A.R. Its, A.V. Kitaev, and A.S. Fokas,  ``An  
isomonodromic Approach in the  Theory of Two-Dimensional Quantum Gravity'', {\em
Usp. Matem. Nauk}, {\bf 45}, 6 (276 ),  135-136 (1990), (Russian), translation
in {\em Russian Math. Surveys},  {\bf 45}, no. 6, 155-157 (1990). 
 
\bibitem{kapa} A. A. Kapaev, ``The Riemann--Hilbert problem for the 
bi-orthogonal polynomials'', nlin.SI/0207036. 

\bibitem{Kazakov} V.A. Kazakov, ``Ising model on a dynamical planar random 
lattice: exact solution'', {\em Phys Lett.} {\bf A119}, 140-144 (1986).  
 
\bibitem{matytsin} A. Matytsin, ``on the large $N$ limit of the Itzykson Zuber
 Integral'', {\em Nuc. Phys.} {\bf B411}, 805 (1994), hep-th/9306077.

\bibitem{Mehta} M.L. Mehta, ``Random Matrices, second edition'', Academic
Press, New York, 1991.


\bibitem{Szego} G. Szeg\"o ``Orthogonal Polynomials'', AMS, Providence,  
Rhode Island, (1939). 
 
\bibitem{UT} K. Ueno and K. Takasaki, ``Toda Lattice Hierarchy'', 
{\it Adv. Studies Pure Math.} {\bf 4}, 1--95 (1984). 
 
\end{thebibliography}
\end{document}